\documentclass[fleqn,10pt]{wlscirep}
\usepackage[utf8]{inputenc}
\usepackage[T1]{fontenc}
\title{Prompt Mechanisms in Medical Imaging: A Comprehensive Survey}

\author[1,+]{Hao Yang}
\author[2,3,+]{Xinlong Liang}
\author[4,+]{Zhang Li}

\author[1]{Yue Sun}
\author[5,6]{Zheyu Hu}
\author[1]{Xinghe Xie}
\author[2,7]{Behdad Dashtbozorg}
\author[1]{Jincheng Huang}
\author[1]{Shiwei Zhu}
\author[2,3]{Luyi Han}
\author[8]{Jiong Zhang}
\author[9]{Shanshan Wang}

\author[2,3,*]{Ritse Mann}
\author[4,*]{Qifeng Yu}
\author[1,*]{Tao Tan}

\affil[1]{Faculty of Applied Sciences, Macao Polytechnic University, Rua de Luis Gonzaga Gomes, Macao, China}
\affil[2]{Netherlands Cancer Institute, Department of Radiology, Amsterdam, 1066 CX, The Netherlands}
\affil[3]{Radboud University Medical Centre, Department of Radiology and Nuclear Medicine, Nijmegen, 6525 GA, The Netherlands}

\affil[4]{College of Aerospace Science and Engineering, National University of Defense Technology, Changsha, China}

\affil[5]{Medical Department of Breast Cancer, Hunan Cancer Hospital, Changsha, China}
\affil[6]{Medical Department of Breast Cancer, the Affiliated Cancer Hospital of Xiangya School of Medicine, Central South University, Changsha, China.}

\affil[7]{Faculty of Biomedical Engineering, Eindhoven University of Technology,
Antonie van Leeuwenhoek, Plesmanlaan 121, Eindhoven, Noord Brabant,
NL}

\affil[8]{Laboratory of Advanced Theranostic Materials and Technology, University of Chinese Academy of Sciences, China}

\affil[9]{Paul C. Lauterbur Research Center for Biomedical Imaging, Shenzhen Institute of Advanced Technology, Chinese Academy of Sciences, China}

\affil[*]{corresponding authors: Ritse.Mann@radboudumc.nl; yuqifeng@nudt.edu.cn; taotanjs@gmail.com}

\affil[+]{these authors contributed equally to this work}


\begin{abstract}
Deep learning offers transformative potential in medical imaging, yet its clinical adoption is frequently hampered by challenges such as data scarcity, distribution shifts, and the need for robust task generalization. Prompt-based methodologies have emerged as a pivotal strategy to guide deep learning models, providing flexible, domain-specific adaptations that significantly enhance model performance and adaptability without extensive retraining. This systematic review critically examines the burgeoning landscape of prompt engineering in medical imaging. We dissect diverse prompt modalities, including textual instructions, visual prompts, and learnable embeddings, and analyze their integration for core tasks such as image generation, segmentation, and classification. Our synthesis reveals how these mechanisms improve task-specific outcomes by enhancing accuracy, robustness, and data efficiency and reducing reliance on manual feature engineering while fostering greater model interpretability by making the model's guidance explicit. Despite substantial advancements, we identify persistent challenges, particularly in prompt design optimization, data heterogeneity, and ensuring scalability for clinical deployment. Finally, this review outlines promising future trajectories, including advanced multimodal prompting and robust clinical integration, underscoring the critical role of prompt-driven AI in accelerating the revolution of diagnostics and personalized treatment planning in medicine.
\end{abstract}
\begin{document}

\flushbottom
\maketitle
%
%
\thispagestyle{empty}


\section*{Introduction}
The integration of Artificial Intelligence (AI), intense learning models such as convolutional neural networks (CNNs), has revolutionized medical image analysis, producing remarkable advancements in image classification, segmentation, and generation \cite{mattjie2023zero, akrout2023diffusion, pinaya2022brain, fischer2024prompt}. Despite these successes, clinical translation and widespread adoption of these models are often impeded by persistent challenges such as data scarcity and distribution shifts. Among these, distribution shifts between training datasets and various real-world clinical data present a significant obstacle, often leading to a degradation in model performance and reliability \cite{yan2024prompt}. These shifts arise from inherent dataset biases, inter-scanner variability, and diverse image acquisition protocols. Consequently, improving model adaptability and robustness to such heterogeneous data distributions has become a significant research imperative in medical image processing \cite{chambon2022adapting, liu2023clip, chen2024sam, gao2023desam, cheng2023sam, zhang2023text, zhu2024memory, silva2025foundation}.

The recent advent of Foundation Models (FMs), large-scale models pre-trained on extensive and diverse datasets, underpinned by architectures such as the Transformer and encompassing architectures such as Large Language Models (LLMs), and crucially for this field, Vision-Language Models (VLMs), offers a transformative new paradigm for AI in medical imaging \cite{ye2023uniseg}. These FMs exhibit superior versatility and transfer learning capabilities, demonstrating immense potential in few-shot learning, cross-modal information fusion, and robust performance in complex data-scarce scenarios. However, effectively harnessing the power of these generalist FMs for specialized medical tasks requires sophisticated adaptation strategies beyond traditional fine-tuning. In this context, prompt-based mechanisms have rapidly emerged as a key and powerful approach to steering FMs. Concurrently, the underlying principles of prompting-guiding model behavior with external contextual information have also been increasingly explored and adapted for a broader range of deep learning architectures in medical imaging, extending beyond large-scale FMs. These mechanisms are proving particularly effective in addressing the aforementioned distribution shifts, managing complex specialized medical knowledge, and unlocking the full potential of FM and other customized models for applications in precision medicine and personalized diagnostics.



\begin{figure}[ht]
\centering
\includegraphics[width=\linewidth]{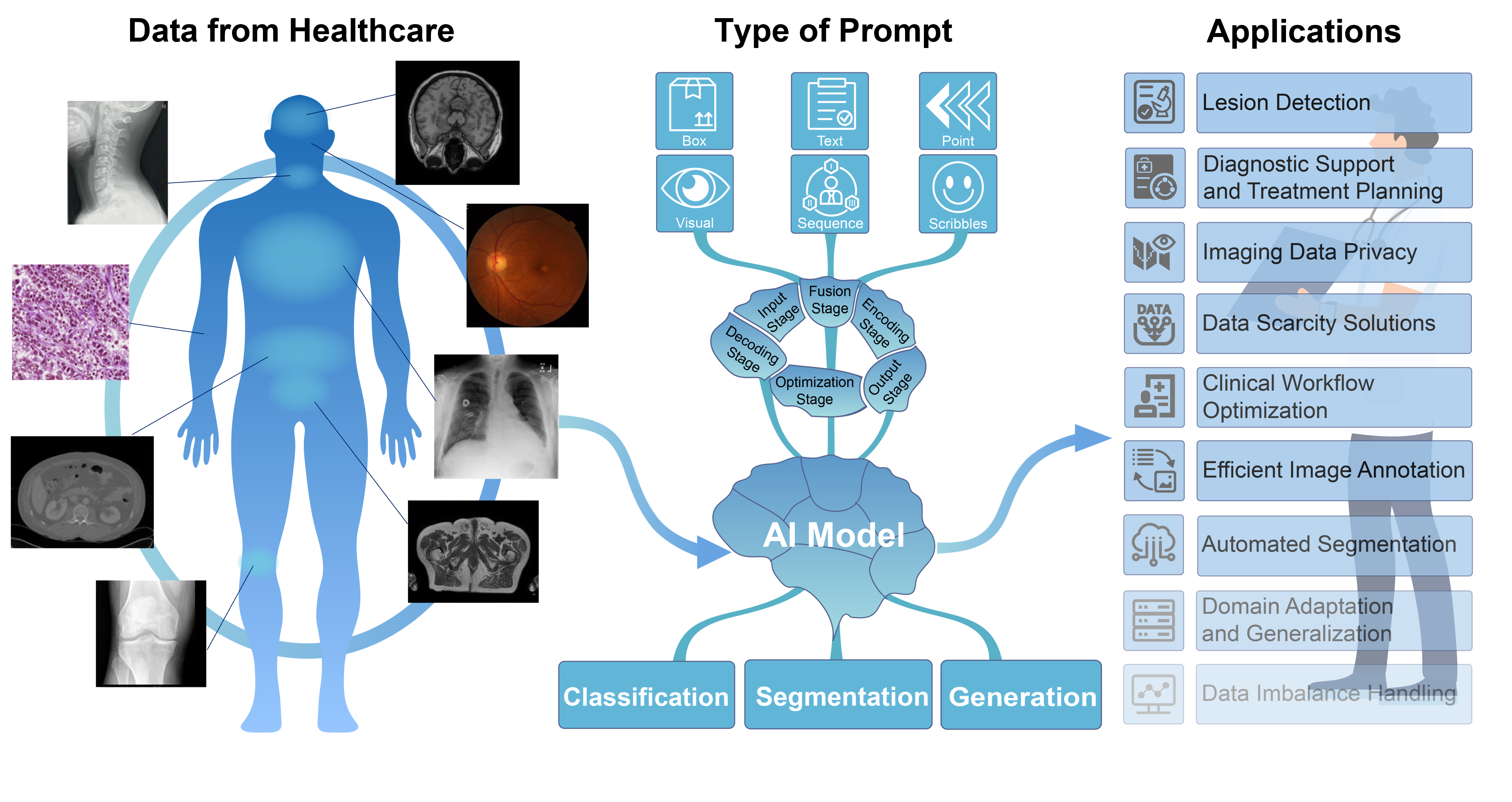}
\caption{An overview of the framework for prompt-based AI in medical imaging. The process begins with various medical imaging data sources, which are processed by an AI model capable of accepting multiple prompt types (visual, text, sequence, etc.). By performing classification, segmentation, or generation tasks, the model ultimately serves a range of clinical applications to improve diagnostic efficiency and accuracy.}
\label{fig:Overview}
\end{figure}

\begin{figure}[ht]
\centering
\includegraphics[width=0.6\linewidth]{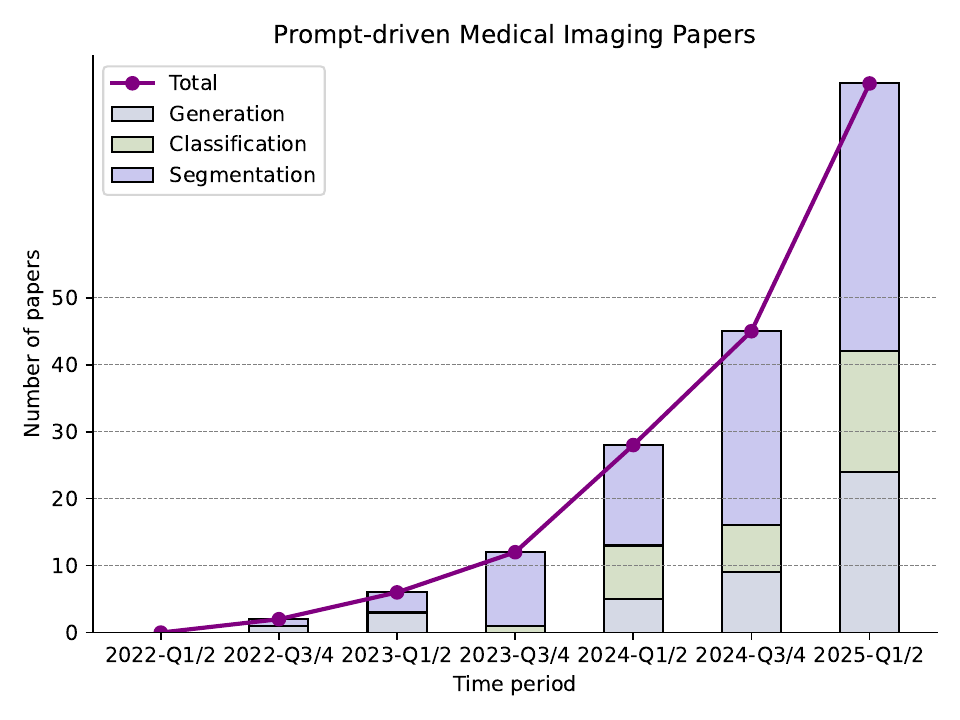}
\caption{Rapid increase of the number of Prompt-driven Medical imaging papers. Generation, Classification, and Segmentation are the three main taxonomy categories introduced in this survey.}
\label{fig:papers_num}
\end{figure}

The core tenet of prompting is to guide and optimize a model's performance on specific downstream tasks by introducing targeted external contextual information. As illustrated in the general framework of figure~\ref{fig:Overview}, this process begins with various sources of medical imaging data. An AI model, designed to be receptive to multiple prompt types (visual, text, etc.), is then guided by these targeted inputs. The prompt leads the model to perform a specific task, such as classification, segmentation, or generation, ultimately serving a range of clinical applications by enhancing diagnostic accuracy and efficiency. This information can be encoded as textual instructions \cite{chambon2022adapting}, visual prompts (e.g., points, boxes) \cite{cao2024domain}, or learnable embeddings. This strategy aims to significantly augment the model's capacity to interpret and analyze medical images with high fidelity, especially when annotated data is scarce or expensive. A key technical advantage of prompting lies in its remarkable parameter efficiency: it often necessitates fine-tuning only a small subset of the model's parameters-or none at all in zero-shot prompting scenarios-to adapt the pre-trained model to new tasks, thereby obviating the need for substantial architectural modifications or complete retraining from scratch \cite{fischer2024prompt}. Furthermore, prompting offers exceptional flexibility and adaptability, with diverse guidance strategies enabling models to tackle a broad spectrum of medical imaging tasks and their associated challenges \cite{zhou2024medsam}. By facilitating the integration of additional knowledge and context, prompting can significantly enhance the generalization and robustness of the model, particularly when faced with limited sample sizes \cite{chambon2022adapting, liu2023clip, chen2024sam, yan2024prompt}. Recent advances in this growing field showcase increasingly sophisticated techniques, including multimodal prompts that fuse information from various channels (e.g., image and text) to tackle complex problems \cite{singhal2022large}, adaptive prompting algorithms that dynamically tailor prompt content based on specific task requirements and data characteristics \cite{zhou2024medsam}, and personalized prompting schemes that customize inputs for individual images to mitigate issues arising from sample distribution discrepancies, thereby enhancing overall system performance \cite{wahd2024sam2rad}.

Despite these promising developments and the rapid proliferation of research, the practical application and optimal design of prompt mechanisms in medical imaging are still confronted by several critical limitations deeply intertwined with the nature of prompting. While prompting is often positioned as a solution for data scarcity~\cite{zhang2023text}, the design and validation of high-quality prompts introduce their data-dependent challenges. Crafting effective textual prompts that capture nuanced clinical knowledge or generating representative learnable embeddings requires a sufficient volume of diverse and meticulously annotated data~\cite{chambon2022adapting, liu2023clip, chen2024sam, yan2024prompt}, the availability of which is curtailed by stringent privacy regulations. The complexities of medical annotation further undermine the quality and reliability of prompts; the process demands specialized expertise and is prone to inter-observer variability. This subjectivity translates directly into inconsistent or ambiguous prompts-whether textual, visual, or learned-which can mislead the model and degrade performance, complicating the development of standardized and reproducible prompting protocols.

Crucially, the challenge of distribution shift~\cite{yan2024prompt} extends beyond the model's core to the prompts themselves. A prompt that is effective for one data distribution (e.g., images from a specific scanner or institution) may fail dramatically on another, a phenomenon that can be described as prompt brittleness. This lack of robustness is a significant barrier, as prompts must be generalizable across different patient populations and the notable domain gaps among imaging modalities (e.g., CT, MRI, X-ray, Ultrasound). These multifaceted factors-spanning prompt design, quality, and robustness-collectively constrain the generalizability and clinical applicability of prompt-guided models, including FMs. This underscores the urgent need for research into more adaptive, robust, and systematically designed prompting strategies that can overcome these inherent obstacles.

Given the rapid advancements and the critical need for consolidated knowledge in this burgeoning area, this paper reviews prompt engineering tailored explicitly for medical imaging. The urgency and relevance of this review are underscored by the exponential growth in related academic publications, as depicted in figure~\ref{fig:papers_num}. The data clearly shows a steep increase in the number of papers on prompt-driven medical imaging, with significant contributions across the primary tasks of generation, classification, and segmentation. This trend highlights the intense research interest in the field and necessitates a timely, structured synthesis of existing work. We aim to comprehensively analyze the diverse methodologies and widespread applications of prompt mechanisms across various model architectures. Our approach involves an extensive literature search to gather and synthesize relevant studies, followed by developing and applying a novel classification system. This system categorizes existing research into two primary dimensions: the core technologies underpinning prompt mechanisms (encompassing design, generation, integration strategies, and their synergy with transfer learning and multimodal learning) and their clinical application paradigms (evaluating effectiveness in classification, segmentation, generation, and their impact on diagnostic accuracy and treatment planning). This review summarizes the distinct advantages and inherent limitations of various prompting methods through a detailed, multi-faceted analysis of the collated literature. It critically evaluates their performance across multiple crucial dimensions, including accuracy, robustness, and data utilization efficiency. Our significant contributions are:
\begin{itemize}
\item \textbf{Systematic and Comprehensive Review:} Providing a structured and in-depth survey of the application and methodology of prompt mechanisms in medical imaging, integrating theoretical advancements with practical applications, and underscoring their transformative potential in image generation, segmentation, and classification tasks.
\item \textbf{Novel Classification Framework and Trend Analysis:} Introducing an innovative classification system that categorizes research into core technologies and clinical applications. This framework delineates current trends and innovations in prompt design, generation, and embedding techniques, offering clear and actionable directions for future research endeavors.
\item \textbf{Identification of Current Challenges and Future Trajectories:} Delivering a thorough summary of the extant research challenges, including issues related to prompt quality, diversity, model scalability, and seamless integration into clinical workflows. Alongside this, we propose promising future development paths, such as advanced multimodal data integration, the pursuit of personalized medical solutions through tailored prompting, and breakthroughs in few-shot or zero-shot learning capabilities.
\end{itemize}

\section*{Theoretical Foundations and Taxonomies of Prompt Mechanisms}
Prompt-based mechanisms represent a significant paradigm shift in guiding deep learning models, particularly for nuanced tasks in medical imaging where precision and domain-specific knowledge are paramount. Understanding these mechanisms' foundational principles and systematic classification is crucial for their practical design and application. This section outlines the core theoretical underpinnings of prompt engineering and presents a taxonomy for categorizing the diverse array of prompting strategies.

\subsection*{Theoretical Foundations}

\subsubsection*{Core Theoretical Principles}
Fundamentally, prompts act as conditioning signals that modulate neural network behavior, enabling precise output control, often without requiring extensive architectural modifications or complete retraining. From an information-theoretic point of view, prompts can be seen as injections of prior knowledge, such as domain-specific terminology, spatial cues, or task objectives, that constrain the hypothesis space, thus enhancing task-specific performance while aiming to maintain generalizability. The theoretical basis for prompts draws from established concepts such as transfer learning, where prompts facilitate knowledge transfer from pre-trained models; a Bayesian inference perspective, viewing prompts as informative priors guiding model learning; and manifold learning, where prompts help navigate the learned latent spaces of models towards desired outputs.

\subsubsection*{Key Operational Mechanisms}
The efficacy of prompts comes from several key operational mechanisms. Attention modulation is primary, where prompts guide the model's internal attention to focus on task-relevant features while suppressing irrelevant information. Feature space transformation occurs when prompts induce changes in the model's representational geometry, projecting inputs into regions corresponding to desired outputs, often via parameter-efficient fine-tuning methods like prompt tuning, prefix tuning, or Low-Rank Adaptation, which apply low-rank updates to the model's weight matrices. Finally, conditional computation pathways may emerge in advanced architectures, where prompts activate specific computational subgraphs, enabling task-specialized processing.

\subsection*{Prompt Mechanisms in Medical Image Generation} 

Medical image generation, aiming to synthesize clinically and anatomically realistic images via deep learning, frequently necessitates meticulous conditioning to align outputs with specific clinical requirements or anatomical verisimilitude. Prompt mechanisms have become indispensable for achieving this alignment, effectively guiding generative models toward producing clinically relevant and structurally sound medical imagery. This subsection details the primary categories of prompts employed in this domain, highlighting how the general principles discussed previously are instantiated.

\subsubsection*{Text-Guided Synthesis: Leveraging Semantic Descriptions} 


Textual prompts are critical conditional controls in medical image generation, particularly for steering diffusion models and other generative architectures. These prompts, often derived from unstructured clinical narratives, professional medical reports (e.g., pathology reports~\cite{yellapragada2024pathldm}), or specific textual annotations detailing disease backgrounds~\cite{chambon2022adapting} or specific pathological features~\cite{xu2024medsyn}, encapsulate vital pathological details and contextual information. This semantic guidance ensures that generated images possess practical clinical significance and anatomical plausibility, which are crucial in fields like dermatology, chest X-ray synthesis, and 3D/4D volumetric data generation. Textual conditions furnish essential prior knowledge and empower developers to precisely direct the generation process by articulating specific disease characteristics. Text-driven synthesis typically relies on cross-attention mechanisms integrated within diffusion models to achieve high-fidelity medical image outputs.

\paragraph{Core Technical Pipeline for Textual Prompting}

\begin{figure}[ht]
\centering
\includegraphics[width=0.8\linewidth]{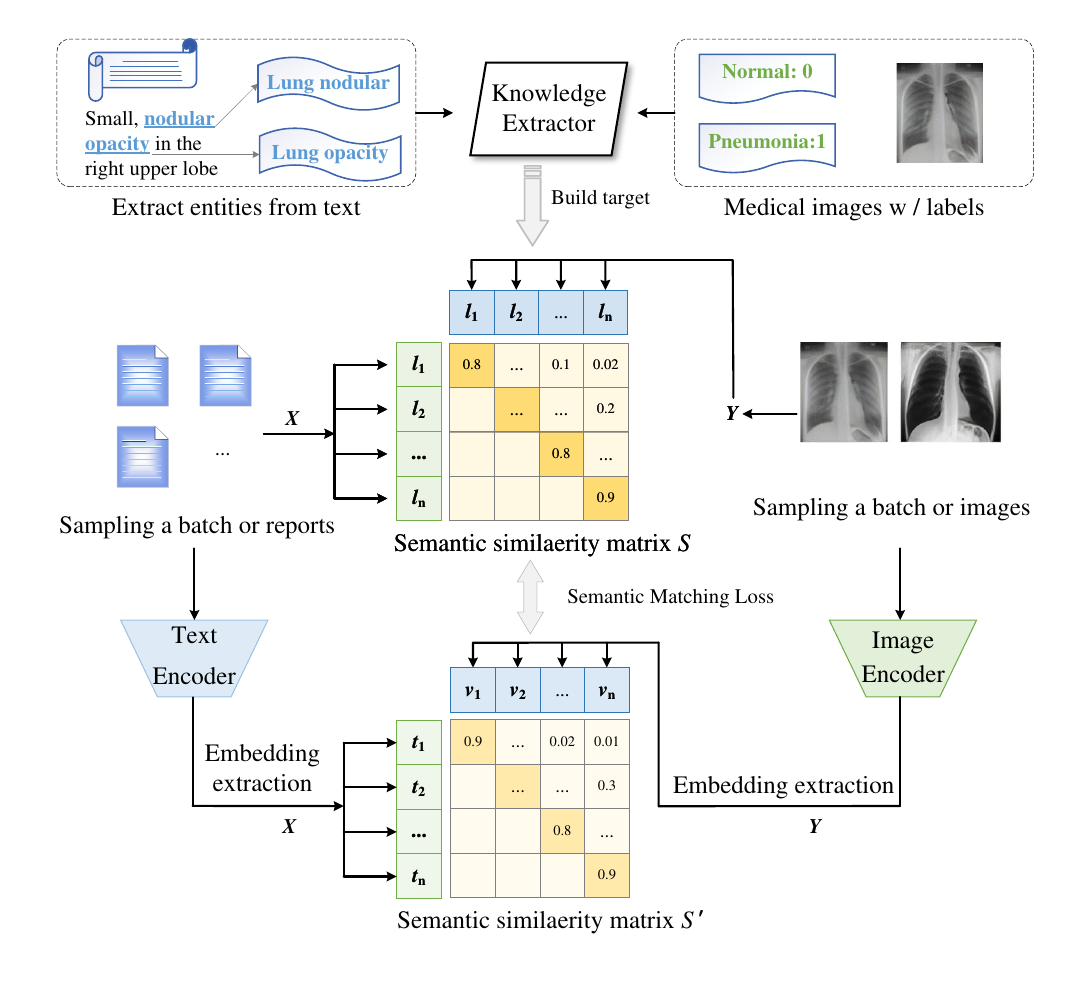}
\caption{The MedCLIP~\cite{wang2022medclip} framework for vision-language pre-training. It aligns medical image and text representations by training encoders to match a fine-grained semantic similarity matrix (bottom) derived from medical knowledge extraction (top).}
\label{fig:1}
\end{figure}

From a technical standpoint, the core of this process often lies in powerful VLMs. These systems, exemplified by architectures such as BiomedCLIP~\cite{zhang2023large} and MedCLIP~\cite{wang2022medclip} (as illustrated in figure~\ref{fig:1}), are fundamentally designed to align representations from medical images and textual descriptions, typically through a contrastive loss function that maximizes the similarity between corresponding image-text pairs. A key component of these architectures is a dedicated text encoder, which is responsible for parsing specialized medical terminology (e.g., "prurigo nodularis" or "squamous cell carcinoma"). To achieve high fidelity, these text encoders are often based on domain-specific language models such as PubMedBERT~\cite{gu2021domain}, ClinicalBERT~\cite{huang2019clinicalbert}, or other specialized Medical BERT variants~\cite{xu2024medsyn}. The primary function of these encoders, whether as part of a larger VLM or used independently, is to transform the text into high-dimensional embedding vectors that capture its rich semantic content. Subsequently, these embeddings are integrated into the generative model, typically through its layers of cross-attention~\cite{huang2019ccnet,chen2021crossvit,petit2021u,lin2022cat}. Within diffusion models, for instance, these textual embeddings play a pivotal role during the iterative denoising process~\cite{akrout2023diffusion,chambon2022adapting,dai2024guidegen,bluethgen2024vision,chambon2022roentgen,hashmi2024xreal,liang2024covid,han2024advancing,liu2024texdc}, guiding the network to construct images that faithfully align with the provided textual descriptions. Researchers can precisely manipulate the generated image characteristics through detailed textual prompts by enabling fine-grained control over aspects such as lesion size, color, or anatomical location.

\begin{figure}[ht]
\centering
\includegraphics[width=0.9\linewidth]{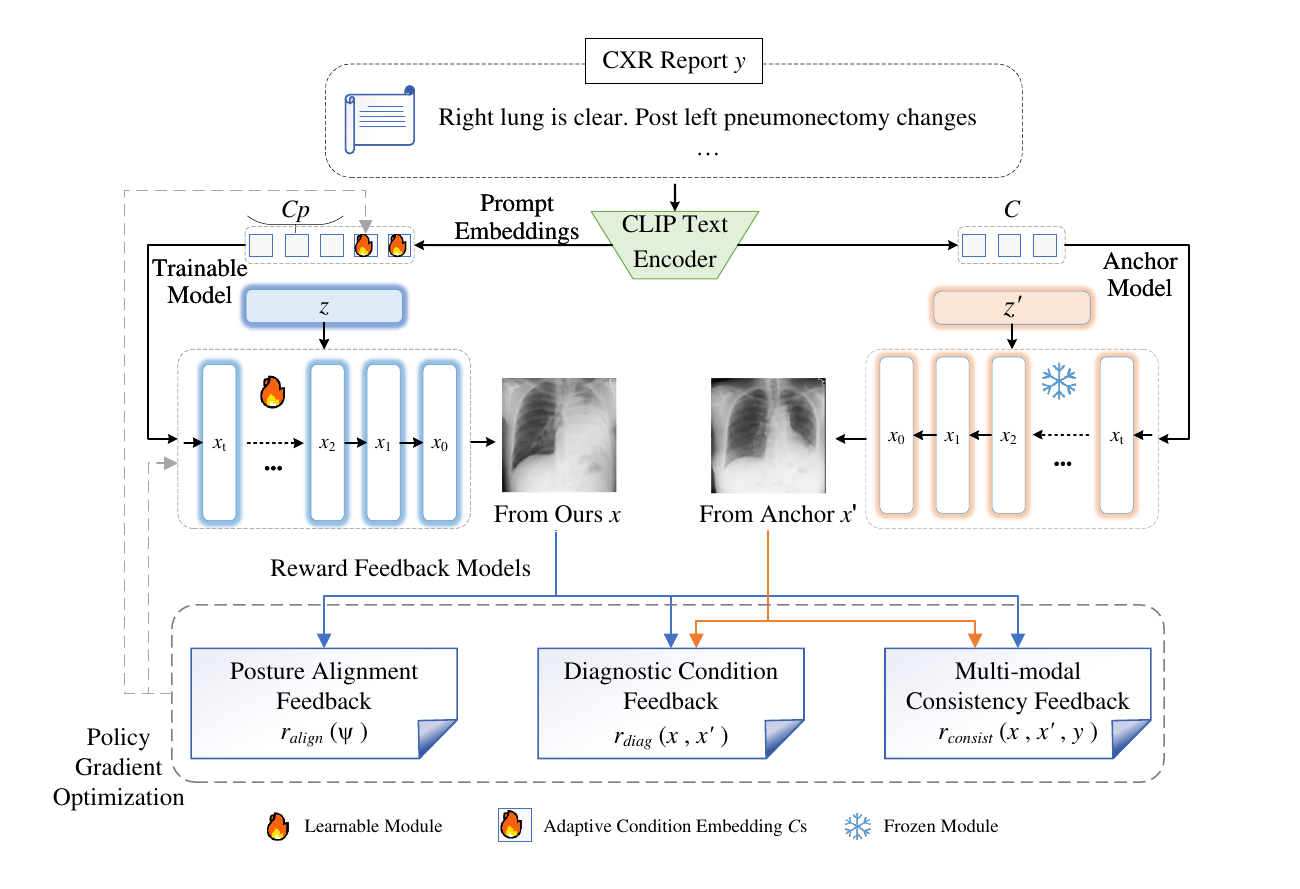}
\caption{The CXRL framework~\cite{han2024advancing}, which utilizes textual reports as prompts to condition CXR generation. A key feature is using a reward feedback loop to refine the learnable prompt embeddings, ensuring alignment between the generated image and the diagnostic information in the report.}
\label{fig:3}
\end{figure}

\begin{figure}[ht]
\centering
\includegraphics[width=0.9\linewidth]{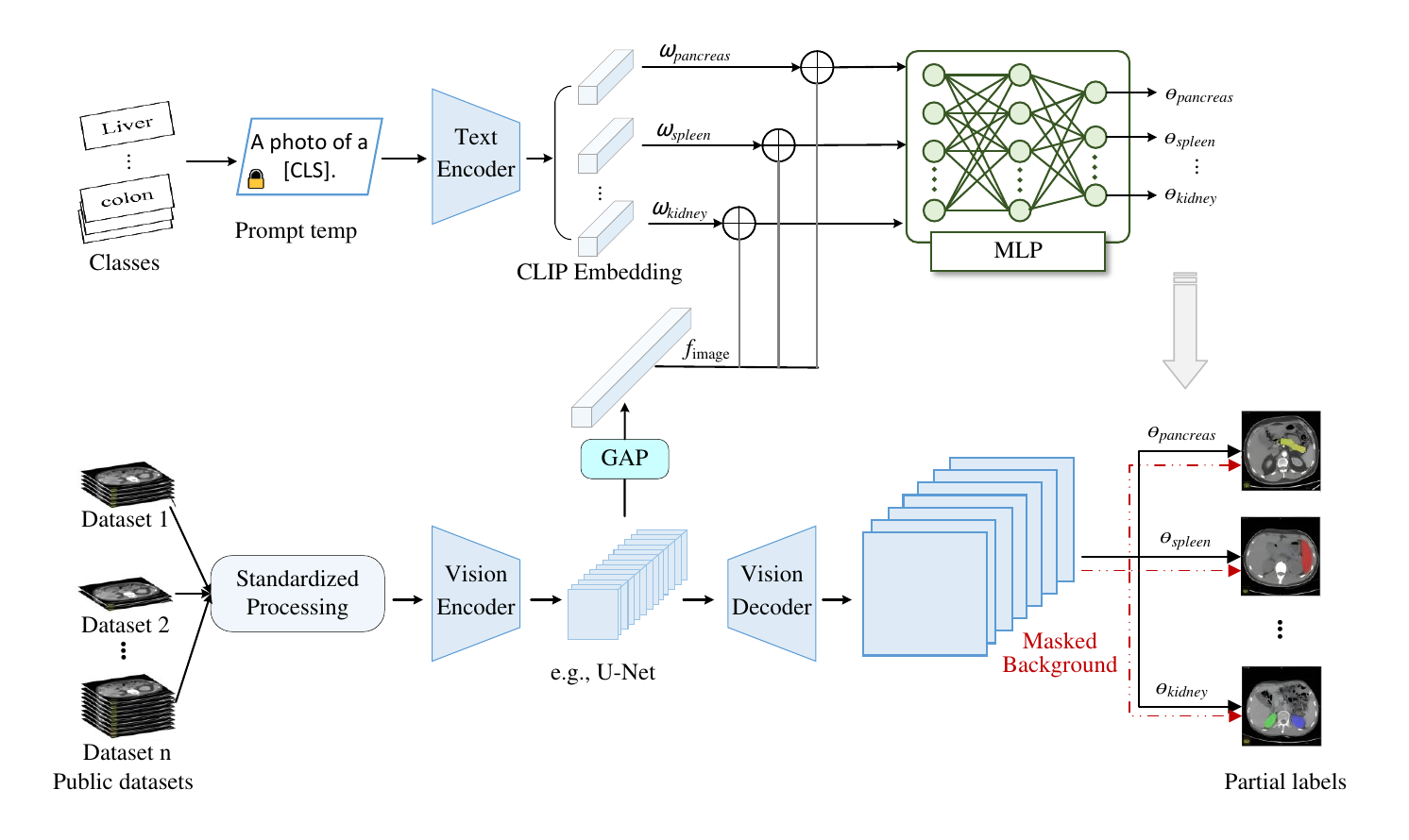}
\caption{A CLIP-driven universal segmentation model~\cite{liu2023clip} where text prompts, generated from class names, are used to dynamically steer the network towards segmenting specific organs and tumors.}
\label{fig:4}
\end{figure}

\paragraph{Strategies for Optimizing Text-Conditioned Generation}
Various advanced strategies and auxiliary modules are often incorporated to enhance textual conditional control further and ensure the domain-specificity and realism of generated images. For example, hierarchical text prompt systems can be constructed using GPT-based summarizers \cite{yellapragada2024pathldm} or localized lesion description modules, facilitating the development of more robust disease representations within the generative model and improving the fidelity of images synthesized for rare or subtle lesions. Concurrently, many approaches integrate sophisticated alignment and consistency enforcement techniques. Some generation pipelines employ a pre-alignment step \cite{liu2024texdc}, where text tokens are first matched with corresponding disease-related regions in latent feature maps before proceeding with diffusion or GAN synthesis. Other methodologies leverage domain adaptation techniques \cite{chambon2022roentgen} or concepts such as Textual Inversion \cite{akrout2023diffusion}. Textual Inversion, for instance, allows the model to learn a new pseudo-word embedding that represents a novel medical concept from just a few image examples, which can then be used in textual prompts to generate precise corresponding pathological features. The synergy between appropriate text encoders, advanced attention mechanisms, and these optimization strategies significantly elevates text-guided medical image synthesis's realism, diversity, and controllability.

\subsubsection*{Non-Textual and Structurally-Aware Prompts in Generation}

In addition to semantic guidance from text, the generation of medical images often benefits from prompts that provide structural, sequential, or other forms of explicit prior information. These non-textual or structurally-aware prompts enable finer-grained and more multidimensional control over the synthesis process, guiding the model from anatomical, temporal, or biological perspectives. Such prompts not only enhance the realism of image details and the consistency of anatomical structures but also help address clinical challenges like data scarcity and inter-sequence or inter-patient variability.

\paragraph{Anatomical Prompts for Structural Fidelity}

Anatomical prompts directly leverage key structural information, such as the delineations of lung lobes, airways, or blood vessels. This information is typically extracted from real CT or MRI images using pre-trained segmentation tools (e.g., lungmask \cite{hofmanninger2020automatic}, NaviAirway \cite{wang2022naviairway}, and TotalSegmentator \cite{wasserthal2023totalsegmentator}). During the generation process, these anatomical maps or constraints are injected into the generator, often through multichannel concatenation with latent representations~\cite{xu2024medsyn} at various generator stages, such as along the input noise or at intermediate layers of a U-Net-like architecture. The generator is then tasked with producing the target intensity image (e.g., a single-channel CT slice) and ensuring consistency with the provided anatomical information. This imposes explicit structural constraints during both low-resolution generation and high-resolution super-resolution stages, effectively suppressing structural hallucinations and ensuring a higher degree of anatomical consistency and realistic detail in the synthesized images.

\paragraph{Sequence Prompts for Multi-Contrast MRI Synthesis}
Particularly relevant for MRI, sequence prompts guide the synthesis of specific image contrasts (e.g., T1-weighted, T1Gd, T2-weighted, FLAIR). These are typically represented using one-hot encodings or learnable embedding vectors that indicate the target MRI sequence. These prompts are injected as dynamic conditions into the generative network. For example, in \cite{10.1007/978-3-031-72120-5_45}, sequence prompts, in conjunction with structural features extracted by an encoder, drive a dynamic decoder. This allows a shared discrete latent representation to be mapped to the specific target sequence, facilitating cross-sequence generation from a common underlying anatomical representation. In another approach \cite{han2024synthesis}, predefined sequence codes are transformed into control vectors via a multilayer perceptron (MLP). These vectors then dynamically modulate a shared weight library within the HyperConv layers of a HyperDecoder module. This mechanism uses a smaller network (the hyper-network) to generate the weights (convolutional kernels) for the leading network (the HyperDecoder) based on the sequence prompt, allowing for dynamic, sequence-specific convolutions. This precisely generates the convolutional kernels specific to each target sequence while capturing shared (redundant) and unique (distinct) information across different MR sequences. Such mechanisms improve generative models' robustness and semantic expressiveness, especially under unsupervised or data-scarce conditions, and provide adequate support for handling missing clinical data or synthesizing entire multi-sequence MRI protocols.

\paragraph{Multi-Variable Prompts for Conditioned Phenotype Generation}

Multi-variable prompts have shown significant utility in generating images conditioned on specific patient characteristics or quantitative biomarkers, such as brain MRI generation. As shown in~\cite{pinaya2022brain}, multiple normalized variables, including patient age, sex, ventricle volume, or normalized brain volume-can be injected into the latent space of a generative model, often via cascading them with latent vectors or using cross-attention mechanisms to allow the model to condition its output on these continuous or categorical variables at different levels of the generative process. This allows for dynamic control over the generation process, allowing the synthesis of brain images that reflect specific demographic or pathological states. Such methods exhibit strong controllability for directed generation (e.g., simulating aging effects). They can show excellent extrapolation capabilities, producing plausible image structures even for combinations of variables not explicitly seen during training.

\paragraph{Label Prompts for Genotype-Phenotype Association in Synthesis}
Label prompts can convert discrete categorical information, such as genotypic data (e.g., IDH mutation status, 1p19q co-deletion status in gliomas), into conditional vectors. These vectors are then injected into the conditioning mechanism of generative models, often by modulating the time or class embeddings within diffusion models \cite{moghadam2023morphology}, to guide the synthesis process. This allows the network to learn and capture subtle morphological differences in images associated explicitly with these genotypes, influencing the denoising and reconstruction stages. This approach has proven effective for pathological image synthesis, producing high-quality images aligned with clinical or molecular characteristics. It offers considerable value for targeted data augmentation, especially for rare genetic subtypes, and for visually exploring genotype-phenotype correlations.

The sophisticated integration of these diverse prompt mechanisms at various stages of generative models is crucial. It enhances the structural consistency and realistic detail of synthesized medical images and significantly strengthens the fine-grained controllability of the generation process, catering to a wide array of complex clinical conditions and research inquiries.

\begin{figure}[ht]
\centering
\includegraphics[width=0.8\linewidth]{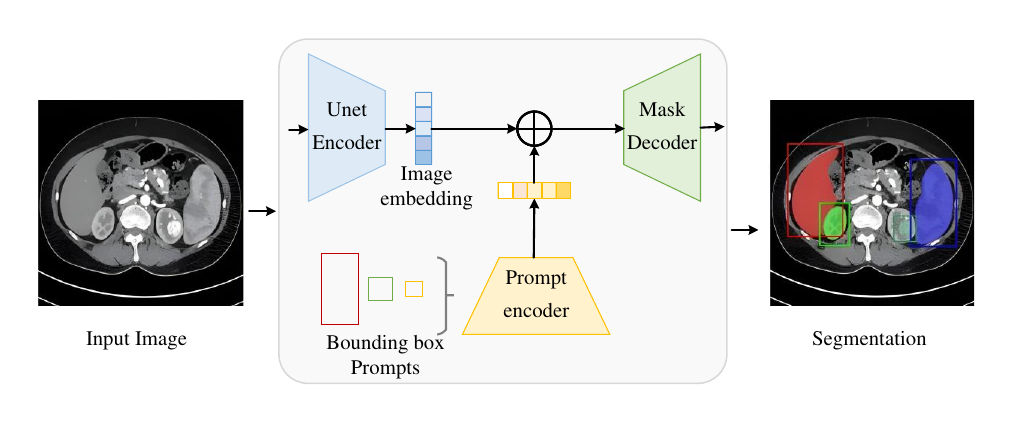}
\caption{ The architecture of MedSAM~\cite{ma2024segment}, which utilizes bounding boxes as prompts to guide the segmentation process. The prompts are encoded and fused with the image embedding, directing the mask decoder to generate precise segmentations for the specified target regions.}
\label{fig:2}
\end{figure}

\subsection*{Prompt Mechanisms in Medical Image Segmentation}

In medical image segmentation, prompts play an essential role by guiding models to accurately focus on specific anatomical regions or pathological structures of interest, ultimately enhancing segmentation accuracy, robustness, and efficiency. Integrated at various stages of the model architecture, from initial input conditioning to feature extraction, attention modulation, and output refinement, prompts significantly influence segmentation outcomes. The effectiveness of a given prompt type is contingent upon the specific segmentation task, the characteristics of the available data (e.g., scarcity, annotation quality), and the nature of user interaction required or desired. This section classifies and examines the principal categories of prompts employed in contemporary medical image segmentation research. It focuses on their integration mode, impact on model performance, and strategic placement within the segmentation pipeline.

\subsubsection*{Visual Prompts for Interactive and Targeted Segmentation} 

Visual prompts have emerged as a potent tool for improving the accuracy and robustness of medical image segmentation, particularly valuable for interactive or semi-automated segmentation workflows. By providing explicit spatial prior information about the target regions, visual prompts direct the model's attention and facilitate precise delineation, especially in scenarios involving complex anatomical structures, low-contrast boundaries, or ambiguous image features. These prompts are pivotal across multiple stages of the segmentation model:

\paragraph{Input Stage: Initial Spatial Guidance}

Users typically provide simple visual prompts, such as point clicks (indicating foreground/background), scribbles, or bounding boxes that roughly encapsulate the region of interest. These sparse inputs are processed by a dedicated prompt encoder—ranging from a simple MLP to a more complex Transformer encoder—to transform them into high-dimensional embedding vectors suitable for conditioning the segmentation model \cite{mattjie2023zero,gao2023desam}. These prompt embeddings are fused with the image features, often through cross-attention mechanisms. This fusion allows the model to take advantage of the user's global image context and local spatial information, thus providing crucial prior knowledge of the target's location and broad semantics for the segmentation task \cite{li2024promise,deng2023sam}. This initial guidance is critical for disambiguating targets from a complex background.

\paragraph{Feature Fusion Stage: Refining Focus in Complex Scenarios}
During the feature fusion stage, visual prompts enhance the model's focus on the intended target regions, particularly when dealing with intricate or poorly defined structures in medical images. They guide the model in concentrating its representational capacity on critical areas. For instance, in the MedSAM model \cite{ma2024segment} (as illustrated in figure~\ref{fig:2}), bounding box prompts have been shown to effectively reduce ambiguity by providing strong spatial priors, helping the model maintain segmentation accuracy even in low-contrast or structurally complex images. The DeSAM framework \cite{gao2023desam} employs an innovative decoupling design, combining a Prompt-Relevant IoU Module with multi-scale image features; this architecture ensures that the model can still function stably and generate accurate segmentation results even when faced with inaccurate or noisy visual prompts. ProMISe \cite{li2024promise} improves a model’s ability to capture 3D spatial context in volumetric medical images by introducing deep embedding layers and self-attention mechanisms specifically for processing 3D visual prompts, ensuring high efficiency and accuracy when handling complex 3D data.

\paragraph{Decoding Stage: Optimizing Mask Generation}
In the decoding stage, visual prompts continue to guide the mask decoder, ensuring it focuses on the precise target regions and thus further optimizing the generation of the final segmentation masks. After the prompt-conditioned image features are processed through the encoder and fusion stages, self-attention mechanisms within the decoder, influenced by the prompt information, direct the model to concentrate on specific regions. For example, bounding box prompts help the model to more accurately define the boundaries of target regions, which is especially critical in medical images where precise delineation impacts clinical decisions \cite{ma2024segment}. Through the coordinated application of these techniques across different stages, visual prompts not only enhance the model's initial understanding of the target in the input stage but also progressively optimize the feature fusion and decoding processes, leading to more accurate, robust, and often more efficient solutions for medical image segmentation.

\subsubsection*{Text-Guided Semantic Segmentation: Leveraging Language for Precision}

Text prompts are increasingly crucial for enhancing medical image segmentation accuracy by providing rich semantic guidance. By supplying structured or natural language descriptions of anatomical structures, pathological lesions, or target characteristics, text prompts help the model better understand and precisely localize these regions of interest. This semantic conditioning typically involves several key technical stages:

\paragraph{Semantic Encoding of Text Prompts}
The process begins with converting text prompts into meaningful semantic embeddings using a text encoder. This provides strong, high-level input information to the segmentation model. In SAM-Med3D-MoE \cite{wang2024sam}, for example, the encoder transforms a text prompt describing a target anatomical region into a feature vector, guiding the model to focus on that specific region. Similarly, SegVol \cite{du2023segvol} utilizes a CLIP-based text encoder to convert an input textual description (e.g., "liver") into high-dimensional vectors. These vectors are subsequently prepared for fusion with image features, laying a solid foundation for effective multimodal integration and semantic understanding.

\paragraph{Cross-Modal Fusion of Text and Image Features}
A crucial step involves effectively fusing these textual semantic embeddings with visual features extracted from the image. In the CLIP-driven Universal Model \cite{liu2023clip} (see figure~\ref{fig:4}), text embeddings generated by CLIP, when integrated with image features (often via cross-attention), help the model to understand complex anatomical relationships (e.g., the spatial relationship between the liver and associated liver tumors). This fusion of semantic and visual information enables the model to segment specific structures or regions based on textual descriptions precisely. In LuGSAM \cite{rameshlugsam}, text prompts are fused with visual features extracted by a vision backbone like Grounding DINO (which itself might use a Swin Transformer), further optimizing the model's cross-modal feature integration capabilities. Through mechanisms like cross-attention, text prompts guide the model to selectively focus on specific target regions (e.g., distinguishing the right lung from the left based on textual input), thereby improving segmentation accuracy for complex or similar-looking anatomical structures.

\paragraph{Decoder Optimization and Output Refinement via Text}
Text prompts also contribute to optimizing the segmentation output during the decoder stage. In frameworks like SegVol \cite{du2023segvol}, after initial fusion with image features, the text-derived information (or the fused representation) is input into the Mask decoder. This helps generate a final segmentation mask that is visually coherent and semantically aligned with the textual description, ensuring the mask accurately corresponds to the target region. In Medclip-samv2 \cite{koleilat2024medclip}, text prompts are ingeniously used to generate intermediate visual prompts (such as bounding boxes or point prompts inferred from the text, often via a separate localization model or attention map analysis). These visual prompts guide an underlying segmentation model such as Segment Anything (SAM)~\cite{kirillov2023segment}, further enhancing segmentation accuracy. This indicates that text prompts can be multifaceted, as direct semantic guides and generators of more explicit spatial prompts for subsequent stages.

\paragraph{Spatial Localization through Hierarchical Textual prompts}
Beyond providing general semantic information, text prompts can significantly enhance model performance by offering precise spatial localization prompts. As demonstrated in Ariadne's thread \cite{zhong2023ariadne}, text prompts can be structured to gradually provide location information about the target region hierarchically, progressing from coarse to fine descriptions. This might include broad regional descriptions (e.g., "left lung"), more specific location descriptions (e.g., "upper lobe infection of the left lung"), and finally, finely detailed lesion characteristics (e.g., "infection area within the superior segment of the upper lobe"). These multi-level spatial prompts, embedded in text, help the model more accurately identify and localize the target region, thereby addressing precision issues in image segmentation that often arise from insufficient contextual information or ambiguity.

\paragraph{Facilitating Weakly Supervised Learning and Pseudo-Label Generation}
Text prompts play a vital role in weakly supervised learning paradigms and the generation of pseudo-labels, reducing the reliance on extensively manually annotated data. In Simtxtseg \cite{xie2024simtxtseg}, text prompts are used to help generate initial visual prompts or attention maps, which are then further processed to create pseudo-masks. These pseudo-masks serve as training targets for the segmentation network, which is particularly valuable in scenarios where pixel-level annotations are scarce. Similarly, TP-DRSeg \cite{li2024tp} employs an explicit prior encoder to transform text descriptions of diabetic retinopathy lesions into visual priors. These priors are then fused with image features to enhance segmentation accuracy. This approach, which improves the model’s generalization ability through pseudo-label generation and weakly supervised training, further underscores the multifaceted utility of text prompts in modern segmentation networks, especially for specialized medical tasks.

\subsubsection*{Advanced Prompting Strategies: Learnable, Adaptive, and Unsupervised Approaches} 
\label{sssec:other_prompts_segmentation}

Beyond direct visual and textual inputs, medical image segmentation has witnessed the development of more sophisticated prompting strategies. These often involve learnable components, adaptive mechanisms, or approaches tailored for unsupervised or domain-adaptation challenges, aiming further to boost segmentation accuracy, robustness, and applicability.

\paragraph{Learnable Prompts for Task-Specific Adaptation}
The core idea behind "learnable prompts" is to obtain a set of parameterized vectors through end-to-end training. These vectors implicitly encode task- and modality-specific prior knowledge, guiding network behavior more effectively than fixed prompts for specific segmentation tasks. In \cite{fischer2024prompt}, learnable prompt vectors are injected into UNet-style "promptable blocks," cascaded with windowed image embeddings, and jointly process image features and task information via multi-head attention mechanisms. The Meduniseg model \cite{ye2024meduniseg} separately employs modality prompts (injected at early encoder stages to help differentiate multimodal input data) and task prompts (fused with sample features via a FUSE module at the encoder's terminal to provide task-specific priors for the decoder). Uniseg \cite{ye2023uniseg} focuses on capturing intrinsic correlations between different segmentation tasks by integrating universal prompts with encoder outputs, delivering task-specific priors at the initial decoder stage. The DCTP-Net \cite{liu2024dctp} introduces a Learnable Prompt block within its prompt-aware branch specifically to extract brain prior knowledge from CT images, which is then fused with image features to reduce structural detail interference during acute ischemic stroke lesion segmentation. The FedLPPA framework \cite{lin2024fedlppa} for federated learning constructs a "triple prompt" system comprising globally shared universal knowledge prompts, locally distributed data-aware prompts, and predefined annotation-sparsity prompts. These diverse prompts are fused with encoder outputs through a Tri-prompt Dual-attention Fusion module and injected with rich contextual information before decoding, enabling personalized and efficient federated weakly-supervised segmentation.

\paragraph{Prompts for Unsupervised Learning and Domain Adaptation}
Several studies have introduced novel prompting strategies to tackle unsupervised annotation or domain adaptation challenges in medical image segmentation.
Chen et al. \cite{chen2024sam} propose the Self-prompt mechanism, which employs a multi-scale self-prompt generation module. By fusing features from various layers of a domain-adaptive encoder with a Feature Pyramid Network for foreground prediction and subsequent screening, high-confidence (self-generated) prompts are element-wise added to the final image features. These are fed into a domain query-enhanced decoder, achieving precise segmentation of nuclear regions without manual annotation.
Lin et al. \cite{lin2023multi} introduce a Domain-specific prompt embedded within a feature transfer module. This module automatically extracts and fuses unique information pertinent to the current target domain, guiding the network to generate domain-aware features and enabling robust feature alignment and adaptation across multiple target domains.
Na et al. \cite{na2024segment} present the Auto-prompt system, which utilizes an independent auxiliary network to generate initial prediction masks. After Sigmoid activation, high-confidence positive and negative prompt points are automatically extracted from these masks and fused with SAM’s Prompt Encoder. This guides the Mask Decoder in completing the cell nucleus segmentation task more accurately, automating the prompt generation process.
Luo et al. \cite{luo2023universal} propose the Task-Specific Prompt as a trainable prompt vector embedded at the back end of a Vision Transformer (ViT) based segmentation model. Together with the image features extracted by the encoder, this prompt enters the Transformer decoder to provide task-specific prior information, tailoring the generic ViT to specific segmentation needs.
Meanwhile, Chen et al. \cite{chen2024each} introduce the Low-frequency prompt. This technique modulates the low-frequency amplitude of test images during the preprocessing stage. The goal is to adjust the image style and texture in the frequency domain, making the test images appear closer to the characteristics of the source domain on which a pre-trained, frozen segmentation network was trained, thus mitigating the detrimental impact of domain shift.

\subsection*{Prompt Mechanisms in Medical Image Classification}

Prompt-based methodologies have become essential for enhancing the performance, generalization capabilities, and interpretability of models engaged in medical image classification. By effectively leveraging prompts, these models can better navigate the inherent complexities of medical data, thereby addressing critical challenges such as domain adaptation, the scarcity of labeled data (enabling few-shot or zero-shot learning), and the increasing demand for explainable AI in clinical decision support. This section details the primary categories of prompts employed for these classification tasks.

\subsubsection*{Text-Driven Classification: From Static Labels to Dynamic Semantic Guidance} 

Text prompts integrated within medical image classification models, predominantly processed by the text encoder component of vision-language architectures, have evolved significantly. They have transitioned from simple static class descriptors to dynamic, trainable, and knowledge-rich constructs, substantially enhancing domain specificity, model interpretability, and overall classification accuracy, particularly in zero-shot and few-shot learning paradigms.

\paragraph{Foundational Approaches with Static and Pre-trained Prompts}
Early and foundational frameworks, such as that detailed in \cite{zhang2023text}, utilize static text prompts, typically defined by class names (e.g., ``well differentiated tubular adenocarcinoma'') sourced directly from dataset labels. These prompts are fed into a frozen biomedical language model (e.g., BioLinkBERT), which generates fixed class embeddings \( x_{cl} \). These embeddings remain invariant during training (due to a "stop-grad" operation) and serve as stable semantic reference points. They are then compared (often via cosine similarity) with image embeddings \( x_v' \), which are derived from a vision encoder (e.g., ViT-B/16) and projected into a shared latent space. The model is optimized using a cross-entropy loss to align these visual features with their corresponding textual semantics, facilitating classification without altering the pre-trained weights of the text encoder.

This static prompting approach is further refined and scaled in influential models like BiomedCLIP \cite{zhang2023biomedclip} and PubMedCLIP \cite{eslami2023pubmedclip}. In these contexts, text prompts often consist of richer textual data, such as PubMed captions (e.g., ``Histology of metastatic amelanotic melanoma...'' \cite{zhang2023large}) or excerpts from medical reports (e.g., ``CT of the chest showing multiple thick-walled cavities...'' \cite{eslami2021does}). These are processed by specialized medical text encoders (e.g., PubMedBERT \cite{gu2021domain}, BioClinicalBERT \cite{wang2022medclip}). During large-scale pre-training, these prompts drive contrastive learning (e.g., using InfoNCE loss) to align the embeddings from the text encoder with those from the vision encoder by iteratively optimizing their similarity scores \cite{zhang2023large}. For downstream zero-shot classification tasks, the pre-trained text encoder converts new class name prompts into target embeddings \( t_p \). These are then matched against the vision encoder's output \( v_p \) for a given image using cosine similarity, enabling classification inference on unseen classes without task-specific retraining \cite{wang2022medclip}. The text encoder is a crucial conduit for injecting domain-specific prior knowledge. In contrast, the vision encoder adapts to extract relevant visual features, a synergy often harmonized by a projection head that aligns the dimensions of the different modalities.

\paragraph{Advanced Dynamic and Learnable Text Prompts for Enhanced Adaptability}
Moving beyond static inputs, more recent frameworks like XCoOp \cite{bie2024xcoop} and MSCPT \cite{han2024mscpt} introduce dynamic and tunable (learnable) text prompts, significantly increasing their operational complexity and adaptability within the text encoder.
In XCoOp, text prompts comprise a combination of soft prompts (learnable vectors, often initialized with a template like ``a photo of a [disease name]'') and clinical challenging prompts (fixed textual descriptions derived from experts, for example, ``a photo of melanoma, with an irregular pigment network...''), both of which are processed by the text encoder. The soft prompts are optimized at both token and prompt levels, aligning their learned embeddings \( V \) with fixed clinical embeddings \( Q \) (derived from challenging prompts) using contrastive and cross-entropy losses. This process enhances interpretability by explicitly tying each learnable token to a specific clinical concept. These rich prompt embeddings then interact with the vision encoder's outputs—both global image features \( p \) and local patch features \( F = \{ f_1, f_2, \ldots, f_M \} \)-through a global-local alignment loss, refining the classification process by mimicking expert diagnostic reasoning which often involves both holistic assessment and attention to local details.
MSCPT \cite{han2024mscpt} further leverages a dual-path text encoder system for multi-scale analysis of Whole Slide Images (WSIs). Frozen low-level encoders process prompts relevant to low-magnification views (e.g., 5\(\times\), with prompts like ``Sparse stroma among tumor cells''), while learnable (prompted) high-level encoders handle prompts for high-magnification views (e.g., 20\(\times\), with prompts like ``Distinct cellular borders...''). The resulting multi-scale textual embeddings are then integrated using transformer layers. This hierarchical textual processing feeds into graph-based and non-parametric pooling mechanisms, aligning with vision encoder outputs to capture contextual and fine-grained multi-scale features essential for accurate WSI classification.

\paragraph{Text Prompts in Multiple Instance Learning Paradigms}
In the context of Multiple Instance Learning (MIL) \cite{qu2024rise,chikontwe2024low}, which is prevalent in computational pathology where an entire WSI (a "bag") is classified based on its constituent patches ("instances"), text prompts extend their influence beyond simple text encoding to orchestrate both instance-level feature extraction and bag-level aggregation.
\cite{qu2024rise} employs composite prompts: these include task labels (e.g., ``a WSI of [Lung adenocarcinoma]''), detailed descriptions generated by models like GPT-4 (e.g., ``glandular or acinar formations''), and learnable vectors. The text encoder processes all these textual components to generate instance prototypes \( P \) and bag-level tokens \( B_i \). These textual embeddings then guide the aggregation of patch features \( Z_i \) (from the vision encoder) via MIL pooling mechanisms, culminating in a bag-level similarity matching for final classification.
Similarly, \cite{chikontwe2024low} optimizes textual prompts (e.g., ``An H\&E image of [breast adenocarcinoma]'') within the text encoder to produce embeddings \( \hat{z}_t \). These embeddings \( \hat{z}_t \) are then used to weight the vision encoder outputs \( \hat{z}_v \) from individual patches in a context-driven pooling scheme, often enhanced by a residual visual adapter to capture visual nuances better. In both these MIL frameworks, the embeddings derived from text prompts dynamically steer the vision encoder's feature extraction and, more critically, the aggregation process, adapting effectively to low-shot learning scenarios and improving the classification of complex, heterogeneous medical images like WSIs.

\paragraph{Specialized Frameworks Balancing Adaptability and Simplicity}
The operational depth and utility of text prompts are further exemplified in dedicated frameworks like FLAIR \cite{silva2023foundation} and MI-Zero \cite{lu2023visual}, which strive to balance model adaptability with implementational simplicity for specific medical imaging domains. FLAIR utilizes expert-derived, descriptive textual prompts (e.g., ``only a few microaneurysms are present'' for retinal images) encoded by a BioClinicalBERT text encoder. These embeddings are used for contrastive pre-training and subsequent zero-shot inference, aligning with vision encoder outputs via similarity scores to perform classification \cite{silva2025foundation}. Conversely, MI-Zero employs static class-name templates (e.g., ``an image showing adenocarcinoma'') within a standard text encoder like PubMedBERT \cite{gu2021domain}. This produces fixed embeddings \( w_m \) used for patch-level similarity computation, aggregating results via MIL operators such as topK pooling for WSI classification \cite{lu2023visual}. While FLAIR's approach allows for tunable knowledge integration through rich, expert-crafted prompts, MI-Zero's strength lies in its simplicity and reliance on robust pre-trained alignments, with both methods effectively interfacing with vision encoders to drive classification.

\subsubsection*{Diverse Non-Textual and Structurally-Informed Prompts for Classification} 

Beyond text-based semantic guidance, various innovative non-textual and structurally-informed prompt mechanisms have been developed to address specific challenges in medical image classification. These often target domain generalization, memory efficiency, leveraging spatial or structural image information, or enhancing model reasoning capabilities.

\paragraph{Learnable Prompts for Domain Generalization and Efficiency}
To tackle the critical problem of domain generalization, \cite{yan2024prompt} proposes Collaborative Domain Prompts. These are lightweight, learnable vectors integrated into the input layer of ViT, typically prepended to the CLS token and patch embeddings. By leveraging a shared prompt component and a domain-specific prompt generator, these prompts facilitate collaboration and knowledge sharing between representations learned for different domains, enabling robust generalization to unseen domains in medical image classification, a crucial capability when explicit domain labels are unavailable or undefined.
Addressing memory efficiency in incremental learning scenarios, \cite{zhu2024memory} introduces Domain Prompts for histopathology classification. Applied within ViT’s multi-head self-attention layers, these prompts are designed to decouple into domain-specific and domain-invariant components. This allows the model to capture unique domain features and shared knowledge across incrementally learned tasks with minimal memory overhead, preventing catastrophic forgetting.

\paragraph{Structural and Embedding-Based Prompts for Fine-Grained Analysis}
For tasks requiring fine-grained analysis, such as nucleus detection and classification in pathology, \cite{huang2023prompt} presents nucleus group embeddings. These are learnable embeddings added to the input space of Swin Transformers, serving as "grouping prompts." They provide initial representations within a grouping transformer classifier to capture semantic similarities among nuclei, enhancing detection and subsequent classification accuracy.

Zhu et al. \cite{zhu2024segprompt} utilize Segmentation Map Prompts for classification tasks where spatial context is important. A dedicated encoder encodes pre-computed or predicted segmentation maps (e.g., identifying kidney stones) into token representations. These tokens are then fed alongside the image tokens into a frozen ViT. The segmentation map prompts interact with image features through self-attention mechanisms, guiding the model to focus on relevant regions delineated by the map, thereby improving classification accuracy by explicitly incorporating spatial context.
Further improving adaptability in nucleus classification, \cite{huang2024unicell} integrates Label Prompts within a dynamic prompt module between the feature encoder and the decoder. These prompts, derived from class labels, help to refine multi-scale features, improving the adaptability and accuracy of nucleus classification across diverse datasets.

\paragraph{Generative, Prototypical, and Inferential Prompts for Advanced Classification}
Leveraging generative capabilities, \cite{ye2024pseudo} proposes Pseudo-Prompts, which constitute dynamically generated class-specific embeddings produced by a dedicated prompt decoder module. These pseudo-prompts are subsequently fed into a frozen text encoder to facilitate multi-label zero-shot learning. The frozen text encoder operates without learnable embedding layers for these specific inputs, effectively leveraging the priors learned by pre-trained vision-language models. This approach enables the system to harness established knowledge representations while maintaining the flexibility to generate task-specific prompt embeddings for novel classification scenarios.

To better capture diverse pathological characteristics in WSIs, \cite{lin2024prompt} develops Prototypical Visual Prompts. These are cluster-center-based prompts derived from representative image patches identified before feature extraction. By conditioning the model on these visual prototypes, they enrich WSI classification.
Enhancing model reasoning, \cite{sanchez2024exploring} combines Few-shot Learning with Chain of Thought Prompts at the input level of large vision-language models (e.g., GPT4) for tasks like blood cell classification. This approach aims to improve classification through enhanced contextual learning and inferential capabilities by providing examples and explicit reasoning steps. However, performance can vary significantly depending on the base model's capacity.

\section*{THE APPLICATION OF THE PROMPT MECHANISM}
\begin{figure}[htbp]
\centering
\includegraphics[width=\linewidth]{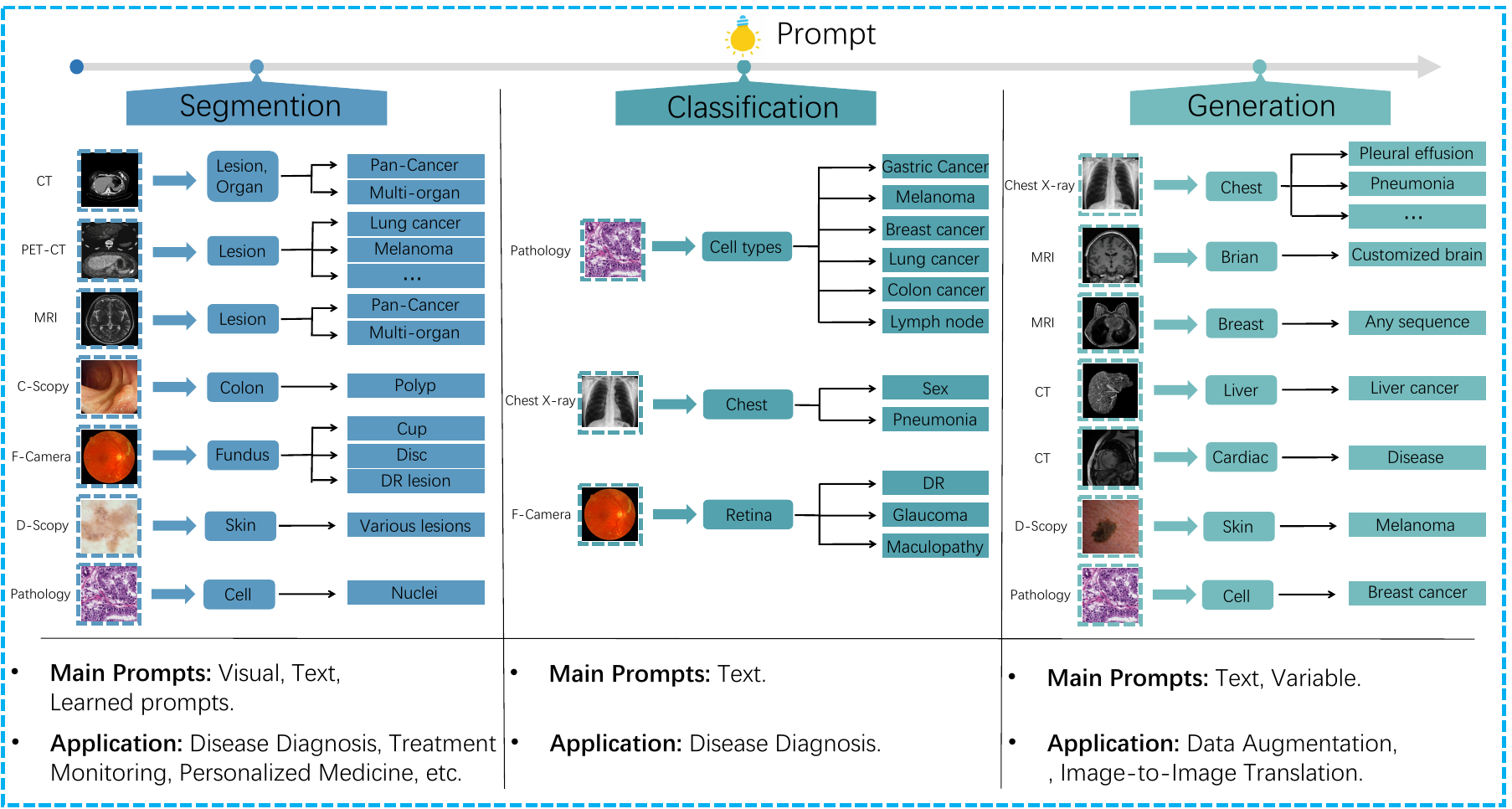}
\caption{Typical applications of prompts in three major medical imaging tasks: segmentation, classification, and generation. Note: P represents positive cases, N represents negative cases, C-scopy refers to colonoscopy, DR refers to Diabetic Retinopathy and D-scope refers to dermatoscope.}
\label{fig:application}
\end{figure}

Prompt engineering enables large vision models to achieve a wide range of AI tasks in the medical field with high performance, as shown in figure~\ref {fig:application}. With customized medical prompts, models can solve complex medical image problems that were previously difficult without giant medical data sets and resources. By designing appropriate prompts, developers can quickly obtain target data without training or fine-tuning the model, which is time-consuming and costly. In the foreseeable future, prompting engineering (especially text prompts) will be further developed to utilize the potential of large medical imaging models fully. In the medical image field, the applications of prompt engineering include:
\begin{itemize}
  \item Generation: Introducing prompt engineering into large-scale medical image models offers boundless potential for generating images from textual or report prompts. These advancements hold promise for addressing the challenge of limited medical imaging data in deep learning and applications in medical education and training.
  \item Segmentation: Medical imaging datasets often suffer from small size and partially labeled problems. However, by leveraging prompts, models can be trained on large amounts of diverse medical data and learn the associations between data, such as anatomical structures, enabling the development of powerful and universal medical segmentation models. 
   \item Classification: In medical image classification tasks, prompts are commonly employed to address the issue of sample scarcity, such as in zero-shot and few-shot learning.  Furthermore, prompts, especially learning-based prompts, are employed to enhance the classification performance of models.
\end{itemize}
We also provided a more comprehensive overview of applications of prompt engineering in medical images, as summarized in Table \ref{GENERATION}, Table \ref{prompt_engineering_applications_1}, Table \ref{prompt_engineering_applications_2}, Table \ref{prompt_engineering_applications_3}, and Table \ref{classification}. 


\begin{table*}[htbp]
    \centering
    \caption{Applications of Prompt Engineering in Generation}
    \begin{tabular}{p{2.5cm}p{1.2cm}p{4cm}p{3cm}p{6cm}} 
        \specialrule{1pt}{0pt}{1pt} 
        Reference & Year & Task & Prompt Type & Link \\
        \midrule
        Akrout et al.\cite{akrout2023diffusion}            & 2023 & Skin                       & Text                                  & - \\
        Pinaya et al.\cite{pinaya2022brain}               & 2022 & Brain MRI                  & Multivariable (age, sex, and brain information) & - \\
        Chambon et al.\cite{chambon2022adapting}          & 2022 & Chest X-ray                & Text                                  & - \\
        Xu et al.\cite{xu2024medsyn}                     & 2024 & Chest CT                   & Text                                  & - \\
        Yellapragada et al.\cite{yellapragada2024pathldm} & 2024 & Histopathology             & Text                                  & \url{https://github.com/cvlab-stonybrook/PathLDM} \\
        Moghadam et al.\cite{moghadam2023morphology}      & 2023 & Histopathology             & Label                                 & - \\
        Hamamci et al.\cite{hamamci2025generatect}        & 2025 & Chest CT                   & Text                                  & \url{https://github.com/ibrahimethemhamamci/GenerateCT} \\
        Dai et al.\cite{dai2024guidegen}                  & 2024 & Joint CT                   & Text                                  & - \\
        Wang et al.\cite{wang2024towards}                 & 2024 & Brain MRI                  & MR imaging parameters                 & - \\
        Shi et al.\cite{shi2025semantic}                  & 2025 & Breast Ultrasound          & Texture structure and lesion information & - \\
        Dahan et al.\cite{dahan2024csg}                   & 2024 & Musculoskeletal Ultrasound & Image                                 & - \\
        Bluethgen et al.\cite{bluethgen2024vision}        & 2024 & Chest X-ray                & Text                                  & - \\
        Yu et al.\cite{yu2024ct}                          & 2024 & Abdominal Lymph Node CT    & Mask                                  & - \\
        Chambon et al.\cite{chambon2022roentgen}          & 2022 & Chest X-ray                & Text                                  & - \\
        Xiao et al.\cite{xiao2023end}                     & 2023 & Liver CT                   & Mask                                  & - \\
        Weber et al.\cite{weber2023cascaded}              & 2023 & Chest X-ray                & Text                                  & - \\
        Hashmi et al.\cite{hashmi2024xreal}               & 2024 & Chest X-ray                & Text                                  & \url{https://github.com/BioMedIA-MBZUAI/XReal} \\
        Liang et al.\cite{liang2024covid}                 & 2024 & Chest X-ray                & Text                                  & - \\
        Han et al.\cite{han2024advancing}                 & 2024 & Chest X-ray                & Text                                  & \url{https://micv-yonsei.github.io/cxrl2024/} \\
        Shentu et al.\cite{shentu2024cxr}                 & 2024 & Chest X-Ray                & Combining text embedding and image embedding & \url{https://github.com/junjie-shentu/CXR-IRGen} \\
        Liu et al.\cite{liu2024texdc}                     & 2024 & 4D Cardiac Cine MRI        & Text                                  & \url{https://github.com/me-congliu/TexDC} \\
        Borghesi et al.\cite{borghesi2024generation}      & 2024 & Skin                       & Mask                                  & \url{https://github.com/aequitas-aod/experiment-gen-skin-images} \\
        Fang et al.\cite{fang2024conditional}             & 2024 & Chest X-Ray                & Edge maps, depth maps, Masks, or CLIP image embeddings & - \\
        Sagers et al.\cite{sagers2022improving}           & 2022 & Skin                       & Text                                  & - \\
        Wang et al.\cite{wang2025toward}           & 2025 &  Brain MRI                       & Text                                  & \url{https://github.com/Wangyulin-user/TUMSyn} \\
        
        Li et al.\cite{li2025interactive}           & 2025 &  Brain MRI                       & Localization Prompt                                  & \url{https://github.com/ChanghuiSu/TLP} \\
        Duan et al.\cite{duan2025fetalflex}           & 2025 &  Fetal Ultrasound                       & Anatomical Prompt                                  & \url{https://dyf1023.github.io/FetalFlex/} \\
        Liu et al.\cite{liu2025treatment}           & 2025 &  Brain MRI                       & multi-parametric MRI and treatment information                                  & - \\
        
        \specialrule{1pt}{0pt}{1pt} 
    \end{tabular}
    \label{GENERATION}
\end{table*}

\begin{table*}[htbp]
    \centering
    \caption{Applications of Prompt Engineering in Segmentation (Part 1)}
    \label{prompt_engineering_applications_1}
    \begin{tabular}{p{2.5cm}p{1cm}p{4cm}p{3cm}p{6cm}} 
        \specialrule{1pt}{0pt}{1pt} 
        Reference & Year & Task & Prompt Type & Link \\
        \midrule
        Mattjie et al.\cite{mattjie2023zero}     & 2023 & Skin Lesions, Lung, Polyps, Breast Tumor, Femur and Ilium            & Visual                       & \url{https://github.com/Malta-Lab/SAM-zero-shot-in-Medical-Imaging} \\
        Fischer et al.\cite{fischer2024prompt}   & 2024 & Pancreas‑CT, Cranial Vault abdomen CT                                 & Learnable prompt             & \url{https://github.com/marcdcfischer/PUNETR} \\
        Liu et al.\cite{liu2023clip}             & 2023 & Organ and Tumor Detection                                            & Text                         & \url{https://github.com/ljwztc/CLIP-Driven-Universal-Model} \\
        Chen et al.\cite{chen2024sam}            & 2024 & Nuclei                                                              & Self-prompt                  & \url{https://github.com/CUHK-AIM-Group/UN-SAM} \\
        Gao et al.\cite{gao2023desam}            & 2023 & Prostate                                                            & Visual                       & \url{https://github.com/yifangao112/DeSAM} \\
        Cheng et al.\cite{cheng2023sam}          & 2023 & Universal Segmentation                                              & Visual, Masks                & \url{https://github.com/uni-medical/SAM-Med2D} \\
        Ye et al.\cite{ye2024meduniseg}          & 2024 & Universal Segmentation                                              & Learnable prompt             & \url{https://github.com/yeerwen/UniSeg} \\
        Wu et al.\cite{wu2024one}                & 2024 & Organ                                                               & Visual, Mask                 & \url{https://github.com/KidsWithTokens/one-prompt} \\
        Li et al.\cite{li2024promise}            & 2024 & CT pancreas and colon                                               & Visual                       & \url{https://github.com/MedICL-VU/ProMISe} \\
        Lin et al.\cite{lin2023multi}            & 2023 & Infant MRI of different ages, high‑ and low‑grade gliomas           & Domain-specific prompt       & \url{https://github.com/MurasakiLin/prompt-DA} \\
        Chang et al.\cite{chang2023pe}           & 2023 & Organ                                                               & Visual                       & - \\
        Bai et al.\cite{bai2023slpt}             & 2023 & CT liver tumor                                                      & Visual                       & - \\
        Wu et al.\cite{wu2024efficient}          & 2024 & Dermatology image, optical disc, Chest X‑ray                        & Image                        & - \\
        Ma et al.\cite{ma2024segment}            & 2024 & Universal Segmentation                                              & Visual                       & \url{https://github.com/bowang-lab/MedSAM} \\
        Xu et al.\cite{xu2023eviprompt}          & 2023 & Universal Segmentation                                              & Pixel Similarities           & - \\
        Xie et al.\cite{xie2024masksam}          & 2024 & CT organ                                                            & Visual, Mask                 & - \\
        Ramesh et al.\cite{rameshlugsam}         & 2024 & Chest X‑Rays                                                        & Text, Visual                 & - \\
        Kato et al.\cite{kato2023one}            & 2023 & Cell                                                                & Visual                       & \url{https://github.com/usagisukisuki/oneshot-part-cellsegmentation} \\
        Zhang et al.\cite{zhang2023continual}    & 2023 & Organ and Tumor Detection                                           & Text                         & \url{https://github.com/MrGiovanni/ContinualLearning} \\
        Ye et al.\cite{ye2023uniseg}             & 2023 & Organ and Tumor                                                     & Learnable Prompt             & \url{https://github.com/yeerwen/UniSeg} \\
        Deng et al.\cite{deng2023sam}            & 2023 & Optic Cup and Disc in Fundus Photographs                            & Visual                       & - \\
        Na et al.\cite{na2024segment}            & 2024 & Nuclei                                                             & Auto-prompt                  & - \\
        Zhong et al.\cite{zhong2023ariadne}      & 2023 & Chest X‑Rays                                                        & Text                         & \url{https://github.com/Junelin2333/LanGuideMedSeg-MICCAI2023} \\
        Du et al.\cite{du2023segvol}             & 2023 & Organ                                                               & Visual, Text                 & \url{https://github.com/BAAI-DCAI/SegVol} \\
        Tomar et al.\cite{tomar2022tganet}       & 2022 & Organ                                                               & Visual, Mask                 & \url{https://github.com/nikhilroxtomar/TGANet} \\
        Xie et al.\cite{xie2024simtxtseg}        & 2024 & Polyp, MRI Brain                                                    & Text, Visual                 & - \\
        Koleilat et al.\cite{koleilat2024medclip}& 2024 & Universal Segmentation                                              & Text, Visual                 & \url{https://github.com/HealthX-Lab/MedCLIP-SAM} \\
        Zhao et al.\cite{zhao2023one}         & 2023 & Universal Segmentation                                              & Text                         & \url{https://github.com/zhaoziheng/SAT} \\
        
        \specialrule{1pt}{0pt}{1pt} 
    \end{tabular}
\end{table*}

\begin{table*}[htbp]
    \centering
    \caption{Applications of Prompt Engineering in Segmentation (Part 2)}
    \label{prompt_engineering_applications_2}
    \begin{tabular}{p{2.5cm}p{1cm}p{4cm}p{3cm}p{6cm}} 
        \specialrule{1pt}{0pt}{1pt} 
        Reference & Year & Task & Prompt Type & Link \\
        \midrule
        Biswas et al.\cite{biswas2023polyp}      & 2023 & Polyp                                                               & Visual, Mask                 & \url{https://github.com/RisabBiswas/Polyp-SAM++} \\
        Chen et al.\cite{chenvp}                 & 2025 & Abdominal and cardiac MRI to CT                                     & Visual                       & - \\
        Han et al.\cite{han2023multiscale}       & 2023 & Nuclear, chest X‑ray, Colon histology                               & Visual                       & - \\
        Saeed et al.\cite{saeed2023prompt}       & 2023 & PET‑CT Head and Neck Cancer                                        & Learnable parameters, Termed prompts & - \\
        Li et al.\cite{li2023multi}              & 2023 & CT scans liver and pancreatic lesions                      & Visual                       & - \\
        Luo et al.\cite{luo2023universal}        & 2023 & Universal Segmentation                                              & Task-Specific Prompt         & - \\
        Xie et al.\cite{xieself}                 & 2023 & Abdominal CT and cardiac MRI                                       & Visual                       & - \\
        Chen et al.\cite{chen2023sppnet}         & 2023 & Nuclei                                                             & Visual                       & \url{https://github.com/xq141839/SPPNet} \\
        
        Sridhar et al.\cite{sridhar2023lung}           & 2023 & Chest X‑ray                                                              & Mask, Visual               & - \\
        Zhang et al.\cite{zhang2023box2pseudo}         & 2023 & CT pulmonary nodule                                                      & Visual                     & - \\
        Glatt et al.\cite{glatt2023topology}           & 2023 & Liver cell                                                               & Visual                     & - \\
        Zhou et al.\cite{zhou2024specific}             & 2024 & Pancreas‑CT, Brain MRI                                                   & Cross‑Prompt               & \url{https://github.com/Fyw1988/MUL_SCIP} \\
        Huang et al.\cite{huang2024robust}             & 2024 & Universal Segmentation                                                   & Visual prompt              & - \\
        Chen et al.\cite{chen2024each}                 & 2024 & Joint optic disc (OD) and cup (OC), polyp                                & Low‑frequency prompt       & \url{https://github.com/Chen-Ziyang/VPTTA} \\
        Ouyang et al.\cite{ouyang2024prompt}           & 2024 & Universal Segmentation                                                   & Text prompt                & - \\
        Chen et al.\cite{chen2024segmentation}         & 2024 & MR femur, tibia, femoral‑ and tibial cartilages                          & Visual                     & \url{https://github.com/chrissyinreallife/KneeSegmentWithSAM.git} \\
        Shaharabany et al.\cite{shaharabany2024zero}   & 2024 & CT organ, nuclear                                                        & Mask, Visual               & \url{https://github.com/talshaharabany/ZeroShotSAM} \\
        Wang et al.\cite{wang2024tp}                   & 2024 & MRI colon and stomach, CT Liver Tumor                                    & Temporal prompt            & - \\
        Adhikari et al.\cite{adhikari2024tunevlseg}    & 2024 & Universal Segmentation                                                   & Visual                     & - \\
        Kong et al.\cite{kong2024swiftmedsam}         & 2024 & Universal Segmentation                                                   & Visual                     & \url{https://github.com/naamiinepal/tunevlseg} \\
        Liu et al.\cite{liu2024feature}               & 2024 & Kidney pathology                                                         & Automatic prompt           & \url{https://github.com/SnowRain510/GBMSeg} \\
        Lin et al.\cite{lin2024fedlppa}                & 2024 & Fundus OD/OC, OCTA FAZ, Endoscopy Polyp, MRI Prostate                    & Learnable prompt           & \url{https://github.com/llmir/FedLPPA} \\
        Chen et al.\cite{chen2024multi}                & 2024 & Universal Ultrasound                                                     & Task Prompt                & - \\
        Cui et al.\cite{cuienhancing}                  & 2024    & Kidney pathology                                                         & Text, Visual               & - \\
        Xie et al.\cite{xie2024promamba}               & 2024 & Polyp                                                                    & Visual                     & - \\
        Xia et al.\cite{xia2024cervical}               & 2024 & MRI Cervical cancer                                                      & Visual                     & - \\
        Zhu et al.\cite{zhu2024segprompt}              & 2024 & Kidney stone                                                             & Mask                       & - \\
        Teng et al.\cite{teng2024knowledge}           & 2024 & MRI brain                                                                & Text                       & \url{https://github.com/TL9792/KGPL} \\
        Chen et al.\cite{chen2024adaptation}          & 2024 & Cine Cardiac MRI                                                         & Visual                     & - \\
        Guan et al.\cite{guanlite}                     & 2024    & Universal Segmentation                                                   & Class and Visual           & - \\
        Song et al.\cite{song2024ep}                   & 2024 & Histopathological breast cancer                                          & Mask                       & \url{https://github.com/QI-NemoSong/EP-SAM} \\
        Khor et al.\cite{khor2024unified}              & 2024 & CT Nasopharyngeal Carcinoma with Prior Anatomical Information            & Visual                     & - \\
        \specialrule{1pt}{0pt}{1pt} 
    \end{tabular}
\end{table*}

\begin{table*}[htbp]
    \centering
    \caption{Applications of Prompt Engineering in Segmentation (Part 3)}
    \label{prompt_engineering_applications_3}
    \begin{tabular}{p{2.5cm}p{1cm}p{4cm}p{3cm}p{6cm}} 
        \specialrule{1pt}{0pt}{1pt} 
        Reference & Year & Task & Prompt Type & Link \\
        \midrule
        Xue et al.\cite{xue2024deep}                   & 2024 & MRI cervical cancer                                                      & Visual                     & - \\
        Cui et al.\cite{cui2024all}                    & 2024 & Nuclei                                                                   & Visual                     & - \\
        Lyu et al.\cite{lyu2024superpixel}             & 2024 & CT liver tumor                                                           & Visual                     & - \\
        Yang et al.\cite{yang2024tavp}                 & 2024 & Chest X‑ray                                                              & Task‑adaptive Visual       & - \\
        Xue et al.\cite{xue2024deep2}                  & 2024 & CT Endometrial cancers                                                   & Visual                     & - \\
        Dai et al.\cite{dai2024sparse}                 & 2024 & CBCT dental images                                                       & Visual                     & - \\
        Cui et al.\cite{cui2024pfps}                   & 2024 & Kidney pathology                                                         & Text                       & - \\
        Hu et al.\cite{hu2024lpam}                     & 2024 & Polyp, colonoscopy, skin injury                                          & Visual                     & - \\
        Liu et al.\cite{liu2024dctp}                   & 2024 & CT Acute Ischemic Stroke Lesion                                          & Learnable prompt           & - \\
        Sun et al.\cite{sun2024aepl}                   & 2024 & MRI Brain Tumor                                                          & Tumor grades               & \url{https://github.com/YonghengSun1997/AEPL} \\
        Song et al.\cite{song2024automatic}            & 2024 & Laryngoscopic Image                                                      & Visual                     & \url{https://github.com/yucongzh} \\
        Cheng et al.\cite{cheng2024frequency}          & 2024 & MRI brain                                                               & Frequency filtering and spatial perturbation prompts & - \\
        Li et al.\cite{li2024centersam}                & 2024 & Nucleus                                                                  & Visual                     & - \\
        Zhang et al.\cite{zhang2024progressive}        & 2024 & Multi‑Class Cell                                                         & Text                       & - \\
        Huang et al.\cite{huang2024iossam}             & 2024 & Tooth                                                                    & Image                      & - \\
        Li et al.\cite{li2024btsspro}                  & 2024 & MRI Breast Cancer Tumor                                                  & Learnable prompt           & - \\
        Shan et al.\cite{shan2025stpnet}                  & 2025 & Polyp, Chest X‑ray, Chest CT                                                  & Text prompt           & \url{https://github.com/HUANGLIZI/STPNet} \\
        
        Wang et al.\cite{wang2025weakmedsam}                  & 2025 & Brain tumor MRI, Abdominal CT, Cardiac MRI                                                  & Mask prompt           & \url{https://github.com/wanghr64/WeakMedSAM} \\

        Yin et al.\cite{yin2025apg}                  & 2025 & breast ultrasound                                                  & Automatic prompt           & - \\

        Liu et al.\cite{liu2025efficient}                  & 2025 & Joint optic disc (OD) and cup (OC), Brain tumor MRI, Hole heart CT and MRI                                                  & Deformable Convolutional Prompt           & - \\

        Yin et al.\cite{yin2025ddfp}                  & 2025 & Multi-organ abdominal CT and MRI                                                  & Frequency prompt           & - \\
        
        Gao et al.\cite{gao2025dual}                  & 2025 & Total-Body PET                                                  & Textual and disentangled organ features           & - \\

        Tian et al.\cite{tian2025self}                  & 2025 & Plaque and vessel carotid ultrasound & Self-prompt          & - \\
        
        Zou et al.\cite{zou2025acea}                  & 2025 & Prostate MRI & Advanced Prompt Points          & - \\

        Chen et al.\cite{chen2025sam}                  & 2025 & Nuclei & Self-prompt          & \url{ https://github.com/CUHK-AIM-Group/UN-SAM} \\

        Zhang et al.\cite{zhang2025category}                  & 2025 & Nuclei & Category Prompt          & \url{ https://github.com/CUHK-AIM-Group/UN-SAM} \\

        Zhao et al.\cite{zhao2025uncertainty}                  & 2025 & Skin Lesion, Thyroid Ultrasound, Spine CT, Cardiac MR & Auto edge prompt          & - \\
        
        \specialrule{1pt}{0pt}{1pt} 
    \end{tabular}
\end{table*}

\begin{table*}[htbp] 
    \centering
    \caption{Applications of Prompt Engineering in Classification}
    \begin{tabular}{p{2.5cm}p{1cm}p{4cm}p{3cm}p{6cm}} 
        \specialrule{1pt}{0pt}{1pt} 
        Reference & Year & Task & Prompt Type & Link \\
        \midrule
        Yan et al.\cite{yan2024prompt}      & 2024 & Melanoma classification and Cancerous tissue detection, Diabetic Retinopathy classification & Collaborative domain prompts      & \url{https://github.com/SiyuanYan1/PLDG/tree/main} \\
        Zhang et al.\cite{zhang2023text}    & 2023 & Pathological image                                             & Text, Learning Visual Prompt      & \url{https://github.com/Yunkun-Zhang/CITE} \\
        Zhu et al.\cite{zhu2024memory}      & 2024 & Breast cancer, epithelium-stroma tissue histopathology  & Domain prompt                     & - \\
        Huang et al.\cite{huang2023prompt}  & 2023 & Nucleus Detection and Classification                             & Nucleus group embeddings           & - \\
        Qu et al.\cite{qu2024rise}          & 2024 & Breast cancer, lung cancer, and cervical cancer Pathology        & Text                               & \url{https://github.com/miccaiif/TOP} \\
        Guo et al.\cite{guo2023multiple}    & 2023 & Lesion                                                          & Multiple text                      & - \\
        Wang et al.\cite{wang2022medclip}   & 2022 & Disease Classification                                           & Text                               & \url{https://github.com/RyanWangZf/MedCLIP} \\
        Lu et al.\cite{lu2023visual}        & 2023 & Histopathology                                                  & Text                               & \url{https://github.com/mahmoodlab/MI-Zero} \\
        Zhu et al.\cite{zhu2024segprompt}   & 2024 & Kidney Stone                                                   & Segmentation Map                   & - \\
        Huang et al.\cite{huang2024unicell} & 2024 & Nucleus Classification                                         & Label prompt                       & \url{https://github.com/lhaof/UniCell} \\
        Eslami et al.\cite{eslami2021does}  & 2021 & Disease type, tumor location, etc.                              & Text                               & \url{https://github.com/sarahESL/PubMedCLIP} \\
        Zhang et al.\cite{zhang2023large}   & 2023 & Lung tissue, colon tissue, pneumonia                            & Text                               & - \\
        Silva et al.\cite{silva2023foundation} & 2023 & Retina                                                     & Text                               & \url{https://github.com/jusiro/FLAIR} \\
        Cao et al.\cite{cao2023domain}      & 2023 & Brain Tumor, Chest disease                                     & Learnable contexts                  & - \\
        Zheng et al.\cite{zheng2024exploring} & 2024 & Chest X-ray disease                                          & Automatic prompt                    & - \\
        Ye et al.\cite{ye2024pseudo}        & 2024 & Chest X‑ray                                                    & Pseudo-Prompt                       & \url{https://github.com/fallingnight/PsPG} \\
        Huang et al.\cite{huang2024fine}    & 2024 & Fundus, dermoscopic, mammography, chest X‑ray                  & Fine-grained prompt from pre-trained models & - \\
        Lin et al.\cite{lin2024prompt}      & 2024 & Breast cancer histopathology                                   & Prototypical Visual Prompt          & - \\
        Chikontwe et al.\cite{chikontwe2024low} & 2024 & Colorectal cancer, breast cancer metastasis detection in lymph nodes histopathology & Text & - \\
        Han et al.\cite{han2024mscpt}       & 2024 & Lung , breast and Kidney cancer histopathology & Text & - \\
        Yang et al.\cite{yang2024using}     & 2024 & Gastric adenocarcinoma histopathology                          & Text-augmented Visual Prompt        & - \\
        Sanchez et al.\cite{sanchez2024exploring} & 2024 & Blood cell                                               & Few-shot learning, Chain of thought & - \\

        Bai et al.\cite{bai2025label} & 2025 & General                                               & Label-Semantic-Based Prompt & - \\
        Koleilat et al.\cite{koleilat2025biomedcoop} & 2025 & General                                               & Learnable prompt & \url{https://github.com/HealthX-Lab/BiomedCoOp} \\
        He et al.\cite{he2025dvpt} & 2025 & General                                               & Dynamic Visual Prompt & - \\
        Luo et al.\cite{luo2025llm} & 2025 & Skin, Gastrointestinal and Respiratory disease                                              & Decoupled Probabilistic Prompt & \url{https://github.com/CUHK-AIM-Group/LDPP} \\
        
        \specialrule{1pt}{0pt}{1pt} 
    \end{tabular}
    \label{classification}
\end{table*}

\subsection*{Medical Image Generation}
One of the significant challenges in developing large vision models in the medical imaging field is the lack of high-quality training data. This is often attributed to the challenges in accessing medical data, which stem from its scarcity and the privacy concerns associated with medical imaging data. In addition, constructing large-scale, accurately annotated datasets requires much effort from experienced radiologists, which is usually impossible. Medical imaging data sets are often imbalanced as pathologic findings are generally rare, which also hurts the training effect of the model \cite{shin2018medical}. The combination of Large-scale Model (Foundation Model) and prompt engineering offers promising prospects for addressing this issue. In addition to developing expert models, the generated image data can be used for medical education and training. Text-prompt image generation could generate realistic training scenarios for medical students and healthcare professionals. By simulating a diverse array of cases, rare conditions, and complex anatomical variations, synthetic medical image generation could enrich educational programs and offer invaluable hands-on experience in a controlled environment \cite{li2023artificial}.

Expanding on the growing success of Stable Diffusion \cite{rombach2022high} and GAN \cite{goodfellow2020generative} in image generation, an increasing number of researchers are integrating these models into the domain of medical image generation to address the inadequacies stemming from limited medical imaging data. While previously, the generating ability of diffusion models was mostly used for unconditional generation of data, such as \cite{bowles2018gan,kwon2019generation,sun2022hierarchical}, these models usually only have the ability to solve a single problem and have low flexibility, making it difficult to transfer to other tasks. The proposal of Foundation Models such as CLIP and DALL-E has greatly promoted the development of text-to-image and image-to-text fields. Benefiting from the development of these multi-modal models, more and more models are being developed specifically for medical image generation, which will help develop more efficient tools to serve radiologists and patients.

Text prompts play a crucial role in image generation by offering flexibility, interpretability, and the ability to define specific disease types, anatomical structures, or imaging features. This enables the generation of highly customized medical images tailored to particular needs. A representative task in this domain is generating Chest X-rays using medical reports \cite{chambon2022adapting, chambon2022roentgen, bluethgen2024vision, weber2023cascaded, hashmi2024xreal, liang2024covid, han2024advancing, shentu2024cxr, fang2024conditional}, with an illustrative figure adapted from \cite{han2024advancing}, as shown in figure \ref{fig:3}. Beyond this, various tasks have also been explored, including skin image generation\cite{akrout2023diffusion, borghesi2024generation}, MRI generation\cite{pinaya2022brain, wang2024towards,liu2024texdc}, histopathology image generation\cite{yellapragada2024pathldm, moghadam2023morphology,sagers2022improving}, and CT image generation \cite{xu2024medsyn, hamamci2025generatect,yu2024ct,xiao2023end}. Apart from text prompts, depending on the specific application, prompts can take various forms such as conditional variables\cite{pinaya2022brain}, masks\cite{xiao2023end}, and labels\cite{moghadam2023morphology}. 

In summary, broad prompts include text and variables, sequences, classes, etc, used to control model output. For example, in \cite{pinaya2022brain}, age, sex, and brain structure volumes are utilized as prompts to generate the expected brain images. \cite{moghadam2023morphology} generating histopathology images with a genotype prompt. However, text prompts remain the primary focus for future development due to their higher scalability and lower usage threshold.

\subsection*{Medical Image Segmentation}
Accurate medical image segmentation can improve diagnostic accuracy, treatment planning, and disease monitoring \cite{ma2024segment}. In image segmentation, textual and visual prompts are among the most widely applied approaches. Textual prompts are commonly transformed into encoded embeddings by language models and then fed into segmentation models. Visual prompts, including points, boxes, and masks, are typically utilized in Visually Prompted Models. A representative work in this area is SAM, which has quickly triggered the development of large models for medical image segmentation. 

The introduction of CLIP \cite{radford2021learning} has highlighted the enormous potential of textual prompts, reshaping how researchers approach multimodal learning and image segmentation tasks. Among the representative works, \cite{liu2023clip} stands out as a landmark study that developed a CLIP-driven universal model for organ and tumor segmentation, as shown in figure~\ref{fig:4}. In this work, CLIP was utilized to generate segmentation prompts, which, compared to traditional one-hot prompts \cite{zhang2021dodnet}, capture anatomical relationships more effectively and significantly expand the dataset. Building on this work, \cite{zhang2023continual} further optimized the model by assigning a separate, independent MLP to each class, which reduced interference between different classes. However, CLIP has limited ability to generalize in medical scenarios due to the differences between natural and medical texts\cite{ye2023uniseg}. Some studies \cite{zhao2023one,zhao2024tg,zhong2023ariadne} employ contrastive learning or utilize specialized text encoders\cite{gu2021domain,huang2019clinicalbert,boecking2022making} specifically designed for medical images.

SAM, as a representative large-scale model in the field of image segmentation, has inspired many studies \cite{mattjie2023zero,deng2023sam,chen2024sam,na2024segment,xu2023eviprompt,gao2023desam} to focus on enhancing visual prompts for directly segmenting medical images using the SAM model. These methods have significantly improved SAM's capabilities in medical image segmentation. Although SAM demonstrates strong segmentation quality and zero-shot generalization to novel scenes and unseen objects, its training data does not include medical images. Consequently, its performance in most medical image segmentation tasks is often unsatisfactory \cite{mazurowski2023segment,hu2023sam,deng2023segment,roy2023sam}. Therefore, many researchers have focused on developing large models specifically for medical image segmentation. For example, \cite{ma2024segment,cheng2023sam_} collected large volumes of medical image data and fine-tuned it based on SAM, resulting in a large-scale promotable model designed for medical image segmentation. The MedSam was shown in figure \ref{fig:2}. It also explored the effects of different prompts on segmentation performance. Box-based prompts provided more explicit guidance, whereas point-based prompts were more prone to ambiguity in medical image segmentation. In \cite{wang2023sam}, the SAM-Med2D model was extended to SAM-Med3D, where sparse prompts were enhanced with 3D position encodings to capture spatial nuances, while dense prompts were processed using 3D convolutions.

There is also work combining text and visual prompts. For instance, \cite{rameshlugsam} extracted text features from text prompts using BERT \cite{devlin2018bert}, and then created bounding boxes that were used as prompts for SAM. Similarly, \cite{koleilat2024medclip} proposed a novel framework called MedCLIP-SAM, which combines CLIP and SAM models to generate segmentations of clinical scans. They fine-tuned the BiomedCLIP model, and the visual prompts were generated using image and text, post-processed with gScoreCAM \cite{chen2022gscorecam}. Other similar approaches, such as \cite{xie2024simtxtseg,li2024tp,biswas2023polyp}, leverage textual prompts to generate visual prompts that enhance segmentation tasks in medical imaging. Some works, like \cite{du2023segvol}, directly fuse textual and visual prompts to improve segmentation. In addition to visual and textual prompts, other types of prompts have been explored. For example, \cite{fischer2024prompt} proposed a promptable UNet architecture that adapts to segmentation tasks using class-dependent, learnable prompt tokens. Similarly, \cite{ye2023uniseg,ye2024meduniseg} designed a learnable universal prompt that captures the relationships among tasks and converts it into task-specific prompts, which are fed into the decoder as part of its input. 

Visual prompts are often applied to large foundation models such as SAM and its derivatives. These prompts can directly indicate areas of interest in the image, providing a more intuitive and precise approach than textual descriptions. Text prompts, on the other hand, can sometimes lead to inconsistencies in segmentation outcomes due to ambiguous descriptions or variations in interpretation. Placing visual prompts directly on the image is often more efficient than composing detailed text prompts, especially for complex medical images like MRI or CT scans. This approach allows for better adaptation to the complexity and diversity of such images. However, the effectiveness of visual prompts depends on the user's ability to accurately place them, which requires a certain level of professional expertise. Additionally, visual prompts are not easily scalable for batch segmentation of medical images, as each target may have a different location. Text prompts, by contrast, are often integrated into expert models for specific tasks and offer unique value in particular scenarios. These prompts can take various forms, such as reports, templates, or labels, and usually require an encoder to convert them into vectors before input into the segmentation model. On one hand, text-prompt-based models can be trained with multiple types and modalities of data simultaneously, making the segmentation model more versatile. On the other hand, text-prompt-based models can be trained with various types and modalities of data simultaneously, making the segmentation model more versatile and capable of segmenting different organs and lesions. On the other hand, text prompts can capture anatomical relationships, improving segmentation performance and enhancing transfer learning on novel tasks.

\subsection*{Medical Image Classification}
In medical imaging, classification tasks typically include disease classification, identifying benign and malignant tumors, molecular subtyping, lesion detection, and nucleus detection and classification. Prompt-based methods for medical image classification are primarily focused on foundational models, particularly CLIP, where classification is often performed as a downstream task. These foundational models are trained using contrastive learning, which has proven to be a highly effective and scalable strategy \cite{lu2023visual}.

As a leading foundational model, CLIP \cite{radford2021learning} has indirectly driven advancements in medical image classification. Several medical-specific variants of CLIP \cite{eslami2021does,wang2022medclip,zhang2023large,hamamci2024foundation,lu2024radclip} have been developed, with an illustrative figure adapted from \cite{wang2022medclip}, as shown in figure. \ref{fig:1}. These models demonstrate robust zero-shot transfer capabilities through prompts, allowing for strong classification performance with only brief text prompts describing the target classes. As \cite{zhou2022learning} confirmed, performance can be further enhanced by carefully designing appropriate text prompts.

Beyond foundational models, some expert models for medical image classification have also incorporated prompt mechanisms to improve performance. For example, \cite{lu2023visual} developed a prompt template and class name pool, randomly sampling 50 prompts to predict three image categories—introducing the first zero-shot transfer framework for histopathology whole-slide image classification. Similarly, \cite{silva2023foundation} incorporated expert domain knowledge through descriptive textual prompts, enriching the limited categorical supervision typically found in medical datasets. Additionally, \cite{qu2024rise} used GPT-4 to acquire language-based prior knowledge at both the instance and bag levels, effectively addressing the challenge of Few-shot Weakly Supervised Whole-Slide Image Classification. In another approach, \cite{guo2023multiple} used multiple prompts to describe medical lesions, enabling the detection of zero-shot lesions.

Moreover, several studies focus on learnable prompts. \cite{zhang2023text} achieved leading performance by introducing learnable visual prompt tokens. In contrast, \cite{huang2023prompt} designed group prompts as learnable parameters to avoid the inefficiencies of fine-tuning the backbone for nucleus detection and classification. \cite{yan2024prompt} proposed a novel framework called Prompt-driven Latent Domain Generalization to address domain generalization in medical image classification without explicitly relying on domain labels, using a set of learnable token prompts. Similarly, \cite{huang2024unicell} introduced a Dynamic Prompt Module, which dynamically adapts intermediate representations to different dataset sources by leveraging dataset names and label properties as learnable prompts. This module uniformly predicts the corresponding categories of pathological images across various datasets. 

Overall, prompt design is flexible and diverse, ranging from text and labels to learnable prompts. Using prompts helps alleviate the challenge of small data volumes while also improving model generalizability and adaptability. However, designing effective text prompts remains challenging, as prompt designs often lack interpretability. Additionally, foundational models may require substantial computational resources for training and fine-tuning.

\subsection*{Medical Image Foundation Models}
As an essential method for interacting with foundation models, prompts play a crucial role in unlocking the potential of these models. This section summarizes the applications of medical foundation models in medical imaging in recent years. Foundation models—the latest generation of AI models—are trained on massive, diverse datasets and can be applied to numerous downstream tasks \cite{bommasani2021opportunities}. Driven by growing datasets, increases in model size, and advances in model architectures, foundation models offer previously unseen abilities that promise to solve more diverse and challenging tasks than current medical AI models, even while requiring little to no labels for specific tasks \cite{moor2023foundation}. To provide a comprehensive perspective, this section highlights the three most widely used applications of foundation models in medical imaging: histopathology, radiology, and ophthalmology.

Histopathology image evaluation is indispensable for cancer diagnoses and subtype classification\cite{wang2024pathology}. Pathology foundation models typically utilize contrastive learning methods, leveraging large numbers of whole-slide images for self-supervised learning. These pre-trained models can extract pathology imaging features for systematic cancer evaluation. Representative works in this area include references \cite{wang2024pathology,vorontsov2024foundation,vorontsov2024foundation,xiang2025vision,xiang2025vision,huang2023visual}. These foundation models have significantly outperformed state-of-the-art deep learning methods in tasks such as cancer detection, tumor origin characterization, genomic mutation identification, and survival prediction.

Compared to pathology images, radiology images present unique challenges for training foundation models due to the diverse formats and dimensions of data across different imaging modalities. Most foundation models in radiology are developed based on a single imaging modality. For example, in Chest X-ray imaging, representative foundation models such as MedCLIP\cite{wang2022medclip}, CheXzero\cite{tiu2022expert}, CXR-CLIP\cite{you2023cxr}, UniChest\cite{dai2024unichest}, and MedKLIP\cite{wu2023medklip} have been developed for tasks such as disease classification. However, developing foundation models is highly challenging for 3D radiology data such as CT and MRI. Researchers often split volumes into slices or sub-volumes\cite{lei2023clip,niu2023medical}. Nevertheless, some studies \cite{wu2023towards,bai2024m3d,zhao2023one,lu2024radclip,hamamci2024foundation,blankemeier2024merlin} have explored 3D foundation models for tasks such as zero-shot findings classification, phenotype classification, zero-shot cross-modal retrieval, disease prediction, radiology report generation, and segmentation. Although these models demonstrate significant progress, they also highlight that comprehensive 3D foundation models for radiology are still being developed.

Another widely used type of foundation model in medicine is the Ophthalmology foundation model. Not only are these models \cite{zhou2023foundation,engelmann2024training,silva2025foundation,men2023drstagenet} used to diagnose common conditions such as diabetic retinopathy, glaucoma, and age-related macular degeneration, but they can also reveal indicators of systemic conditions like early-onset Parkinson’s disease and cardiovascular disorders. In addition to fundus photography and OCT, other imaging modalities such as angiography, slit-lamp imaging, and ultrasound are also used to develop foundation models\cite{qiu2023visionfm,shi2024eyefound}. These models support a wider range of tasks and outperform expert-designed models.

\section*{Future Directions and Open Challenges}
\begin{figure}[ht]
\centering
\includegraphics[width=\linewidth]{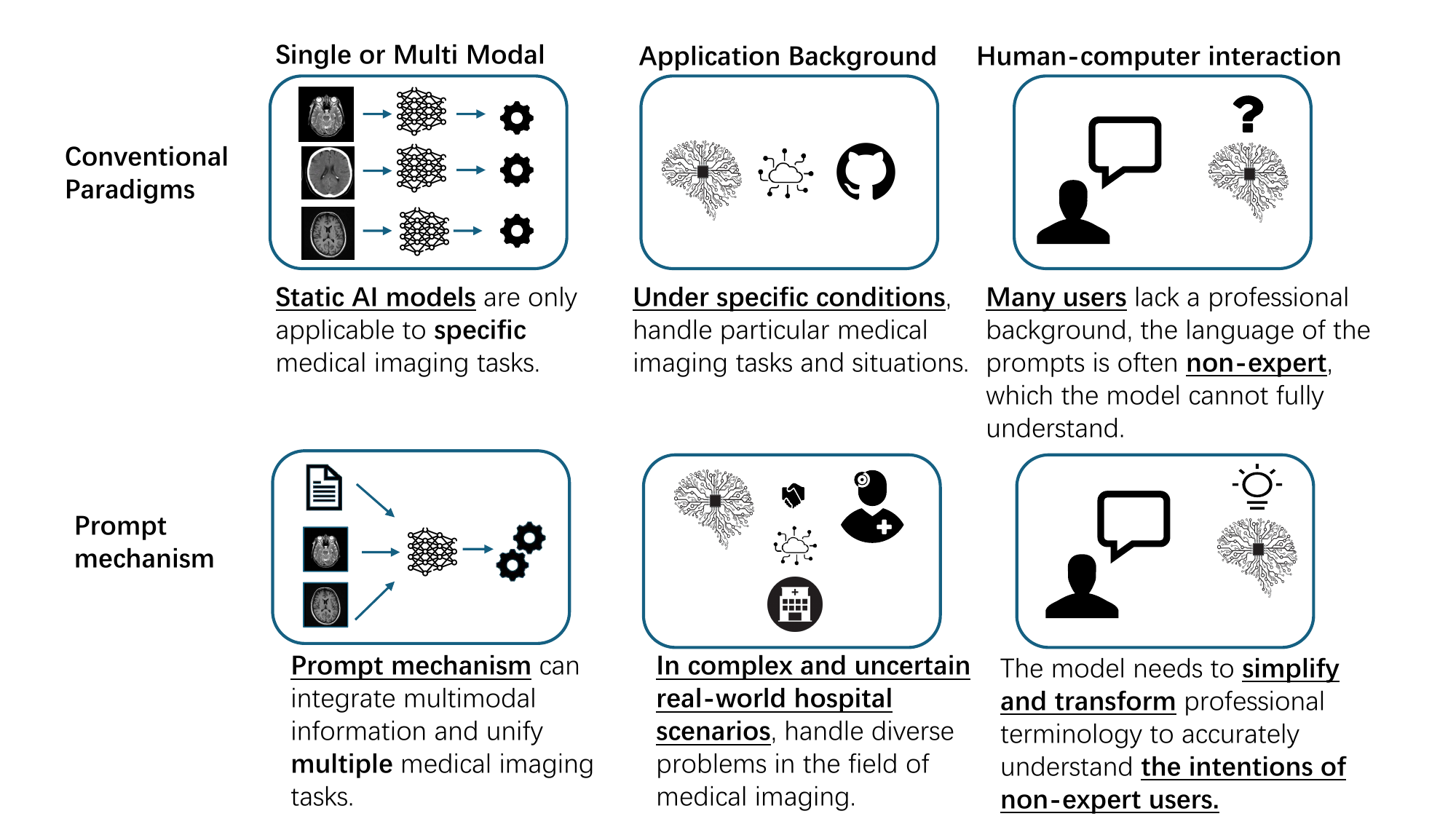}
\caption{The future development of traditional deep learning and the future development of prompt-based deep learning.}
\label{fig:future direction}
\end{figure}

The advent of prompt-based mechanisms signifies a paradigm shift from traditional, often static, AI models towards more dynamic, interactive, and context-aware frameworks in medical imaging. As illustrated in figure~\ref{fig:future direction}, while traditional deep learning models often operate as isolated, task-specific entities, prompt-driven AI evolves towards systems that can flexibly adapt to the complex demands of diverse medical imaging tasks and dynamic clinical environments. This inherent dynamism and the ability to integrate contextual information enhance model performance, generalization, and interpretability. However, realizing the full potential of prompt-driven AI in medicine necessitates addressing several open challenges and exploring promising future research trajectories.


\subsection*{Enhancing Model Capabilities through Advanced Prompting}

Future research will focus on evolving prompt mechanisms to further unlock the capabilities of underlying AI models, particularly large foundation models.

\subsubsection*{Towards Unified Multi-Task and Multi-Modal Generalists}
The complexity of real-world medical practice demands AI systems capable of handling many tasks and data types \cite{moor2023foundation}. Prompt engineering is pivotal in transitioning from single-task, unimodal models to unified multi-task generalists. For example, a single foundation model, guided by different prompts, could perform target detection, image synthesis, and segmentation in various organs and imaging modalities (e.g., CT, MRI, PET), as demonstrated by early efforts such as seq2seq models for synthesis and segmentation \cite{10.1007/978-3-031-72120-5_45} and UMS-Rep for shared knowledge across tasks \cite{zamzmi2020unified}. Future work should aim to develop more sophisticated prompt-based architectures that enhance cross-task knowledge transfer and parameter efficiency, enabling robust performance across a broader spectrum of clinical scenarios.

Concurrently, the transition from unimodal to multimodal prompting holds immense potential \cite{ye2024meduniseg}. Clinical decision making inherently relies on integrating information from various sources (e.g., images, text-based electronic health records, genomic data, and patient history). Multimodal prompts, which condition a model on inputs from several modalities simultaneously, can provide richer contextual information than unimodal prompts, leading to superior performance. For example, TSF-Seq2Seq utilizes image and text information from other modalities as prompts to synthesize images of a specified target modality \cite{han2023explainable}. SimTxtSeg leverages the cross-modal fusion of images and text to generate high-quality pseudo-labels in segmentation tasks \cite{xie2024simtxtseg}. Future research should explore more effective strategies to combine and align multimodal prompt information, developing models that can seamlessly reason across and integrate these diverse data streams to improve diagnostic accuracy and treatment planning.

\subsubsection*{Designing Superior, Robust, and Standardized Prompting Methodologies}
The efficacy of current prompt-based systems heavily relies on the quality and design of the input prompts \cite{10.1007/978-3-031-72120-5_45,han2023explainable,lu2023visual}. However, existing prompt engineering practices face significant challenges:
\begin{itemize}
    \item \textbf{High Sensitivity and Fragility:} Models often exhibit high sensitivity to minor variations in prompt phrasing or structure, often due to the vast, high-dimensional input space they operate in and the specific patterns they learned during pre-training, impacting output consistency and reliability.
    \item \textbf{Lack of Standardization and Automation:} Prompt design is frequently an empirical, manual, and experience-driven process, lacking unified optimization standards, robust automated generation techniques, or systematic evaluation frameworks, leading to variability in performance.
    \item \textbf{Limited Support for Complex Reasoning Tasks:} Current prompts often struggle to effectively guide models through complex, multi-stage reasoning or multi-role interactions, as seen in the need for specialized techniques like Chain-of-Thought prompting, potentially leading to information loss or misinterpretation.
\end{itemize}
Addressing these limitations is crucial. Inspired by advances like CoCoOp \cite{zhou2022conditional}, which enhances prompt generalization by learning conditional input tokens, and SegVol \cite{du2023segvol}, which demonstrates segmentation of numerous anatomical categories using semantic and spatial prompts on a large scale, future research must focus on developing more robust, generalizable, and standardized prompting methods. This includes creating automated prompt optimization techniques, establishing benchmarks for prompt evaluation, developing formalisms for complex task decomposition via prompts, and designing prompts less susceptible to adversarial perturbations or unintended biases.

\subsection*{Seamless Clinical Integration and Practical Deployment}
For prompt-driven AI to make a tangible impact, seamless integration into existing clinical workflows and practical deployment considerations are paramount.

\subsubsection*{Responsive and Adaptive Clinical Prompting Systems}
Current medical ecosystems involve diverse information systems (e.g., EMR, PACS) storing heterogeneous data (including structured data such as lab results and vital signs, alongside unstructured clinical notes and reports). To effectively integrate prompt models, it is essential to overcome system incompatibility and data diversification challenges to enable real-time analysis and interpretation that supports clinical decision making \cite{jiang2017artificial}. Future efforts should therefore focus on building highly responsive and adaptive prompt system frameworks. This involves developing architectures capable of processing complex, real-time clinical data streams, dynamically generating precise prompts tailored to evolving clinical scenarios, and ensuring seamless, interoperable connections with existing medical information systems to reduce data silos.

\subsubsection*{Hardware-Agnostic and Efficient Model Deployment}
The scalability of prompt-based models for deployment on existing hospital hardware, which may be computationally constrained despite being expensive, is critical. Running large foundation models, even with efficient prompting, can be challenging. Therefore, it is necessary to develop effective model scaling and compression techniques-such as knowledge distillation \cite{sinha2020multi,le2025moma,huang2025learnable}, which transfers knowledge from a large model to a smaller one, weight quantization \cite{xie2024mh}, parameter sharing \cite{shi2023deep,liu2023enhancing}, and efficient attention mechanisms \cite{qiu2024learning} tailored to prompted FMs. These methods should enable dynamic adaptation to diverse computational environments, facilitating efficient inference on resource-limited devices prevalent in clinical settings.

\subsubsection*{Optimizing Human–AI Collaboration in Clinical Workflows}
Prompt models can substantially improve collaboration efficiency between clinicians and AI systems. Using human-provided prompts, AI can summarize complex medical data or highlight critical findings, thereby supporting clinical diagnosis. However, determining the optimal division of labor between humans and AI remains a significant challenge. Studies have shown that suboptimal human-AI interaction can sometimes result in performance that is worse than AI operating alone\cite{kim2020changes}; nevertheless, AI assistance often provides significant benefits for junior clinicians\cite{tschandl2020human}. Future work should focus on designing intuitive and efficient prompting interfaces and interaction protocols, with optimizations for different clinical tasks (e.g., real-time surgical guidance versus image‐based diagnostic review). This requires moving beyond generic interaction methods, such as simple clicks or bounding‐box selections common in SAM, and toward developing multi-modal prompting tools (e.g., text, voice, images, masks) that optimize human–AI interaction and enhance performance in clinical settings.

\subsection*{Advancing Human-Computer Interaction for Broader Accessibility}

Significant advances in human-computer interaction are required to make prompt-driven AI accessible and effective for a wider range of users, including those without deep technical or medical experience.

\subsubsection*{Natural Language Interaction for Non-Expert Users}
Current prompt models primarily rely on LLMs to process specialized natural language. However, in real-world medical settings, users (such as patients or administrative staff) often lack professional backgrounds and use lay expressions. Consequently, models must be strengthened in their ability to understand, interpret, and respond to non-expert language. Future research should focus on optimizing conversational AI to convert complex medical terminology into accessible language, accurately capture the intent behind non-expert queries, and deliver practical guidance clearly and intuitively. This approach can draw on systems such as Mani-GPT\cite{zhang2023mani} and DiagGPT\cite{cao2023diaggpt}, which emphasize advanced natural interaction and dynamic dialogue management, including context tracking and user intent recognition. The ultimate goal is to broaden the user base and enhance user experience across a wider range of medical applications through more intuitive, adaptive dialogue strategies while ensuring robust model performance.

\subsubsection*{Co-Adaptive Learning and Personalized Prompts}

Future systems will evolve towards co-adaptive learning, creating a synergistic loop where the AI and its user learn from one another. On one hand, the system will personalize its outputs by understanding a user’s interaction patterns, such as adapting to a radiologist who prefers concise, finding-focused summaries over one requiring detailed anatomical descriptions~\cite{shi2024general}. On the other hand, more effective prompts optimize the human-AI partnership. This powerful adaptive capability extends beyond user preferences to patient-specific data; by generating highly personalized prompts based on an individual's medical history, genetic predispositions, and ongoing treatments, these systems can deliver precise, tailored support for diagnosis and treatment planning, truly embodying the principles of personalized medicine~\cite{shenfeld2025language}.

\section*{Conclusion}
Prompt-based mechanisms represent a transformative advance in medical imaging, significantly enhancing deep learning models' adaptability, precision, and utility, especially for large foundation models. By enabling the effective integration of contextual information and expert knowledge through textual, visual, or multimodal prompts, these methodologies are pivotal in addressing long-standing challenges such as limited annotated data, data heterogeneity, and the complexity of medical scenarios. This review has systematically surveyed the landscape, highlighting how prompts facilitate more accurate image generation, precise delineation of anatomical structures in segmentation, and robust and interpretable classification, even in challenging zero-shot and few-shot learning settings. The continued synergy between advanced prompt engineering and the evolving capabilities of foundation models holds immense potential to revolutionize medical imaging. This will not only advance diagnostic accuracy and optimize treatment outcomes but will also enhance the scalability and accessibility of cutting-edge AI-powered healthcare systems globally, ultimately contributing to more efficient, equitable, and patient-centered medicine.

\bibliography{main}

\begin{thebibliography}{100}
\urlstyle{rm}
\expandafter\ifx\csname url\endcsname\relax
  \def\url#1{\texttt{#1}}\fi
\expandafter\ifx\csname urlprefix\endcsname\relax\def\urlprefix{URL }\fi
\expandafter\ifx\csname doiprefix\endcsname\relax\def\doiprefix{DOI: }\fi
\providecommand{\bibinfo}[2]{#2}
\providecommand{\eprint}[2][]{\url{#2}}

\bibitem{mattjie2023zero}
\bibinfo{author}{Mattjie, C.} \emph{et~al.}
\newblock \bibinfo{title}{Zero-shot performance of the segment anything model (sam) in 2d medical imaging: A comprehensive evaluation and practical guidelines}.
\newblock In \emph{\bibinfo{booktitle}{2023 IEEE 23rd International Conference on Bioinformatics and Bioengineering (BIBE)}}, \bibinfo{pages}{108--112} (\bibinfo{organization}{IEEE Computer Society}, \bibinfo{year}{2023}).

\bibitem{akrout2023diffusion}
\bibinfo{author}{Akrout, M.} \emph{et~al.}
\newblock \bibinfo{journal}{\bibinfo{title}{Diffusion-based data augmentation for skin disease classification: Impact across original medical datasets to fully synthetic images}}.
\newblock {\emph{\JournalTitle{arXiv preprint arXiv:2301.04802}}}  (\bibinfo{year}{2023}).

\bibitem{pinaya2022brain}
\bibinfo{author}{Pinaya, W.~H.} \emph{et~al.}
\newblock \bibinfo{title}{Brain imaging generation with latent diffusion models}.
\newblock In \emph{\bibinfo{booktitle}{MICCAI Workshop on Deep Generative Models}}, \bibinfo{pages}{117--126} (\bibinfo{organization}{Springer}, \bibinfo{year}{2022}).

\bibitem{fischer2024prompt}
\bibinfo{author}{Fischer, M.}, \bibinfo{author}{Bartler, A.} \& \bibinfo{author}{Yang, B.}
\newblock \bibinfo{journal}{\bibinfo{title}{Prompt tuning for parameter-efficient medical image segmentation}}.
\newblock {\emph{\JournalTitle{Medical Image Analysis}}} \textbf{\bibinfo{volume}{91}}, \bibinfo{pages}{103024} (\bibinfo{year}{2024}).

\bibitem{yan2024prompt}
\bibinfo{author}{Yan, S.} \emph{et~al.}
\newblock \bibinfo{journal}{\bibinfo{title}{Prompt-driven latent domain generalization for medical image classification}}.
\newblock {\emph{\JournalTitle{IEEE Transactions on Medical Imaging}}}  (\bibinfo{year}{2024}).

\bibitem{chambon2022adapting}
\bibinfo{author}{Chambon, P.}, \bibinfo{author}{Bluethgen, C.}, \bibinfo{author}{Langlotz, C.~P.} \& \bibinfo{author}{Chaudhari, A.}
\newblock \bibinfo{journal}{\bibinfo{title}{Adapting pretrained vision-language foundational models to medical imaging domains}}.
\newblock {\emph{\JournalTitle{arXiv preprint arXiv:2210.04133}}}  (\bibinfo{year}{2022}).

\bibitem{liu2023clip}
\bibinfo{author}{Liu, J.} \emph{et~al.}
\newblock \bibinfo{title}{Clip-driven universal model for organ segmentation and tumor detection}.
\newblock In \emph{\bibinfo{booktitle}{Proceedings of the IEEE/CVF International Conference on Computer Vision}}, \bibinfo{pages}{21152--21164} (\bibinfo{year}{2023}).

\bibitem{chen2024sam}
\bibinfo{author}{Chen, Z.}, \bibinfo{author}{Xu, Q.}, \bibinfo{author}{Liu, X.} \& \bibinfo{author}{Yuan, Y.}
\newblock \bibinfo{journal}{\bibinfo{title}{Un-sam: Universal prompt-free segmentation for generalized nuclei images}}.
\newblock {\emph{\JournalTitle{arXiv preprint arXiv:2402.16663}}}  (\bibinfo{year}{2024}).

\bibitem{gao2023desam}
\bibinfo{author}{Gao, Y.}, \bibinfo{author}{Xia, W.}, \bibinfo{author}{Hu, D.} \& \bibinfo{author}{Gao, X.}
\newblock \bibinfo{journal}{\bibinfo{title}{Desam: Decoupling segment anything model for generalizable medical image segmentation}}.
\newblock {\emph{\JournalTitle{arXiv preprint arXiv:2306.00499}}}  (\bibinfo{year}{2023}).

\bibitem{cheng2023sam}
\bibinfo{author}{Cheng, D.} \emph{et~al.}
\newblock \bibinfo{journal}{\bibinfo{title}{Sam on medical images: A comprehensive study on three prompt modes}}.
\newblock {\emph{\JournalTitle{arXiv preprint arXiv:2305.00035}}}  (\bibinfo{year}{2023}).

\bibitem{zhang2023text}
\bibinfo{author}{Zhang, Y.} \emph{et~al.}
\newblock \bibinfo{title}{Text-guided foundation model adaptation for pathological image classification}.
\newblock In \emph{\bibinfo{booktitle}{International Conference on Medical Image Computing and Computer-Assisted Intervention}}, \bibinfo{pages}{272--282} (\bibinfo{organization}{Springer}, \bibinfo{year}{2023}).

\bibitem{zhu2024memory}
\bibinfo{author}{Zhu, Y.}, \bibinfo{author}{Li, K.}, \bibinfo{author}{Yu, L.} \& \bibinfo{author}{Heng, P.~A.}
\newblock \bibinfo{title}{Memory-efficient prompt tuning for incremental histopathology classification}.
\newblock In \emph{\bibinfo{booktitle}{Proceedings of the AAAI Conference on Artificial Intelligence}}, vol.~\bibinfo{volume}{38}, \bibinfo{pages}{7802--7810} (\bibinfo{year}{2024}).

\bibitem{silva2025foundation}
\bibinfo{author}{Silva-Rodriguez, J.}, \bibinfo{author}{Chakor, H.}, \bibinfo{author}{Kobbi, R.}, \bibinfo{author}{Dolz, J.} \& \bibinfo{author}{Ayed, I.~B.}
\newblock \bibinfo{journal}{\bibinfo{title}{A foundation language-image model of the retina (flair): Encoding expert knowledge in text supervision}}.
\newblock {\emph{\JournalTitle{Medical Image Analysis}}} \textbf{\bibinfo{volume}{99}}, \bibinfo{pages}{103357} (\bibinfo{year}{2025}).

\bibitem{ye2023uniseg}
\bibinfo{author}{Ye, Y.}, \bibinfo{author}{Xie, Y.}, \bibinfo{author}{Zhang, J.}, \bibinfo{author}{Chen, Z.} \& \bibinfo{author}{Xia, Y.}
\newblock \bibinfo{title}{Uniseg: A prompt-driven universal segmentation model as well as a strong representation learner}.
\newblock In \emph{\bibinfo{booktitle}{International Conference on Medical Image Computing and Computer-Assisted Intervention}}, \bibinfo{pages}{508--518} (\bibinfo{organization}{Springer}, \bibinfo{year}{2023}).

\bibitem{cao2024domain}
\bibinfo{author}{Cao, Q.}, \bibinfo{author}{Xu, Z.}, \bibinfo{author}{Chen, Y.}, \bibinfo{author}{Ma, C.} \& \bibinfo{author}{Yang, X.}
\newblock \bibinfo{title}{Domain prompt learning with quaternion networks}.
\newblock In \emph{\bibinfo{booktitle}{Proceedings of the IEEE/CVF Conference on Computer Vision and Pattern Recognition}}, \bibinfo{pages}{26637--26646} (\bibinfo{year}{2024}).

\bibitem{zhou2024medsam}
\bibinfo{author}{Zhou, N.} \emph{et~al.}
\newblock \bibinfo{journal}{\bibinfo{title}{Medsam-u: Uncertainty-guided auto multi-prompt adaptation for reliable medsam}}.
\newblock {\emph{\JournalTitle{arXiv preprint arXiv:2409.00924}}}  (\bibinfo{year}{2024}).

\bibitem{singhal2022large}
\bibinfo{author}{Singhal, K.} \emph{et~al.}
\newblock \bibinfo{journal}{\bibinfo{title}{Large language models encode clinical knowledge}}.
\newblock {\emph{\JournalTitle{arXiv preprint arXiv:2212.13138}}}  (\bibinfo{year}{2022}).

\bibitem{wahd2024sam2rad}
\bibinfo{author}{Wahd, A.~S.} \emph{et~al.}
\newblock \bibinfo{journal}{\bibinfo{title}{Sam2rad: A segmentation model for medical images with learnable prompts}}.
\newblock {\emph{\JournalTitle{arXiv preprint arXiv:2409.06821}}}  (\bibinfo{year}{2024}).

\bibitem{yellapragada2024pathldm}
\bibinfo{author}{Yellapragada, S.} \emph{et~al.}
\newblock \bibinfo{title}{Pathldm: Text conditioned latent diffusion model for histopathology}.
\newblock In \emph{\bibinfo{booktitle}{Proceedings of the IEEE/CVF Winter Conference on Applications of Computer Vision}}, \bibinfo{pages}{5182--5191} (\bibinfo{year}{2024}).

\bibitem{xu2024medsyn}
\bibinfo{author}{Xu, Y.} \emph{et~al.}
\newblock \bibinfo{journal}{\bibinfo{title}{Medsyn: Text-guided anatomy-aware synthesis of high-fidelity 3d ct images}}.
\newblock {\emph{\JournalTitle{IEEE Transactions on Medical Imaging}}}  (\bibinfo{year}{2024}).

\bibitem{zhang2023large}
\bibinfo{author}{Zhang, S.} \emph{et~al.}
\newblock \bibinfo{journal}{\bibinfo{title}{Large-scale domain-specific pretraining for biomedical vision-language processing}}.
\newblock {\emph{\JournalTitle{arXiv preprint arXiv:2303.00915}}}  (\bibinfo{year}{2023}).

\bibitem{wang2022medclip}
\bibinfo{author}{Wang, Z.}, \bibinfo{author}{Wu, Z.}, \bibinfo{author}{Agarwal, D.} \& \bibinfo{author}{Sun, J.}
\newblock \bibinfo{title}{Medclip: Contrastive learning from unpaired medical images and text}.
\newblock In \emph{\bibinfo{booktitle}{Proceedings of the 2022 Conference on Empirical Methods in Natural Language Processing}}, \bibinfo{pages}{3876--3887} (\bibinfo{year}{2022}).

\bibitem{gu2021domain}
\bibinfo{author}{Gu, Y.} \emph{et~al.}
\newblock \bibinfo{journal}{\bibinfo{title}{Domain-specific language model pretraining for biomedical natural language processing}}.
\newblock {\emph{\JournalTitle{ACM Transactions on Computing for Healthcare (HEALTH)}}} \textbf{\bibinfo{volume}{3}}, \bibinfo{pages}{1--23} (\bibinfo{year}{2021}).

\bibitem{huang2019clinicalbert}
\bibinfo{author}{Huang, K.}, \bibinfo{author}{Altosaar, J.} \& \bibinfo{author}{Ranganath, R.}
\newblock \bibinfo{journal}{\bibinfo{title}{Clinicalbert: Modeling clinical notes and predicting hospital readmission}}.
\newblock {\emph{\JournalTitle{arXiv preprint arXiv:1904.05342}}}  (\bibinfo{year}{2019}).

\bibitem{huang2019ccnet}
\bibinfo{author}{Huang, Z.} \emph{et~al.}
\newblock \bibinfo{title}{Ccnet: Criss-cross attention for semantic segmentation}.
\newblock In \emph{\bibinfo{booktitle}{Proceedings of the IEEE/CVF international conference on computer vision}}, \bibinfo{pages}{603--612} (\bibinfo{year}{2019}).

\bibitem{chen2021crossvit}
\bibinfo{author}{Chen, C.-F.~R.}, \bibinfo{author}{Fan, Q.} \& \bibinfo{author}{Panda, R.}
\newblock \bibinfo{title}{Crossvit: Cross-attention multi-scale vision transformer for image classification}.
\newblock In \emph{\bibinfo{booktitle}{Proceedings of the IEEE/CVF international conference on computer vision}}, \bibinfo{pages}{357--366} (\bibinfo{year}{2021}).

\bibitem{petit2021u}
\bibinfo{author}{Petit, O.} \emph{et~al.}
\newblock \bibinfo{title}{U-net transformer: Self and cross attention for medical image segmentation}.
\newblock In \emph{\bibinfo{booktitle}{Machine Learning in Medical Imaging: 12th International Workshop, MLMI 2021, Held in Conjunction with MICCAI 2021, Strasbourg, France, September 27, 2021, Proceedings 12}}, \bibinfo{pages}{267--276} (\bibinfo{organization}{Springer}, \bibinfo{year}{2021}).

\bibitem{lin2022cat}
\bibinfo{author}{Lin, H.}, \bibinfo{author}{Cheng, X.}, \bibinfo{author}{Wu, X.} \& \bibinfo{author}{Shen, D.}
\newblock \bibinfo{title}{Cat: Cross attention in vision transformer}.
\newblock In \emph{\bibinfo{booktitle}{2022 IEEE international conference on multimedia and expo (ICME)}}, \bibinfo{pages}{1--6} (\bibinfo{organization}{IEEE}, \bibinfo{year}{2022}).

\bibitem{dai2024guidegen}
\bibinfo{author}{Dai, L.}, \bibinfo{author}{Zhang, R.}, \bibinfo{author}{Huang, Z.} \& \bibinfo{author}{Zhang, X.}
\newblock \bibinfo{journal}{\bibinfo{title}{Guidegen: A text-guided framework for joint ct volume and anatomical structure generation}}.
\newblock {\emph{\JournalTitle{arXiv preprint arXiv:2403.07247}}}  (\bibinfo{year}{2024}).

\bibitem{bluethgen2024vision}
\bibinfo{author}{Bluethgen, C.} \emph{et~al.}
\newblock \bibinfo{journal}{\bibinfo{title}{A vision--language foundation model for the generation of realistic chest x-ray images}}.
\newblock {\emph{\JournalTitle{Nature Biomedical Engineering}}} \bibinfo{pages}{1--13} (\bibinfo{year}{2024}).

\bibitem{chambon2022roentgen}
\bibinfo{author}{Chambon, P.} \emph{et~al.}
\newblock \bibinfo{journal}{\bibinfo{title}{Roentgen: vision-language foundation model for chest x-ray generation}}.
\newblock {\emph{\JournalTitle{arXiv preprint arXiv:2211.12737}}}  (\bibinfo{year}{2022}).

\bibitem{hashmi2024xreal}
\bibinfo{author}{Hashmi, A. U.~R.} \emph{et~al.}
\newblock \bibinfo{journal}{\bibinfo{title}{Xreal: Realistic anatomy and pathology-aware x-ray generation via controllable diffusion model}}.
\newblock {\emph{\JournalTitle{arXiv preprint arXiv:2403.09240}}}  (\bibinfo{year}{2024}).

\bibitem{liang2024covid}
\bibinfo{author}{Liang, Z.}, \bibinfo{author}{Xue, Z.}, \bibinfo{author}{Rajaraman, S.} \& \bibinfo{author}{Antani, S.}
\newblock \bibinfo{title}{Covid-19 pneumonia chest x-ray pattern synthesis by stable diffusion}.
\newblock In \emph{\bibinfo{booktitle}{2024 IEEE Southwest Symposium on Image Analysis and Interpretation (SSIAI)}}, \bibinfo{pages}{21--24} (\bibinfo{organization}{IEEE}, \bibinfo{year}{2024}).

\bibitem{han2024advancing}
\bibinfo{author}{Han, W.}, \bibinfo{author}{Kim, C.}, \bibinfo{author}{Ju, D.}, \bibinfo{author}{Shim, Y.} \& \bibinfo{author}{Hwang, S.~J.}
\newblock \bibinfo{title}{Advancing text-driven chest x-ray generation with policy-based reinforcement learning}.
\newblock In \emph{\bibinfo{booktitle}{International Conference on Medical Image Computing and Computer-Assisted Intervention}}, \bibinfo{pages}{56--66} (\bibinfo{organization}{Springer}, \bibinfo{year}{2024}).

\bibitem{liu2024texdc}
\bibinfo{author}{Liu, C.}, \bibinfo{author}{Yuan, X.}, \bibinfo{author}{Yu, Z.} \& \bibinfo{author}{Wang, Y.}
\newblock \bibinfo{title}{Texdc: Text-driven disease-aware 4d cardiac cine mri images generation}.
\newblock In \emph{\bibinfo{booktitle}{Proceedings of the Asian Conference on Computer Vision}}, \bibinfo{pages}{3005--3021} (\bibinfo{year}{2024}).

\bibitem{hofmanninger2020automatic}
\bibinfo{author}{Hofmanninger, J.} \emph{et~al.}
\newblock \bibinfo{journal}{\bibinfo{title}{Automatic lung segmentation in routine imaging is primarily a data diversity problem, not a methodology problem}}.
\newblock {\emph{\JournalTitle{European Radiology Experimental}}} \textbf{\bibinfo{volume}{4}}, \bibinfo{pages}{1--13} (\bibinfo{year}{2020}).

\bibitem{wang2022naviairway}
\bibinfo{author}{Wang, A.}, \bibinfo{author}{Tam, T. C.~C.}, \bibinfo{author}{Poon, H.~M.}, \bibinfo{author}{Yu, K.-C.} \& \bibinfo{author}{Lee, W.-N.}
\newblock \bibinfo{journal}{\bibinfo{title}{Naviairway: a bronchiole-sensitive deep learning-based airway segmentation pipeline}}.
\newblock {\emph{\JournalTitle{arXiv preprint arXiv:2203.04294}}}  (\bibinfo{year}{2022}).

\bibitem{wasserthal2023totalsegmentator}
\bibinfo{author}{Wasserthal, J.} \emph{et~al.}
\newblock \bibinfo{journal}{\bibinfo{title}{Totalsegmentator: robust segmentation of 104 anatomic structures in ct images}}.
\newblock {\emph{\JournalTitle{Radiology: Artificial Intelligence}}} \textbf{\bibinfo{volume}{5}} (\bibinfo{year}{2023}).

\bibitem{10.1007/978-3-031-72120-5_45}
\bibinfo{author}{Han, L.} \emph{et~al.}
\newblock \bibinfo{title}{Non-adversarial learning: Vector-quantized common latent space for multi-sequence mri}.
\newblock In \bibinfo{editor}{Linguraru, M.~G.} \emph{et~al.} (eds.) \emph{\bibinfo{booktitle}{Medical Image Computing and Computer Assisted Intervention -- MICCAI 2024}}, \bibinfo{pages}{481--491} (\bibinfo{publisher}{Springer Nature Switzerland}, \bibinfo{address}{Cham}, \bibinfo{year}{2024}).

\bibitem{han2024synthesis}
\bibinfo{author}{Han, L.} \emph{et~al.}
\newblock \bibinfo{journal}{\bibinfo{title}{Synthesis-based imaging-differentiation representation learning for multi-sequence 3d/4d mri}}.
\newblock {\emph{\JournalTitle{Medical Image Analysis}}} \textbf{\bibinfo{volume}{92}}, \bibinfo{pages}{103044} (\bibinfo{year}{2024}).

\bibitem{moghadam2023morphology}
\bibinfo{author}{Moghadam, P.~A.} \emph{et~al.}
\newblock \bibinfo{title}{A morphology focused diffusion probabilistic model for synthesis of histopathology images}.
\newblock In \emph{\bibinfo{booktitle}{Proceedings of the IEEE/CVF Winter Conference on Applications of Computer Vision}}, \bibinfo{pages}{2000--2009} (\bibinfo{year}{2023}).

\bibitem{ma2024segment}
\bibinfo{author}{Ma, J.} \emph{et~al.}
\newblock \bibinfo{journal}{\bibinfo{title}{Segment anything in medical images}}.
\newblock {\emph{\JournalTitle{Nature Communications}}} \textbf{\bibinfo{volume}{15}}, \bibinfo{pages}{654} (\bibinfo{year}{2024}).

\bibitem{li2024promise}
\bibinfo{author}{Li, H.}, \bibinfo{author}{Liu, H.}, \bibinfo{author}{Hu, D.}, \bibinfo{author}{Wang, J.} \& \bibinfo{author}{Oguz, I.}
\newblock \bibinfo{title}{Promise: Prompt-driven 3d medical image segmentation using pretrained image foundation models}.
\newblock In \emph{\bibinfo{booktitle}{2024 IEEE International Symposium on Biomedical Imaging (ISBI)}}, \bibinfo{pages}{1--5} (\bibinfo{organization}{IEEE}, \bibinfo{year}{2024}).

\bibitem{deng2023sam}
\bibinfo{author}{Deng, G.} \emph{et~al.}
\newblock \bibinfo{title}{Sam-u: Multi-box prompts triggered uncertainty estimation for reliable sam in medical image}.
\newblock In \emph{\bibinfo{booktitle}{International Conference on Medical Image Computing and Computer-Assisted Intervention}}, \bibinfo{pages}{368--377} (\bibinfo{organization}{Springer}, \bibinfo{year}{2023}).

\bibitem{wang2024sam}
\bibinfo{author}{Wang, G.} \emph{et~al.}
\newblock \bibinfo{title}{Sam-med3d-moe: Towards a non-forgetting segment anything model via mixture of experts for 3d medical image segmentation}.
\newblock In \emph{\bibinfo{booktitle}{International Conference on Medical Image Computing and Computer-Assisted Intervention}}, \bibinfo{pages}{552--561} (\bibinfo{organization}{Springer}, \bibinfo{year}{2024}).

\bibitem{du2023segvol}
\bibinfo{author}{Du, Y.}, \bibinfo{author}{Bai, F.}, \bibinfo{author}{Huang, T.} \& \bibinfo{author}{Zhao, B.}
\newblock \bibinfo{journal}{\bibinfo{title}{Segvol: Universal and interactive volumetric medical image segmentation}}.
\newblock {\emph{\JournalTitle{arXiv preprint arXiv:2311.13385}}}  (\bibinfo{year}{2023}).

\bibitem{rameshlugsam}
\bibinfo{author}{Ramesh, D.~B.}, \bibinfo{author}{Iytha~Sridhar, R.}, \bibinfo{author}{Upadhyaya, P.} \& \bibinfo{author}{Kamaleswaran, R.}
\newblock \bibinfo{journal}{\bibinfo{title}{Lugsam: A novel framework for integrating text prompts to segment anything model (sam) for segmentation tasks of icu chest x-rays}}.
\newblock {\emph{\JournalTitle{Pulakesh and Kamaleswaran, Rishikesan, Lugsam: A Novel Framework for Integrating Text Prompts to Segment Anything Model (Sam) for Segmentation Tasks of Icu Chest X-Rays}}} .

\bibitem{koleilat2024medclip}
\bibinfo{author}{Koleilat, T.}, \bibinfo{author}{Asgariandehkordi, H.}, \bibinfo{author}{Rivaz, H.} \& \bibinfo{author}{Xiao, Y.}
\newblock \bibinfo{journal}{\bibinfo{title}{Medclip-samv2: Towards universal text-driven medical image segmentation}}.
\newblock {\emph{\JournalTitle{arXiv preprint arXiv:2409.19483}}}  (\bibinfo{year}{2024}).

\bibitem{kirillov2023segment}
\bibinfo{author}{Kirillov, A.} \emph{et~al.}
\newblock \bibinfo{journal}{\bibinfo{title}{Segment anything}}.
\newblock {\emph{\JournalTitle{arXiv preprint arXiv:2304.02643}}}  (\bibinfo{year}{2023}).

\bibitem{zhong2023ariadne}
\bibinfo{author}{Zhong, Y.}, \bibinfo{author}{Xu, M.}, \bibinfo{author}{Liang, K.}, \bibinfo{author}{Chen, K.} \& \bibinfo{author}{Wu, M.}
\newblock \bibinfo{title}{Ariadne’s thread: Using text prompts to improve segmentation of infected areas from chest x-ray images}.
\newblock In \emph{\bibinfo{booktitle}{International Conference on Medical Image Computing and Computer-Assisted Intervention}}, \bibinfo{pages}{724--733} (\bibinfo{organization}{Springer}, \bibinfo{year}{2023}).

\bibitem{xie2024simtxtseg}
\bibinfo{author}{Xie, Y.}, \bibinfo{author}{Zhou, T.}, \bibinfo{author}{Zhou, Y.} \& \bibinfo{author}{Chen, G.}
\newblock \bibinfo{journal}{\bibinfo{title}{Simtxtseg: Weakly-supervised medical image segmentation with simple text cues}}.
\newblock {\emph{\JournalTitle{arXiv preprint arXiv:2406.19364}}}  (\bibinfo{year}{2024}).

\bibitem{li2024tp}
\bibinfo{author}{Li, W.}, \bibinfo{author}{Xiong, X.}, \bibinfo{author}{Xia, P.}, \bibinfo{author}{Ju, L.} \& \bibinfo{author}{Ge, Z.}
\newblock \bibinfo{journal}{\bibinfo{title}{Tp-drseg: Improving diabetic retinopathy lesion segmentation with explicit text-prompts assisted sam}}.
\newblock {\emph{\JournalTitle{arXiv preprint arXiv:2406.15764}}}  (\bibinfo{year}{2024}).

\bibitem{ye2024meduniseg}
\bibinfo{author}{Ye, Y.}, \bibinfo{author}{Chen, Z.}, \bibinfo{author}{Zhang, J.}, \bibinfo{author}{Xie, Y.} \& \bibinfo{author}{Xia, Y.}
\newblock \bibinfo{journal}{\bibinfo{title}{Meduniseg: 2d and 3d medical image segmentation via a prompt-driven universal model}}.
\newblock {\emph{\JournalTitle{arXiv preprint arXiv:2410.05905}}}  (\bibinfo{year}{2024}).

\bibitem{liu2024dctp}
\bibinfo{author}{Liu, J.} \emph{et~al.}
\newblock \bibinfo{journal}{\bibinfo{title}{Dctp-net: Dual-branch clip-enhance textual prompt-aware network for acute ischemic stroke lesion segmentation from ct image}}.
\newblock {\emph{\JournalTitle{IEEE Journal of Biomedical and Health Informatics}}}  (\bibinfo{year}{2024}).

\bibitem{lin2024fedlppa}
\bibinfo{author}{Lin, L.} \emph{et~al.}
\newblock \bibinfo{journal}{\bibinfo{title}{Fedlppa: Learning personalized prompt and aggregation for federated weakly-supervised medical image segmentation}}.
\newblock {\emph{\JournalTitle{arXiv preprint arXiv:2402.17502}}}  (\bibinfo{year}{2024}).

\bibitem{lin2023multi}
\bibinfo{author}{Lin, Y.} \emph{et~al.}
\newblock \bibinfo{title}{Multi-target domain adaptation with prompt learning for medical image segmentation}.
\newblock In \emph{\bibinfo{booktitle}{International Conference on Medical Image Computing and Computer-Assisted Intervention}}, \bibinfo{pages}{717--727} (\bibinfo{organization}{Springer}, \bibinfo{year}{2023}).

\bibitem{na2024segment}
\bibinfo{author}{Na, S.}, \bibinfo{author}{Guo, Y.}, \bibinfo{author}{Jiang, F.}, \bibinfo{author}{Ma, H.} \& \bibinfo{author}{Huang, J.}
\newblock \bibinfo{journal}{\bibinfo{title}{Segment any cell: A sam-based auto-prompting fine-tuning framework for nuclei segmentation}}.
\newblock {\emph{\JournalTitle{arXiv preprint arXiv:2401.13220}}}  (\bibinfo{year}{2024}).

\bibitem{luo2023universal}
\bibinfo{author}{Luo, W.} \emph{et~al.}
\newblock \bibinfo{title}{Universal medical image segmentation with task-specific prompt-guided transformer model}.
\newblock In \emph{\bibinfo{booktitle}{2023 International Annual Conference on Complex Systems and Intelligent Science (CSIS-IAC)}}, \bibinfo{pages}{569--575} (\bibinfo{organization}{IEEE}, \bibinfo{year}{2023}).

\bibitem{chen2024each}
\bibinfo{author}{Chen, Z.}, \bibinfo{author}{Pan, Y.}, \bibinfo{author}{Ye, Y.}, \bibinfo{author}{Lu, M.} \& \bibinfo{author}{Xia, Y.}
\newblock \bibinfo{title}{Each test image deserves a specific prompt: Continual test-time adaptation for 2d medical image segmentation}.
\newblock In \emph{\bibinfo{booktitle}{Proceedings of the IEEE/CVF Conference on Computer Vision and Pattern Recognition}}, \bibinfo{pages}{11184--11193} (\bibinfo{year}{2024}).

\bibitem{zhang2023biomedclip}
\bibinfo{author}{Zhang, S.} \emph{et~al.}
\newblock \bibinfo{journal}{\bibinfo{title}{Biomedclip: a multimodal biomedical foundation model pretrained from fifteen million scientific image-text pairs}}.
\newblock {\emph{\JournalTitle{arXiv preprint arXiv:2303.00915}}}  (\bibinfo{year}{2023}).

\bibitem{eslami2023pubmedclip}
\bibinfo{author}{Eslami, S.}, \bibinfo{author}{Meinel, C.} \& \bibinfo{author}{De~Melo, G.}
\newblock \bibinfo{title}{Pubmedclip: How much does clip benefit visual question answering in the medical domain?}
\newblock In \emph{\bibinfo{booktitle}{Findings of the Association for Computational Linguistics: EACL 2023}}, \bibinfo{pages}{1181--1193} (\bibinfo{year}{2023}).

\bibitem{eslami2021does}
\bibinfo{author}{Eslami, S.}, \bibinfo{author}{de~Melo, G.} \& \bibinfo{author}{Meinel, C.}
\newblock \bibinfo{journal}{\bibinfo{title}{Does clip benefit visual question answering in the medical domain as much as it does in the general domain?}}
\newblock {\emph{\JournalTitle{arXiv preprint arXiv:2112.13906}}}  (\bibinfo{year}{2021}).

\bibitem{bie2024xcoop}
\bibinfo{author}{Bie, Y.}, \bibinfo{author}{Luo, L.}, \bibinfo{author}{Chen, Z.} \& \bibinfo{author}{Chen, H.}
\newblock \bibinfo{journal}{\bibinfo{title}{Xcoop: Explainable prompt learning for computer-aided diagnosis via concept-guided context optimization}}.
\newblock {\emph{\JournalTitle{arXiv preprint arXiv:2403.09410}}}  (\bibinfo{year}{2024}).

\bibitem{han2024mscpt}
\bibinfo{author}{Han, M.} \emph{et~al.}
\newblock \bibinfo{journal}{\bibinfo{title}{Mscpt: Few-shot whole slide image classification with multi-scale and context-focused prompt tuning}}.
\newblock {\emph{\JournalTitle{arXiv preprint arXiv:2408.11505}}}  (\bibinfo{year}{2024}).

\bibitem{qu2024rise}
\bibinfo{author}{Qu, L.}, \bibinfo{author}{Fu, K.}, \bibinfo{author}{Wang, M.}, \bibinfo{author}{Song, Z.} \emph{et~al.}
\newblock \bibinfo{journal}{\bibinfo{title}{The rise of ai language pathologists: Exploring two-level prompt learning for few-shot weakly-supervised whole slide image classification}}.
\newblock {\emph{\JournalTitle{Advances in Neural Information Processing Systems}}} \textbf{\bibinfo{volume}{36}} (\bibinfo{year}{2024}).

\bibitem{chikontwe2024low}
\bibinfo{author}{Chikontwe, P.}, \bibinfo{author}{Kang, M.}, \bibinfo{author}{Luna, M.}, \bibinfo{author}{Nam, S.} \& \bibinfo{author}{Park, S.~H.}
\newblock \bibinfo{title}{Low-shot prompt tuning for multiple instance learning based histology classification}.
\newblock In \emph{\bibinfo{booktitle}{International Conference on Medical Image Computing and Computer-Assisted Intervention}}, \bibinfo{pages}{285--295} (\bibinfo{organization}{Springer}, \bibinfo{year}{2024}).

\bibitem{silva2023foundation}
\bibinfo{author}{Silva-Rodriguez, J.}, \bibinfo{author}{Chakor, H.}, \bibinfo{author}{Kobbi, R.}, \bibinfo{author}{Dolz, J.} \& \bibinfo{author}{Ayed, I.~B.}
\newblock \bibinfo{journal}{\bibinfo{title}{A foundation language-image model of the retina (flair): Encoding expert knowledge in text supervision}}.
\newblock {\emph{\JournalTitle{arXiv preprint arXiv:2308.07898}}}  (\bibinfo{year}{2023}).

\bibitem{lu2023visual}
\bibinfo{author}{Lu, M.~Y.} \emph{et~al.}
\newblock \bibinfo{title}{Visual language pretrained multiple instance zero-shot transfer for histopathology images}.
\newblock In \emph{\bibinfo{booktitle}{Proceedings of the IEEE/CVF Conference on Computer Vision and Pattern Recognition}}, \bibinfo{pages}{19764--19775} (\bibinfo{year}{2023}).

\bibitem{huang2023prompt}
\bibinfo{author}{Huang, J.}, \bibinfo{author}{Li, H.}, \bibinfo{author}{Sun, W.}, \bibinfo{author}{Wan, X.} \& \bibinfo{author}{Li, G.}
\newblock \bibinfo{title}{Prompt-based grouping transformer for nucleus detection and classification}.
\newblock In \emph{\bibinfo{booktitle}{International Conference on Medical Image Computing and Computer-Assisted Intervention}}, \bibinfo{pages}{569--579} (\bibinfo{organization}{Springer}, \bibinfo{year}{2023}).

\bibitem{zhu2024segprompt}
\bibinfo{author}{Zhu, W.} \emph{et~al.}
\newblock \bibinfo{title}{Segprompt: Using segmentation map as a better prompt to finetune deep models for kidney stone classification}.
\newblock In \emph{\bibinfo{booktitle}{Medical Imaging with Deep Learning}}, \bibinfo{pages}{1680--1690} (\bibinfo{organization}{PMLR}, \bibinfo{year}{2024}).

\bibitem{huang2024unicell}
\bibinfo{author}{Huang, J.}, \bibinfo{author}{Li, H.}, \bibinfo{author}{Wan, X.} \& \bibinfo{author}{Li, G.}
\newblock \bibinfo{journal}{\bibinfo{title}{Unicell: Universal cell nucleus classification via prompt learning}}.
\newblock {\emph{\JournalTitle{arXiv preprint arXiv:2402.12938}}}  (\bibinfo{year}{2024}).

\bibitem{ye2024pseudo}
\bibinfo{author}{Ye, Y.}, \bibinfo{author}{Zhang, J.} \& \bibinfo{author}{Shi, H.}
\newblock \bibinfo{title}{Pseudo-prompt generating in pre-trained vision-language models for multi-label medical image classification}.
\newblock In \emph{\bibinfo{booktitle}{Chinese Conference on Pattern Recognition and Computer Vision (PRCV)}}, \bibinfo{pages}{279--298} (\bibinfo{organization}{Springer}, \bibinfo{year}{2024}).

\bibitem{lin2024prompt}
\bibinfo{author}{Lin, Y.}, \bibinfo{author}{Zhu, Z.}, \bibinfo{author}{Cheng, K.-T.} \& \bibinfo{author}{Chen, H.}
\newblock \bibinfo{journal}{\bibinfo{title}{Prompt-guided adaptive model transformation for whole slide image classification}}.
\newblock {\emph{\JournalTitle{arXiv preprint arXiv:2403.12537}}}  (\bibinfo{year}{2024}).

\bibitem{sanchez2024exploring}
\bibinfo{author}{Sánchez~Quijada, M.}
\newblock \bibinfo{journal}{\bibinfo{title}{Exploring large vision-language models with prompt engineering for peripheral blood cell image analysis and classification}}.
\newblock {\emph{\JournalTitle{Universitat Oberta de Catalunya (UOC)}}}  (\bibinfo{year}{2024}).

\bibitem{hamamci2025generatect}
\bibinfo{author}{Hamamci, I.~E.} \emph{et~al.}
\newblock \bibinfo{title}{Generatect: Text-conditional generation of 3d chest ct volumes}.
\newblock In \emph{\bibinfo{booktitle}{European Conference on Computer Vision}}, \bibinfo{pages}{126--143} (\bibinfo{organization}{Springer}, \bibinfo{year}{2025}).

\bibitem{wang2024towards}
\bibinfo{author}{Wang, Y.} \emph{et~al.}
\newblock \bibinfo{journal}{\bibinfo{title}{Towards general text-guided image synthesis for customized multimodal brain mri generation}}.
\newblock {\emph{\JournalTitle{arXiv preprint arXiv:2409.16818}}}  (\bibinfo{year}{2024}).

\bibitem{shi2025semantic}
\bibinfo{author}{Shi, S.}, \bibinfo{author}{Li, H.}, \bibinfo{author}{Zhang, Y.} \& \bibinfo{author}{Wang, X.}
\newblock \bibinfo{journal}{\bibinfo{title}{Semantic information-guided attentional gan-based ultrasound image synthesis method}}.
\newblock {\emph{\JournalTitle{Biomedical Signal Processing and Control}}} \textbf{\bibinfo{volume}{102}}, \bibinfo{pages}{107273} (\bibinfo{year}{2025}).

\bibitem{dahan2024csg}
\bibinfo{author}{Dahan, E.} \emph{et~al.}
\newblock \bibinfo{journal}{\bibinfo{title}{Csg: A context-semantic guided diffusion approach in de novo musculoskeletal ultrasound image generation}}.
\newblock {\emph{\JournalTitle{arXiv preprint arXiv:2412.05833}}}  (\bibinfo{year}{2024}).

\bibitem{yu2024ct}
\bibinfo{author}{Yu, Y.} \emph{et~al.}
\newblock \bibinfo{journal}{\bibinfo{title}{Ct synthesis with conditional diffusion models for abdominal lymph node segmentation}}.
\newblock {\emph{\JournalTitle{arXiv preprint arXiv:2403.17770}}}  (\bibinfo{year}{2024}).

\bibitem{xiao2023end}
\bibinfo{author}{Xiao, Q.} \& \bibinfo{author}{Zhao, L.}
\newblock \bibinfo{journal}{\bibinfo{title}{End-to-end 3d liver ct image synthesis from vasculature using a multi-task conditional generative adversarial network}}.
\newblock {\emph{\JournalTitle{Applied Sciences}}} \textbf{\bibinfo{volume}{13}}, \bibinfo{pages}{6784} (\bibinfo{year}{2023}).

\bibitem{weber2023cascaded}
\bibinfo{author}{Weber, T.}, \bibinfo{author}{Ingrisch, M.}, \bibinfo{author}{Bischl, B.} \& \bibinfo{author}{R{\"u}gamer, D.}
\newblock \bibinfo{title}{Cascaded latent diffusion models for high-resolution chest x-ray synthesis}.
\newblock In \emph{\bibinfo{booktitle}{Pacific-Asia Conference on Knowledge Discovery and Data Mining}}, \bibinfo{pages}{180--191} (\bibinfo{organization}{Springer}, \bibinfo{year}{2023}).

\bibitem{shentu2024cxr}
\bibinfo{author}{Shentu, J.} \& \bibinfo{author}{Al~Moubayed, N.}
\newblock \bibinfo{title}{Cxr-irgen: An integrated vision and language model for the generation of clinically accurate chest x-ray image-report pairs}.
\newblock In \emph{\bibinfo{booktitle}{Proceedings of the IEEE/CVF Winter Conference on Applications of Computer Vision}}, \bibinfo{pages}{5212--5221} (\bibinfo{year}{2024}).

\bibitem{borghesi2024generation}
\bibinfo{author}{Borghesi, A.} \& \bibinfo{author}{Calegari, R.}
\newblock \bibinfo{title}{Generation of clinical skin images with pathology with scarce data}.
\newblock In \emph{\bibinfo{booktitle}{AI for Health Equity and Fairness: Leveraging AI to Address Social Determinants of Health}}, \bibinfo{pages}{47--64} (\bibinfo{publisher}{Springer}, \bibinfo{year}{2024}).

\bibitem{fang2024conditional}
\bibinfo{author}{Fang, Z.} \emph{et~al.}
\newblock \bibinfo{journal}{\bibinfo{title}{Conditional diffusion model for x-ray segmentation data generation}}.
\newblock {\emph{\JournalTitle{Journal of Artificial Intelligence Practice}}} \textbf{\bibinfo{volume}{7}}, \bibinfo{pages}{7--10} (\bibinfo{year}{2024}).

\bibitem{sagers2022improving}
\bibinfo{author}{Sagers, L.~W.} \emph{et~al.}
\newblock \bibinfo{journal}{\bibinfo{title}{Improving dermatology classifiers across populations using images generated by large diffusion models}}.
\newblock {\emph{\JournalTitle{arXiv preprint arXiv:2211.13352}}}  (\bibinfo{year}{2022}).

\bibitem{wang2025toward}
\bibinfo{author}{Wang, Y.} \emph{et~al.}
\newblock \bibinfo{journal}{\bibinfo{title}{Toward general text-guided multimodal brain mri synthesis for diagnosis and medical image analysis}}.
\newblock {\emph{\JournalTitle{Cell Reports Medicine}}}  (\bibinfo{year}{2025}).

\bibitem{li2025interactive}
\bibinfo{author}{Li, L.} \emph{et~al.}
\newblock \bibinfo{journal}{\bibinfo{title}{Interactive gadolinium-free mri synthesis: A transformer with localization prompt learning}}.
\newblock {\emph{\JournalTitle{arXiv preprint arXiv:2503.01265}}}  (\bibinfo{year}{2025}).

\bibitem{duan2025fetalflex}
\bibinfo{author}{Duan, Y.} \emph{et~al.}
\newblock \bibinfo{journal}{\bibinfo{title}{Fetalflex: Anatomy-guided diffusion model for flexible control on fetal ultrasound image synthesis}}.
\newblock {\emph{\JournalTitle{arXiv preprint arXiv:2503.14906}}}  (\bibinfo{year}{2025}).

\bibitem{liu2025treatment}
\bibinfo{author}{Liu, Q.} \emph{et~al.}
\newblock \bibinfo{journal}{\bibinfo{title}{Treatment-aware diffusion probabilistic model for longitudinal mri generation and diffuse glioma growth prediction}}.
\newblock {\emph{\JournalTitle{IEEE Transactions on Medical Imaging}}}  (\bibinfo{year}{2025}).

\bibitem{wu2024one}
\bibinfo{author}{Wu, J.} \& \bibinfo{author}{Xu, M.}
\newblock \bibinfo{title}{One-prompt to segment all medical images}.
\newblock In \emph{\bibinfo{booktitle}{Proceedings of the IEEE/CVF Conference on Computer Vision and Pattern Recognition}}, \bibinfo{pages}{11302--11312} (\bibinfo{year}{2024}).

\bibitem{chang2023pe}
\bibinfo{author}{Chang, A.} \emph{et~al.}
\newblock \bibinfo{title}{Pe-med: Prompt enhancement for interactive medical image segmentation}.
\newblock In \emph{\bibinfo{booktitle}{International Workshop on Machine Learning in Medical Imaging}}, \bibinfo{pages}{257--266} (\bibinfo{organization}{Springer}, \bibinfo{year}{2023}).

\bibitem{bai2023slpt}
\bibinfo{author}{Bai, F.} \emph{et~al.}
\newblock \bibinfo{title}{Slpt: Selective labeling meets prompt tuning on label-limited lesion segmentation}.
\newblock In \emph{\bibinfo{booktitle}{International Conference on Medical Image Computing and Computer-Assisted Intervention}}, \bibinfo{pages}{14--24} (\bibinfo{organization}{Springer}, \bibinfo{year}{2023}).

\bibitem{wu2024efficient}
\bibinfo{author}{Wu, C.}, \bibinfo{author}{Restrepo, D.}, \bibinfo{author}{Shuai, Z.}, \bibinfo{author}{Liu, Z.} \& \bibinfo{author}{Shen, L.}
\newblock \bibinfo{title}{Efficient in-context medical segmentation with meta-driven visual prompt selection}.
\newblock In \emph{\bibinfo{booktitle}{International Conference on Medical Image Computing and Computer-Assisted Intervention}}, \bibinfo{pages}{255--265} (\bibinfo{organization}{Springer}, \bibinfo{year}{2024}).

\bibitem{xu2023eviprompt}
\bibinfo{author}{Xu, Y.}, \bibinfo{author}{Tang, J.}, \bibinfo{author}{Men, A.} \& \bibinfo{author}{Chen, Q.}
\newblock \bibinfo{journal}{\bibinfo{title}{Eviprompt: A training-free evidential prompt generation method for segment anything model in medical images}}.
\newblock {\emph{\JournalTitle{arXiv preprint arXiv:2311.06400}}}  (\bibinfo{year}{2023}).

\bibitem{xie2024masksam}
\bibinfo{author}{Xie, B.}, \bibinfo{author}{Tang, H.}, \bibinfo{author}{Duan, B.}, \bibinfo{author}{Cai, D.} \& \bibinfo{author}{Yan, Y.}
\newblock \bibinfo{journal}{\bibinfo{title}{Masksam: Towards auto-prompt sam with mask classification for medical image segmentation}}.
\newblock {\emph{\JournalTitle{arXiv preprint arXiv:2403.14103}}}  (\bibinfo{year}{2024}).

\bibitem{kato2023one}
\bibinfo{author}{Kato, S.} \& \bibinfo{author}{Hotta, K.}
\newblock \bibinfo{title}{One-shot and partially-supervised cell image segmentation using small visual prompt}.
\newblock In \emph{\bibinfo{booktitle}{Proceedings of the IEEE/CVF Conference on Computer Vision and Pattern Recognition}}, \bibinfo{pages}{4295--4304} (\bibinfo{year}{2023}).

\bibitem{zhang2023continual}
\bibinfo{author}{Zhang, Y.} \emph{et~al.}
\newblock \bibinfo{title}{Continual learning for abdominal multi-organ and tumor segmentation}.
\newblock In \emph{\bibinfo{booktitle}{International conference on medical image computing and computer-assisted intervention}}, \bibinfo{pages}{35--45} (\bibinfo{organization}{Springer}, \bibinfo{year}{2023}).

\bibitem{tomar2022tganet}
\bibinfo{author}{Tomar, N.~K.}, \bibinfo{author}{Jha, D.}, \bibinfo{author}{Bagci, U.} \& \bibinfo{author}{Ali, S.}
\newblock \bibinfo{title}{Tganet: Text-guided attention for improved polyp segmentation}.
\newblock In \emph{\bibinfo{booktitle}{International Conference on Medical Image Computing and Computer-Assisted Intervention}}, \bibinfo{pages}{151--160} (\bibinfo{organization}{Springer}, \bibinfo{year}{2022}).

\bibitem{zhao2023one}
\bibinfo{author}{Zhao, Z.} \emph{et~al.}
\newblock \bibinfo{journal}{\bibinfo{title}{One model to rule them all: Towards universal segmentation for medical images with text prompts}}.
\newblock {\emph{\JournalTitle{arXiv preprint arXiv:2312.17183}}}  (\bibinfo{year}{2023}).

\bibitem{biswas2023polyp}
\bibinfo{author}{Biswas, R.}
\newblock \bibinfo{journal}{\bibinfo{title}{Polyp-sam++: Can a text guided sam perform better for polyp segmentation?}}
\newblock {\emph{\JournalTitle{arXiv preprint arXiv:2308.06623}}}  (\bibinfo{year}{2023}).

\bibitem{chenvp}
\bibinfo{author}{Chen, Y.}, \bibinfo{author}{Wang, Y.} \& \bibinfo{author}{Xie, Z.}
\newblock \bibinfo{journal}{\bibinfo{title}{Vp-sfda: Visual prompt source-free domain adaptation for cross-modal medical image}}.
\newblock {\emph{\JournalTitle{Health Data Science}}} .

\bibitem{han2023multiscale}
\bibinfo{author}{Han, X.}, \bibinfo{author}{Chen, Q.}, \bibinfo{author}{Xie, Z.}, \bibinfo{author}{Li, X.} \& \bibinfo{author}{Yang, H.}
\newblock \bibinfo{journal}{\bibinfo{title}{Multiscale progressive text prompt network for medical image segmentation}}.
\newblock {\emph{\JournalTitle{Computers \& Graphics}}} \textbf{\bibinfo{volume}{116}}, \bibinfo{pages}{262--274} (\bibinfo{year}{2023}).

\bibitem{saeed2023prompt}
\bibinfo{author}{Saeed, N.}, \bibinfo{author}{Ridzuan, M.}, \bibinfo{author}{Majzoub, R.~A.} \& \bibinfo{author}{Yaqub, M.}
\newblock \bibinfo{journal}{\bibinfo{title}{Prompt-based tuning of transformer models for multi-center medical image segmentation of head and neck cancer}}.
\newblock {\emph{\JournalTitle{Bioengineering}}} \textbf{\bibinfo{volume}{10}}, \bibinfo{pages}{879} (\bibinfo{year}{2023}).

\bibitem{li2023multi}
\bibinfo{author}{Li, X.}, \bibinfo{author}{Zhang, Y.} \& \bibinfo{author}{Zhao, L.}
\newblock \bibinfo{journal}{\bibinfo{title}{Multi-prompt fine-tuning of foundation models for enhanced medical image segmentation}}.
\newblock {\emph{\JournalTitle{arXiv preprint arXiv:2310.02381}}}  (\bibinfo{year}{2023}).

\bibitem{xieself}
\bibinfo{author}{Xie, B.}, \bibinfo{author}{Tang, H.}, \bibinfo{author}{Cai, D.}, \bibinfo{author}{Yan, Y.} \& \bibinfo{author}{Agam, G.}
\newblock \bibinfo{journal}{\bibinfo{title}{Self-prompt sam: Medical image segmentation via automatic prompt sam adaptation}}.
\newblock {\emph{\JournalTitle{arXiv preprint arXiv:2502.00630}}}  (\bibinfo{year}{2025}).

\bibitem{chen2023sppnet}
\bibinfo{author}{Xu, Q.} \emph{et~al.}
\newblock \bibinfo{title}{Sppnet: A single-point prompt network for nuclei image segmentation}.
\newblock In \bibinfo{editor}{Cao, X.}, \bibinfo{editor}{Xu, X.}, \bibinfo{editor}{Rekik, I.}, \bibinfo{editor}{Cui, Z.} \& \bibinfo{editor}{Ouyang, X.} (eds.) \emph{\bibinfo{booktitle}{Machine Learning in Medical Imaging}}, \bibinfo{pages}{227--236} (\bibinfo{publisher}{Springer Nature Switzerland}, \bibinfo{address}{Cham}, \bibinfo{year}{2024}).

\bibitem{sridhar2023lung}
\bibinfo{author}{Sridhar, R.~I.} \& \bibinfo{author}{Kamaleswaran, R.}
\newblock \bibinfo{journal}{\bibinfo{title}{Lung segment anything model (lusam): A prompt-integrated framework for automated lung segmentation on icu chest x-ray images}}.
\newblock {\emph{\JournalTitle{Authorea Preprints}}}  (\bibinfo{year}{2023}).

\bibitem{zhang2023box2pseudo}
\bibinfo{author}{Zhang, S.}, \bibinfo{author}{Yue, J.}, \bibinfo{author}{Wang, C.}, \bibinfo{author}{Liu, X.} \& \bibinfo{author}{Wang, G.}
\newblock \bibinfo{title}{Box2pseudo: A semi-supervised learning framework for pulmonary nodule segmentation with box-prompt pseudo supervision}.
\newblock In \emph{\bibinfo{booktitle}{2023 IEEE International Conference on Bioinformatics and Biomedicine (BIBM)}}, \bibinfo{pages}{1696--1703} (\bibinfo{organization}{IEEE}, \bibinfo{year}{2023}).

\bibitem{glatt2023topology}
\bibinfo{author}{Glatt, R.} \& \bibinfo{author}{Shusen, L.}
\newblock \bibinfo{title}{Topology data analysis guided prompt optimization of segment anything model for zero-shot segmentation of biological images}.
\newblock \bibinfo{type}{Tech. Rep.}, \bibinfo{institution}{Lawrence Livermore National Laboratory (LLNL), Livermore, CA (United States)} (\bibinfo{year}{2023}).

\bibitem{zhou2024specific}
\bibinfo{author}{Zhou, Q.}, \bibinfo{author}{Feng, Y.}, \bibinfo{author}{Huang, Z.}, \bibinfo{author}{Ding, M.} \& \bibinfo{author}{Zhang, X.}
\newblock \bibinfo{journal}{\bibinfo{title}{Specific instance and cross prompt based robust 3d semi-supervised medical image segmentation}}.
\newblock {\emph{\JournalTitle{IEEE Transactions on Instrumentation and Measurement}}}  (\bibinfo{year}{2024}).

\bibitem{huang2024robust}
\bibinfo{author}{Huang, Y.} \emph{et~al.}
\newblock \bibinfo{title}{Robust box prompt based sam for medical image segmentation}.
\newblock In \emph{\bibinfo{booktitle}{International Workshop on Machine Learning in Medical Imaging}}, \bibinfo{pages}{1--11} (\bibinfo{organization}{Springer}, \bibinfo{year}{2024}).

\bibitem{ouyang2024prompt}
\bibinfo{author}{Ouyang, X.} \emph{et~al.}
\newblock \bibinfo{title}{Prompt-based segmentation model of anatomical structures and lesions in ct images}.
\newblock In \emph{\bibinfo{booktitle}{International Conference on Medical Image Computing and Computer-Assisted Intervention}}, \bibinfo{pages}{522--532} (\bibinfo{organization}{Springer}, \bibinfo{year}{2024}).

\bibitem{chen2024segmentation}
\bibinfo{author}{Chen, Y.} \emph{et~al.}
\newblock \bibinfo{title}{Segmentation by registration-enabled sam prompt engineering using five reference images}.
\newblock In \emph{\bibinfo{booktitle}{International Workshop on Biomedical Image Registration}}, \bibinfo{pages}{241--252} (\bibinfo{organization}{Springer}, \bibinfo{year}{2024}).

\bibitem{shaharabany2024zero}
\bibinfo{author}{Shaharabany, T.} \& \bibinfo{author}{Wolf, L.}
\newblock \bibinfo{title}{Zero-shot medical image segmentation based on sparse prompt using finetuned sam}.
\newblock In \emph{\bibinfo{booktitle}{Medical Imaging with Deep Learning}} (\bibinfo{year}{2024}).

\bibitem{wang2024tp}
\bibinfo{author}{Wang, R.}, \bibinfo{author}{Zhuang, L.}, \bibinfo{author}{Chen, H.}, \bibinfo{author}{Xu, B.} \& \bibinfo{author}{Cai, R.}
\newblock \bibinfo{journal}{\bibinfo{title}{Tp-unet: Temporal prompt guided unet for medical image segmentation}}.
\newblock {\emph{\JournalTitle{arXiv preprint arXiv:2411.11305}}}  (\bibinfo{year}{2024}).

\bibitem{adhikari2024tunevlseg}
\bibinfo{author}{Adhikari, R.}, \bibinfo{author}{Thapaliya, S.}, \bibinfo{author}{Dhakal, M.} \& \bibinfo{author}{Khanal, B.}
\newblock \bibinfo{title}{Tunevlseg: Prompt tuning benchmark for vision-language segmentation models}.
\newblock In \emph{\bibinfo{booktitle}{Proceedings of the Asian Conference on Computer Vision}}, \bibinfo{pages}{126--144} (\bibinfo{year}{2024}).

\bibitem{kong2024swiftmedsam}
\bibinfo{author}{Kong, Y.}, \bibinfo{author}{Kim, K.}, \bibinfo{author}{Jeong, S.}, \bibinfo{author}{Lee, K.~E.} \& \bibinfo{author}{Kong, H.}
\newblock \bibinfo{title}{Swiftmedsam: An ultra-lightweight prompt-based universal medical image segmentation model for highly constrained environments}.
\newblock In \emph{\bibinfo{booktitle}{CVPR 2024: Segment Anything In Medical Images On Laptop}}.

\bibitem{liu2024feature}
\bibinfo{author}{Liu, X.} \emph{et~al.}
\newblock \bibinfo{title}{Feature-prompting gbmseg: One-shot reference guided training-free prompt engineering for glomerular basement membrane segmentation}.
\newblock In \emph{\bibinfo{booktitle}{International Conference on Medical Image Computing and Computer-Assisted Intervention}}, \bibinfo{pages}{276--285} (\bibinfo{organization}{Springer}, \bibinfo{year}{2024}).

\bibitem{chen2024multi}
\bibinfo{author}{Chen, H.} \emph{et~al.}
\newblock \bibinfo{journal}{\bibinfo{title}{Multi-organ foundation model for universal ultrasound image segmentation with task prompt and anatomical prior}}.
\newblock {\emph{\JournalTitle{IEEE Transactions on Medical Imaging}}}  (\bibinfo{year}{2024}).

\bibitem{cuienhancing}
\bibinfo{author}{Cui, C.} \emph{et~al.}
\newblock \bibinfo{title}{Enhancing physician flexibility: Prompt-guided multi-class pathological segmentation for diverse outcomes}.
\newblock In \emph{\bibinfo{booktitle}{IEEE-EMBS International Conference on Biomedical and Health Informatics}}.

\bibitem{xie2024promamba}
\bibinfo{author}{Xie, J.} \emph{et~al.}
\newblock \bibinfo{journal}{\bibinfo{title}{Promamba: Prompt-mamba for polyp segmentation}}.
\newblock {\emph{\JournalTitle{arXiv preprint arXiv:2403.13660}}}  (\bibinfo{year}{2024}).

\bibitem{xia2024cervical}
\bibinfo{author}{Xia, Y.} \emph{et~al.}
\newblock \bibinfo{journal}{\bibinfo{title}{Cervical-yosa: Utilizing prompt engineering and pre-trained large-scale models for automated segmentation of multi-sequence mri images in cervical cancer}}.
\newblock {\emph{\JournalTitle{IET Image Processing}}} \textbf{\bibinfo{volume}{18}}, \bibinfo{pages}{3556--3569} (\bibinfo{year}{2024}).

\bibitem{teng2024knowledge}
\bibinfo{author}{Teng, L.} \emph{et~al.}
\newblock \bibinfo{title}{Knowledge-guided prompt learning for lifespan brain mr image segmentation}.
\newblock In \emph{\bibinfo{booktitle}{International Conference on Medical Image Computing and Computer-Assisted Intervention}}, \bibinfo{pages}{238--248} (\bibinfo{organization}{Springer}, \bibinfo{year}{2024}).

\bibitem{chen2024adaptation}
\bibinfo{author}{Chen, Z.} \emph{et~al.}
\newblock \bibinfo{journal}{\bibinfo{title}{Adaptation of prompt-enabled segment-anything-model enhance the accuracy and generalizability of cine cardiac magnetic resonance segmentation}}.
\newblock {\emph{\JournalTitle{Circulation}}} \textbf{\bibinfo{volume}{150}}, \bibinfo{pages}{A4143921--A4143921} (\bibinfo{year}{2024}).

\bibitem{guanlite}
\bibinfo{author}{Guan, H.}, \bibinfo{author}{Dai, B.} \& \bibinfo{author}{Zhang, J.}
\newblock \bibinfo{title}{Lite class-prompt tiny-vit for multi-modality medical image segmentation}.
\newblock In \bibinfo{editor}{Ma, J.}, \bibinfo{editor}{Zhou, Y.} \& \bibinfo{editor}{Wang, B.} (eds.) \emph{\bibinfo{booktitle}{Medical Image Segmentation Foundation Models. CVPR 2024 Challenge: Segment Anything in Medical Images on Laptop}}, \bibinfo{pages}{151--166} (\bibinfo{publisher}{Springer Nature Switzerland}, \bibinfo{address}{Cham}, \bibinfo{year}{2025}).

\bibitem{song2024ep}
\bibinfo{author}{Song, J.}, \bibinfo{author}{Yun, S.}, \bibinfo{author}{Yoon, S.}, \bibinfo{author}{Kim, J.} \& \bibinfo{author}{Lee, S.}
\newblock \bibinfo{journal}{\bibinfo{title}{Ep-sam: Weakly supervised histopathology segmentation via enhanced prompt with segment anything}}.
\newblock {\emph{\JournalTitle{arXiv preprint arXiv:2410.13621}}}  (\bibinfo{year}{2024}).

\bibitem{khor2024unified}
\bibinfo{author}{Khor, H.~G.} \emph{et~al.}
\newblock \bibinfo{title}{Unified prompt-visual interactive segmentation of clinical target volume in ct for nasopharyngeal carcinoma with prior anatomical information}.
\newblock In \emph{\bibinfo{booktitle}{International Conference on Medical Image Computing and Computer-Assisted Intervention}}, \bibinfo{pages}{659--669} (\bibinfo{organization}{Springer}, \bibinfo{year}{2024}).

\bibitem{xue2024deep}
\bibinfo{author}{Xue, X.} \emph{et~al.}
\newblock \bibinfo{journal}{\bibinfo{title}{Deep learning-based segmentation for high-dose-rate brachytherapy in cervical cancer using 3d prompt-resunet}}.
\newblock {\emph{\JournalTitle{Physics in Medicine \& Biology}}} \textbf{\bibinfo{volume}{69}}, \bibinfo{pages}{195008} (\bibinfo{year}{2024}).

\bibitem{cui2024all}
\bibinfo{author}{Cui, C.} \emph{et~al.}
\newblock \bibinfo{title}{All-in-sam: from weak annotation to pixel-wise nuclei segmentation with prompt-based finetuning}.
\newblock In \emph{\bibinfo{booktitle}{Journal of Physics: Conference Series}}, vol. \bibinfo{volume}{2722}, \bibinfo{pages}{012012} (\bibinfo{organization}{IOP Publishing}, \bibinfo{year}{2024}).

\bibitem{lyu2024superpixel}
\bibinfo{author}{Lyu, F.}, \bibinfo{author}{Xu, J.}, \bibinfo{author}{Zhu, Y.}, \bibinfo{author}{Wong, G. L.-H.} \& \bibinfo{author}{Yuen, P.~C.}
\newblock \bibinfo{title}{Superpixel-guided segment anything model for liver tumor segmentation with couinaud segment prompt}.
\newblock In \emph{\bibinfo{booktitle}{International Conference on Medical Image Computing and Computer-Assisted Intervention}}, \bibinfo{pages}{678--688} (\bibinfo{organization}{Springer}, \bibinfo{year}{2024}).

\bibitem{yang2024tavp}
\bibinfo{author}{Yang, J.}, \bibinfo{author}{Huang, Y.}, \bibinfo{author}{He, X.}, \bibinfo{author}{Shen, L.} \& \bibinfo{author}{Qiu, G.}
\newblock \bibinfo{journal}{\bibinfo{title}{Tavp: Task-adaptive visual prompt for cross-domain few-shot segmentation}}.
\newblock {\emph{\JournalTitle{arXiv preprint arXiv:2409.05393}}}  (\bibinfo{year}{2024}).

\bibitem{xue2024deep2}
\bibinfo{author}{Xue, X.} \emph{et~al.}
\newblock \bibinfo{journal}{\bibinfo{title}{A deep learning-based 3d prompt-nnunet model for automatic segmentation in brachytherapy of postoperative endometrial carcinoma}}.
\newblock {\emph{\JournalTitle{Journal of Applied Clinical Medical Physics}}} \bibinfo{pages}{e14371} (\bibinfo{year}{2024}).

\bibitem{dai2024sparse}
\bibinfo{author}{Dai, P.}, \bibinfo{author}{Ou, Y.}, \bibinfo{author}{Yang, Y.}, \bibinfo{author}{Liu, Y.} \& \bibinfo{author}{Zhao, Y.}
\newblock \bibinfo{title}{Sparse anatomical prompt semi-supervised learning with masked image modeling for cbct tooth segmentation}.
\newblock In \emph{\bibinfo{booktitle}{2024 IEEE International Symposium on Biomedical Imaging (ISBI)}}, \bibinfo{pages}{1--5} (\bibinfo{organization}{IEEE}, \bibinfo{year}{2024}).

\bibitem{cui2024pfps}
\bibinfo{author}{Cui, C.} \emph{et~al.}
\newblock \bibinfo{journal}{\bibinfo{title}{Pfps: Prompt-guided flexible pathological segmentation for diverse potential outcomes using large vision and language models}}.
\newblock {\emph{\JournalTitle{arXiv preprint arXiv:2407.09979}}}  (\bibinfo{year}{2024}).

\bibitem{hu2024lpam}
\bibinfo{author}{Hu, K.} \& \bibinfo{author}{Xu, C.}
\newblock \bibinfo{journal}{\bibinfo{title}{Lpam: A lightweight medical segmentation network based on mamba improved by prompt attention}}.
\newblock {\emph{\JournalTitle{IET Image Processing}}} \textbf{\bibinfo{volume}{18}}, \bibinfo{pages}{3545--3555} (\bibinfo{year}{2024}).

\bibitem{sun2024aepl}
\bibinfo{author}{Sun, Y.}, \bibinfo{author}{Liu, M.} \& \bibinfo{author}{Lian, C.}
\newblock \bibinfo{journal}{\bibinfo{title}{Aepl: Automated and editable prompt learning for brain tumor segmentation}}.
\newblock {\emph{\JournalTitle{arXiv preprint arXiv:2410.19847}}}  (\bibinfo{year}{2024}).

\bibitem{song2024automatic}
\bibinfo{author}{Song, Y.}, \bibinfo{author}{Zhang, Y.} \& \bibinfo{author}{Li, M.}
\newblock \bibinfo{journal}{\bibinfo{title}{An automatic laryngoscopic image segmentation system based on sam prompt engineering: From glottis annotation to vocal fold segmentation}}.
\newblock {\emph{\JournalTitle{Authorea Preprints}}}  (\bibinfo{year}{2024}).

\bibitem{cheng2024frequency}
\bibinfo{author}{Cheng, Y.} \& \bibinfo{author}{Zheng, Y.}
\newblock \bibinfo{journal}{\bibinfo{title}{Frequency filtering prompt tuning for medical image semantic segmentation with missing modalities}}.
\newblock {\emph{\JournalTitle{Big Data and Information Analytics}}} \textbf{\bibinfo{volume}{8}}, \bibinfo{pages}{109--128} (\bibinfo{year}{2024}).

\bibitem{li2024centersam}
\bibinfo{author}{Li, Y.}, \bibinfo{author}{Ren, H.}, \bibinfo{author}{Deng, J.}, \bibinfo{author}{Ma, X.} \& \bibinfo{author}{Xie, X.}
\newblock \bibinfo{title}{Centersam: Fully automatic prompt for dense nucleus segmentation}.
\newblock In \emph{\bibinfo{booktitle}{2024 IEEE International Symposium on Biomedical Imaging (ISBI)}}, \bibinfo{pages}{1--5} (\bibinfo{organization}{IEEE}, \bibinfo{year}{2024}).

\bibitem{zhang2024progressive}
\bibinfo{author}{Zhang, Q.}, \bibinfo{author}{Guo, H.}, \bibinfo{author}{Yang, S.}, \bibinfo{author}{Li, Q.} \& \bibinfo{author}{Wang, Y.}
\newblock \bibinfo{journal}{\bibinfo{title}{Progressive vision-language prompt for multi-organ multi-class cell semantic segmentation with single branch}}.
\newblock {\emph{\JournalTitle{arXiv preprint arXiv:2412.02978}}}  (\bibinfo{year}{2024}).

\bibitem{huang2024iossam}
\bibinfo{author}{Huang, X.}, \bibinfo{author}{He, D.}, \bibinfo{author}{Li, Z.}, \bibinfo{author}{Zhang, X.} \& \bibinfo{author}{Wang, X.}
\newblock \bibinfo{title}{Iossam: Label efficient multi-view prompt-driven tooth segmentation}.
\newblock In \emph{\bibinfo{booktitle}{International Conference on Medical Image Computing and Computer-Assisted Intervention}}, \bibinfo{pages}{632--642} (\bibinfo{organization}{Springer}, \bibinfo{year}{2024}).

\bibitem{li2024btsspro}
\bibinfo{author}{Li, W.} \emph{et~al.}
\newblock \bibinfo{journal}{\bibinfo{title}{Btsspro: Prompt-guided multimodal co-learning for breast cancer tumor segmentation and survival prediction}}.
\newblock {\emph{\JournalTitle{IEEE Journal of Biomedical and Health Informatics}}}  (\bibinfo{year}{2024}).

\bibitem{shan2025stpnet}
\bibinfo{author}{Shan, D.} \emph{et~al.}
\newblock \bibinfo{journal}{\bibinfo{title}{Stpnet: Scale-aware text prompt network for medical image segmentation}}.
\newblock {\emph{\JournalTitle{IEEE Transactions on Image Processing}}}  (\bibinfo{year}{2025}).

\bibitem{wang2025weakmedsam}
\bibinfo{author}{Wang, H.} \emph{et~al.}
\newblock \bibinfo{journal}{\bibinfo{title}{Weakmedsam: Weakly-supervised medical image segmentation via sam with sub-class exploration and prompt affinity mining}}.
\newblock {\emph{\JournalTitle{IEEE Transactions on Medical Imaging}}}  (\bibinfo{year}{2025}).

\bibitem{yin2025apg}
\bibinfo{author}{Yin, D.}, \bibinfo{author}{Zheng, Q.}, \bibinfo{author}{Chen, L.}, \bibinfo{author}{Hu, Y.} \& \bibinfo{author}{Wang, Q.}
\newblock \bibinfo{journal}{\bibinfo{title}{Apg-sam: Automatic prompt generation for sam-based breast lesion segmentation with boundary-aware optimization}}.
\newblock {\emph{\JournalTitle{Expert Systems with Applications}}} \textbf{\bibinfo{volume}{276}}, \bibinfo{pages}{127048} (\bibinfo{year}{2025}).

\bibitem{liu2025efficient}
\bibinfo{author}{Liu, S.}, \bibinfo{author}{Zhang, D.} \& \bibinfo{author}{Hao, X.}
\newblock \bibinfo{title}{Efficient deformable convolutional prompt for continual test-time adaptation in medical image segmentation}.
\newblock In \emph{\bibinfo{booktitle}{Proceedings of the AAAI Conference on Artificial Intelligence}}, vol.~\bibinfo{volume}{39}, \bibinfo{pages}{5550--5557} (\bibinfo{year}{2025}).

\bibitem{yin2025ddfp}
\bibinfo{author}{Yin, S.}, \bibinfo{author}{Liu, S.} \& \bibinfo{author}{Wang, M.}
\newblock \bibinfo{journal}{\bibinfo{title}{Ddfp: Data-dependent frequency prompt for source free domain adaptation of medical image segmentation}}.
\newblock {\emph{\JournalTitle{Knowledge-Based Systems}}} \bibinfo{pages}{113651} (\bibinfo{year}{2025}).

\bibitem{gao2025dual}
\bibinfo{author}{Gao, Y.} \emph{et~al.}
\newblock \bibinfo{journal}{\bibinfo{title}{Dual-prompt-enhanced multiorgan segmentation model for total-body pet images}}.
\newblock {\emph{\JournalTitle{IEEE Transactions on Radiation and Plasma Medical Sciences}}}  (\bibinfo{year}{2025}).

\bibitem{tian2025self}
\bibinfo{author}{Tian, C.} \emph{et~al.}
\newblock \bibinfo{journal}{\bibinfo{title}{Self-prompt contextual learning with axialmamba for multi-label segmentation in carotid ultrasound}}.
\newblock {\emph{\JournalTitle{Expert Systems with Applications}}} \textbf{\bibinfo{volume}{274}}, \bibinfo{pages}{126749} (\bibinfo{year}{2025}).

\bibitem{zou2025acea}
\bibinfo{author}{Zou, J.} \emph{et~al.}
\newblock \bibinfo{journal}{\bibinfo{title}{Acea-net: Weakly supervised prostate 3d mri image segmentation via advanced prompt points}}.
\newblock {\emph{\JournalTitle{IEEE Journal of Biomedical and Health Informatics}}}  (\bibinfo{year}{2025}).

\bibitem{chen2025sam}
\bibinfo{author}{Chen, Z.}, \bibinfo{author}{Xu, Q.}, \bibinfo{author}{Liu, X.} \& \bibinfo{author}{Yuan, Y.}
\newblock \bibinfo{journal}{\bibinfo{title}{Un-sam: Domain-adaptive self-prompt segmentation for universal nuclei images}}.
\newblock {\emph{\JournalTitle{Medical Image Analysis}}} \bibinfo{pages}{103607} (\bibinfo{year}{2025}).

\bibitem{zhang2025category}
\bibinfo{author}{Zhang, Y.} \emph{et~al.}
\newblock \bibinfo{title}{Category prompt mamba network for nuclei segmentation and classification}.
\newblock In \emph{\bibinfo{booktitle}{Proceedings of the AAAI Conference on Artificial Intelligence}}, vol.~\bibinfo{volume}{39}, \bibinfo{pages}{10284--10292} (\bibinfo{year}{2025}).

\bibitem{zhao2025uncertainty}
\bibinfo{author}{Zhao, J.} \emph{et~al.}
\newblock \bibinfo{journal}{\bibinfo{title}{Uncertainty-driven edge prompt generation network for medical image segmentation}}.
\newblock {\emph{\JournalTitle{IEEE Transactions on Medical Imaging}}}  (\bibinfo{year}{2025}).

\bibitem{guo2023multiple}
\bibinfo{author}{Guo, M.} \emph{et~al.}
\newblock \bibinfo{title}{Multiple prompt fusion for zero-shot lesion detection using vision-language models}.
\newblock In \emph{\bibinfo{booktitle}{International Conference on Medical Image Computing and Computer-Assisted Intervention}}, \bibinfo{pages}{283--292} (\bibinfo{organization}{Springer}, \bibinfo{year}{2023}).

\bibitem{cao2023domain}
\bibinfo{author}{Cao, Q.}, \bibinfo{author}{Xu, Z.}, \bibinfo{author}{Chen, Y.}, \bibinfo{author}{Ma, C.} \& \bibinfo{author}{Yang, X.}
\newblock \bibinfo{journal}{\bibinfo{title}{Domain prompt learning with quaternion networks}}.
\newblock {\emph{\JournalTitle{arXiv preprint arXiv:2312.08878}}}  (\bibinfo{year}{2023}).

\bibitem{zheng2024exploring}
\bibinfo{author}{Zheng, F.} \emph{et~al.}
\newblock \bibinfo{journal}{\bibinfo{title}{Exploring low-resource medical image classification with weakly supervised prompt learning}}.
\newblock {\emph{\JournalTitle{Pattern Recognition}}} \textbf{\bibinfo{volume}{149}}, \bibinfo{pages}{110250} (\bibinfo{year}{2024}).

\bibitem{huang2024fine}
\bibinfo{author}{Huang, Y.}, \bibinfo{author}{Cheng, P.}, \bibinfo{author}{Tam, R.} \& \bibinfo{author}{Tang, X.}
\newblock \bibinfo{title}{Fine-grained prompt tuning: A parameter and memory efficient transfer learning method for high-resolution medical image classification}.
\newblock In \emph{\bibinfo{booktitle}{International Conference on Medical Image Computing and Computer-Assisted Intervention}}, \bibinfo{pages}{120--130} (\bibinfo{organization}{Springer}, \bibinfo{year}{2024}).

\bibitem{yang2024using}
\bibinfo{author}{Yang, L.} \& \bibinfo{author}{Qu, W.}
\newblock \bibinfo{title}{Using text-augmented visual prompt learning for histopathology image classification}.
\newblock In \emph{\bibinfo{booktitle}{2024 5th International Conference on Big Data \& Artificial Intelligence \& Software Engineering (ICBASE)}}, \bibinfo{pages}{272--276} (\bibinfo{organization}{IEEE}, \bibinfo{year}{2024}).

\bibitem{bai2025label}
\bibinfo{author}{Bai, Y.}, \bibinfo{author}{Bai, L.}, \bibinfo{author}{Yang, X.} \& \bibinfo{author}{Liang, J.}
\newblock \bibinfo{journal}{\bibinfo{title}{Label-semantic-based prompt tuning for vision transformer adaptation in medical image analysis}}.
\newblock {\emph{\JournalTitle{IEEE Transactions on Circuits and Systems for Video Technology}}}  (\bibinfo{year}{2025}).

\bibitem{koleilat2025biomedcoop}
\bibinfo{author}{Koleilat, T.}, \bibinfo{author}{Asgariandehkordi, H.}, \bibinfo{author}{Rivaz, H.} \& \bibinfo{author}{Xiao, Y.}
\newblock \bibinfo{title}{Biomedcoop: Learning to prompt for biomedical vision-language models}.
\newblock In \emph{\bibinfo{booktitle}{Proceedings of the Computer Vision and Pattern Recognition Conference}}, \bibinfo{pages}{14766--14776} (\bibinfo{year}{2025}).

\bibitem{he2025dvpt}
\bibinfo{author}{He, A.}, \bibinfo{author}{Wu, Y.}, \bibinfo{author}{Wang, Z.}, \bibinfo{author}{Li, T.} \& \bibinfo{author}{Fu, H.}
\newblock \bibinfo{journal}{\bibinfo{title}{Dvpt: Dynamic visual prompt tuning of large pre-trained models for medical image analysis}}.
\newblock {\emph{\JournalTitle{Neural Networks}}} \textbf{\bibinfo{volume}{185}}, \bibinfo{pages}{107168} (\bibinfo{year}{2025}).

\bibitem{luo2025llm}
\bibinfo{author}{Luo, Y.} \emph{et~al.}
\newblock \bibinfo{journal}{\bibinfo{title}{Llm-guided decoupled probabilistic prompt for continual learning in medical image diagnosis}}.
\newblock {\emph{\JournalTitle{IEEE Transactions on Medical Imaging}}}  (\bibinfo{year}{2025}).

\bibitem{shin2018medical}
\bibinfo{author}{Shin, H.-C.} \emph{et~al.}
\newblock \bibinfo{title}{Medical image synthesis for data augmentation and anonymization using generative adversarial networks}.
\newblock In \emph{\bibinfo{booktitle}{Simulation and Synthesis in Medical Imaging: Third International Workshop, SASHIMI 2018, Held in Conjunction with MICCAI 2018, Granada, Spain, September 16, 2018, Proceedings 3}}, \bibinfo{pages}{1--11} (\bibinfo{organization}{Springer}, \bibinfo{year}{2018}).

\bibitem{li2023artificial}
\bibinfo{author}{Li, X.} \emph{et~al.}
\newblock \bibinfo{journal}{\bibinfo{title}{Artificial general intelligence for medical imaging}}.
\newblock {\emph{\JournalTitle{arXiv preprint arXiv:2306.05480}}}  (\bibinfo{year}{2023}).

\bibitem{rombach2022high}
\bibinfo{author}{Rombach, R.}, \bibinfo{author}{Blattmann, A.}, \bibinfo{author}{Lorenz, D.}, \bibinfo{author}{Esser, P.} \& \bibinfo{author}{Ommer, B.}
\newblock \bibinfo{title}{High-resolution image synthesis with latent diffusion models}.
\newblock In \emph{\bibinfo{booktitle}{Proceedings of the IEEE/CVF conference on computer vision and pattern recognition}}, \bibinfo{pages}{10684--10695} (\bibinfo{year}{2022}).

\bibitem{goodfellow2020generative}
\bibinfo{author}{Goodfellow, I.} \emph{et~al.}
\newblock \bibinfo{journal}{\bibinfo{title}{Generative adversarial networks}}.
\newblock {\emph{\JournalTitle{Communications of the ACM}}} \textbf{\bibinfo{volume}{63}}, \bibinfo{pages}{139--144} (\bibinfo{year}{2020}).

\bibitem{bowles2018gan}
\bibinfo{author}{Bowles, C.} \emph{et~al.}
\newblock \bibinfo{journal}{\bibinfo{title}{Gan augmentation: Augmenting training data using generative adversarial networks}}.
\newblock {\emph{\JournalTitle{arXiv preprint arXiv:1810.10863}}}  (\bibinfo{year}{2018}).

\bibitem{kwon2019generation}
\bibinfo{author}{Kwon, G.}, \bibinfo{author}{Han, C.} \& \bibinfo{author}{Kim, D.-s.}
\newblock \bibinfo{title}{Generation of 3d brain mri using auto-encoding generative adversarial networks}.
\newblock In \emph{\bibinfo{booktitle}{International Conference on Medical Image Computing and Computer-Assisted Intervention}}, \bibinfo{pages}{118--126} (\bibinfo{organization}{Springer}, \bibinfo{year}{2019}).

\bibitem{sun2022hierarchical}
\bibinfo{author}{Sun, L.} \emph{et~al.}
\newblock \bibinfo{journal}{\bibinfo{title}{Hierarchical amortized gan for 3d high resolution medical image synthesis}}.
\newblock {\emph{\JournalTitle{IEEE journal of biomedical and health informatics}}} \textbf{\bibinfo{volume}{26}}, \bibinfo{pages}{3966--3975} (\bibinfo{year}{2022}).

\bibitem{radford2021learning}
\bibinfo{author}{Radford, A.} \emph{et~al.}
\newblock \bibinfo{title}{Learning transferable visual models from natural language supervision}.
\newblock In \emph{\bibinfo{booktitle}{International conference on machine learning}}, \bibinfo{pages}{8748--8763} (\bibinfo{organization}{PMLR}, \bibinfo{year}{2021}).

\bibitem{zhang2021dodnet}
\bibinfo{author}{Zhang, J.}, \bibinfo{author}{Xie, Y.}, \bibinfo{author}{Xia, Y.} \& \bibinfo{author}{Shen, C.}
\newblock \bibinfo{title}{Dodnet: Learning to segment multi-organ and tumors from multiple partially labeled datasets}.
\newblock In \emph{\bibinfo{booktitle}{Proceedings of the IEEE/CVF conference on computer vision and pattern recognition}}, \bibinfo{pages}{1195--1204} (\bibinfo{year}{2021}).

\bibitem{zhao2024tg}
\bibinfo{author}{Zhao, Y.} \emph{et~al.}
\newblock \bibinfo{journal}{\bibinfo{title}{Tg-lmm: Enhancing medical image segmentation accuracy through text-guided large multi-modal model}}.
\newblock {\emph{\JournalTitle{arXiv preprint arXiv:2409.03412}}}  (\bibinfo{year}{2024}).

\bibitem{boecking2022making}
\bibinfo{author}{Boecking, B.} \emph{et~al.}
\newblock \bibinfo{title}{Making the most of text semantics to improve biomedical vision--language processing}.
\newblock In \emph{\bibinfo{booktitle}{European conference on computer vision}}, \bibinfo{pages}{1--21} (\bibinfo{organization}{Springer}, \bibinfo{year}{2022}).

\bibitem{mazurowski2023segment}
\bibinfo{author}{Mazurowski, M.~A.} \emph{et~al.}
\newblock \bibinfo{journal}{\bibinfo{title}{Segment anything model for medical image analysis: an experimental study}}.
\newblock {\emph{\JournalTitle{Medical Image Analysis}}} \textbf{\bibinfo{volume}{89}}, \bibinfo{pages}{102918} (\bibinfo{year}{2023}).

\bibitem{hu2023sam}
\bibinfo{author}{Hu, C.} \& \bibinfo{author}{Li, X.}
\newblock \bibinfo{journal}{\bibinfo{title}{When sam meets medical images: An investigation of segment anything model (sam) on multi-phase liver tumor segmentation}}.
\newblock {\emph{\JournalTitle{arXiv preprint arXiv:2304.08506}}}  (\bibinfo{year}{2023}).

\bibitem{deng2023segment}
\bibinfo{author}{Deng, R.} \emph{et~al.}
\newblock \bibinfo{journal}{\bibinfo{title}{Segment anything model (sam) for digital pathology: Assess zero-shot segmentation on whole slide imaging}}.
\newblock {\emph{\JournalTitle{arXiv preprint arXiv:2304.04155}}}  (\bibinfo{year}{2023}).

\bibitem{roy2023sam}
\bibinfo{author}{Roy, S.} \emph{et~al.}
\newblock \bibinfo{journal}{\bibinfo{title}{Sam. md: Zero-shot medical image segmentation capabilities of the segment anything model}}.
\newblock {\emph{\JournalTitle{arXiv preprint arXiv:2304.05396}}}  (\bibinfo{year}{2023}).

\bibitem{cheng2023sam_}
\bibinfo{author}{Cheng, J.} \emph{et~al.}
\newblock \bibinfo{journal}{\bibinfo{title}{Sam-med2d}}.
\newblock {\emph{\JournalTitle{arXiv preprint arXiv:2308.16184}}}  (\bibinfo{year}{2023}).

\bibitem{wang2023sam}
\bibinfo{author}{Wang, H.} \emph{et~al.}
\newblock \bibinfo{journal}{\bibinfo{title}{Sam-med3d}}.
\newblock {\emph{\JournalTitle{arXiv preprint arXiv:2310.15161}}}  (\bibinfo{year}{2023}).

\bibitem{devlin2018bert}
\bibinfo{author}{Devlin, J.}
\newblock \bibinfo{journal}{\bibinfo{title}{Bert: Pre-training of deep bidirectional transformers for language understanding}}.
\newblock {\emph{\JournalTitle{arXiv preprint arXiv:1810.04805}}}  (\bibinfo{year}{2018}).

\bibitem{chen2022gscorecam}
\bibinfo{author}{Chen, P.}, \bibinfo{author}{Li, Q.}, \bibinfo{author}{Biaz, S.}, \bibinfo{author}{Bui, T.} \& \bibinfo{author}{Nguyen, A.}
\newblock \bibinfo{title}{gscorecam: What objects is clip looking at?}
\newblock In \emph{\bibinfo{booktitle}{Proceedings of the Asian Conference on Computer Vision}}, \bibinfo{pages}{1959--1975} (\bibinfo{year}{2022}).

\bibitem{hamamci2024foundation}
\bibinfo{author}{Hamamci, I.~E.} \emph{et~al.}
\newblock \bibinfo{journal}{\bibinfo{title}{A foundation model utilizing chest ct volumes and radiology reports for supervised-level zero-shot detection of abnormalities}}.
\newblock {\emph{\JournalTitle{CoRR}}}  (\bibinfo{year}{2024}).

\bibitem{lu2024radclip}
\bibinfo{author}{Lu, Z.}, \bibinfo{author}{Li, H.}, \bibinfo{author}{Parikh, N.~A.}, \bibinfo{author}{Dillman, J.~R.} \& \bibinfo{author}{He, L.}
\newblock \bibinfo{journal}{\bibinfo{title}{Radclip: Enhancing radiologic image analysis through contrastive language-image pre-training}}.
\newblock {\emph{\JournalTitle{arXiv preprint arXiv:2403.09948}}}  (\bibinfo{year}{2024}).

\bibitem{zhou2022learning}
\bibinfo{author}{Zhou, K.}, \bibinfo{author}{Yang, J.}, \bibinfo{author}{Loy, C.~C.} \& \bibinfo{author}{Liu, Z.}
\newblock \bibinfo{journal}{\bibinfo{title}{Learning to prompt for vision-language models}}.
\newblock {\emph{\JournalTitle{International Journal of Computer Vision}}} \textbf{\bibinfo{volume}{130}}, \bibinfo{pages}{2337--2348} (\bibinfo{year}{2022}).

\bibitem{bommasani2021opportunities}
\bibinfo{author}{Bommasani, R.} \emph{et~al.}
\newblock \bibinfo{journal}{\bibinfo{title}{On the opportunities and risks of foundation models}}.
\newblock {\emph{\JournalTitle{arXiv preprint arXiv:2108.07258}}}  (\bibinfo{year}{2021}).

\bibitem{moor2023foundation}
\bibinfo{author}{Moor, M.} \emph{et~al.}
\newblock \bibinfo{journal}{\bibinfo{title}{Foundation models for generalist medical artificial intelligence}}.
\newblock {\emph{\JournalTitle{Nature}}} \textbf{\bibinfo{volume}{616}}, \bibinfo{pages}{259--265} (\bibinfo{year}{2023}).

\bibitem{wang2024pathology}
\bibinfo{author}{Wang, X.} \emph{et~al.}
\newblock \bibinfo{journal}{\bibinfo{title}{A pathology foundation model for cancer diagnosis and prognosis prediction}}.
\newblock {\emph{\JournalTitle{Nature}}} \textbf{\bibinfo{volume}{634}}, \bibinfo{pages}{970--978} (\bibinfo{year}{2024}).

\bibitem{vorontsov2024foundation}
\bibinfo{author}{Vorontsov, E.} \emph{et~al.}
\newblock \bibinfo{journal}{\bibinfo{title}{A foundation model for clinical-grade computational pathology and rare cancers detection}}.
\newblock {\emph{\JournalTitle{Nature medicine}}} \bibinfo{pages}{1--12} (\bibinfo{year}{2024}).

\bibitem{xiang2025vision}
\bibinfo{author}{Xiang, J.} \emph{et~al.}
\newblock \bibinfo{journal}{\bibinfo{title}{A vision--language foundation model for precision oncology}}.
\newblock {\emph{\JournalTitle{Nature}}} \bibinfo{pages}{1--10} (\bibinfo{year}{2025}).

\bibitem{huang2023visual}
\bibinfo{author}{Huang, Z.}, \bibinfo{author}{Bianchi, F.}, \bibinfo{author}{Yuksekgonul, M.}, \bibinfo{author}{Montine, T.~J.} \& \bibinfo{author}{Zou, J.}
\newblock \bibinfo{journal}{\bibinfo{title}{A visual--language foundation model for pathology image analysis using medical twitter}}.
\newblock {\emph{\JournalTitle{Nature medicine}}} \textbf{\bibinfo{volume}{29}}, \bibinfo{pages}{2307--2316} (\bibinfo{year}{2023}).

\bibitem{tiu2022expert}
\bibinfo{author}{Tiu, E.} \emph{et~al.}
\newblock \bibinfo{journal}{\bibinfo{title}{Expert-level detection of pathologies from unannotated chest x-ray images via self-supervised learning}}.
\newblock {\emph{\JournalTitle{Nature Biomedical Engineering}}} \textbf{\bibinfo{volume}{6}}, \bibinfo{pages}{1399--1406} (\bibinfo{year}{2022}).

\bibitem{you2023cxr}
\bibinfo{author}{You, K.} \emph{et~al.}
\newblock \bibinfo{title}{Cxr-clip: Toward large scale chest x-ray language-image pre-training}.
\newblock In \emph{\bibinfo{booktitle}{International Conference on Medical Image Computing and Computer-Assisted Intervention}}, \bibinfo{pages}{101--111} (\bibinfo{organization}{Springer}, \bibinfo{year}{2023}).

\bibitem{dai2024unichest}
\bibinfo{author}{Dai, T.} \emph{et~al.}
\newblock \bibinfo{journal}{\bibinfo{title}{Unichest: Conquer-and-divide pre-training for multi-source chest x-ray classification}}.
\newblock {\emph{\JournalTitle{IEEE Transactions on Medical Imaging}}}  (\bibinfo{year}{2024}).

\bibitem{wu2023medklip}
\bibinfo{author}{Wu, C.}, \bibinfo{author}{Zhang, X.}, \bibinfo{author}{Zhang, Y.}, \bibinfo{author}{Wang, Y.} \& \bibinfo{author}{Xie, W.}
\newblock \bibinfo{title}{Medklip: Medical knowledge enhanced language-image pre-training for x-ray diagnosis}.
\newblock In \emph{\bibinfo{booktitle}{Proceedings of the IEEE/CVF International Conference on Computer Vision}}, \bibinfo{pages}{21372--21383} (\bibinfo{year}{2023}).

\bibitem{lei2023clip}
\bibinfo{author}{Lei, Y.}, \bibinfo{author}{Li, Z.}, \bibinfo{author}{Shen, Y.}, \bibinfo{author}{Zhang, J.} \& \bibinfo{author}{Shan, H.}
\newblock \bibinfo{title}{Clip-lung: Textual knowledge-guided lung nodule malignancy prediction}.
\newblock In \emph{\bibinfo{booktitle}{International Conference on Medical Image Computing and Computer-Assisted Intervention}}, \bibinfo{pages}{403--412} (\bibinfo{organization}{Springer}, \bibinfo{year}{2023}).

\bibitem{niu2023medical}
\bibinfo{author}{Niu, C.} \emph{et~al.}
\newblock \bibinfo{journal}{\bibinfo{title}{Medical multimodal-multitask foundation model for superior chest ct performance}}.
\newblock {\emph{\JournalTitle{arXiv preprint arXiv:2304.02649}}}  (\bibinfo{year}{2023}).

\bibitem{wu2023towards}
\bibinfo{author}{Wu, C.}, \bibinfo{author}{Zhang, X.}, \bibinfo{author}{Zhang, Y.}, \bibinfo{author}{Wang, Y.} \& \bibinfo{author}{Xie, W.}
\newblock \bibinfo{journal}{\bibinfo{title}{Towards generalist foundation model for radiology}}.
\newblock {\emph{\JournalTitle{arXiv preprint arXiv:2308.02463}}}  (\bibinfo{year}{2023}).

\bibitem{bai2024m3d}
\bibinfo{author}{Bai, F.}, \bibinfo{author}{Du, Y.}, \bibinfo{author}{Huang, T.}, \bibinfo{author}{Meng, M. Q.-H.} \& \bibinfo{author}{Zhao, B.}
\newblock \bibinfo{journal}{\bibinfo{title}{M3d: Advancing 3d medical image analysis with multi-modal large language models}}.
\newblock {\emph{\JournalTitle{arXiv preprint arXiv:2404.00578}}}  (\bibinfo{year}{2024}).

\bibitem{blankemeier2024merlin}
\bibinfo{author}{Blankemeier, L.} \emph{et~al.}
\newblock \bibinfo{journal}{\bibinfo{title}{Merlin: A vision language foundation model for 3d computed tomography}}.
\newblock {\emph{\JournalTitle{Research Square}}} \bibinfo{pages}{rs--3} (\bibinfo{year}{2024}).

\bibitem{zhou2023foundation}
\bibinfo{author}{Zhou, Y.} \emph{et~al.}
\newblock \bibinfo{journal}{\bibinfo{title}{A foundation model for generalizable disease detection from retinal images}}.
\newblock {\emph{\JournalTitle{Nature}}} \textbf{\bibinfo{volume}{622}}, \bibinfo{pages}{156--163} (\bibinfo{year}{2023}).

\bibitem{engelmann2024training}
\bibinfo{author}{Engelmann, J.} \& \bibinfo{author}{Bernabeu, M.~O.}
\newblock \bibinfo{journal}{\bibinfo{title}{Training a high-performance retinal foundation model with half-the-data and 400 times less compute}}.
\newblock {\emph{\JournalTitle{arXiv preprint arXiv:2405.00117}}}  (\bibinfo{year}{2024}).

\bibitem{men2023drstagenet}
\bibinfo{author}{Men, Y.} \emph{et~al.}
\newblock \bibinfo{journal}{\bibinfo{title}{Drstagenet: Deep learning for diabetic retinopathy staging from fundus images}}.
\newblock {\emph{\JournalTitle{arXiv preprint arXiv:2312.14891}}}  (\bibinfo{year}{2023}).

\bibitem{qiu2023visionfm}
\bibinfo{author}{Qiu, J.} \emph{et~al.}
\newblock \bibinfo{journal}{\bibinfo{title}{Visionfm: a multi-modal multi-task vision foundation model for generalist ophthalmic artificial intelligence}}.
\newblock {\emph{\JournalTitle{arXiv preprint arXiv:2310.04992}}}  (\bibinfo{year}{2023}).

\bibitem{shi2024eyefound}
\bibinfo{author}{Shi, D.} \emph{et~al.}
\newblock \bibinfo{journal}{\bibinfo{title}{Eyefound: A multimodal generalist foundation model for ophthalmic imaging}}.
\newblock {\emph{\JournalTitle{arXiv preprint arXiv:2405.11338}}}  (\bibinfo{year}{2024}).

\bibitem{zamzmi2020unified}
\bibinfo{author}{Zamzmi, G.}, \bibinfo{author}{Rajaraman, S.} \& \bibinfo{author}{Antani, S.}
\newblock \bibinfo{journal}{\bibinfo{title}{Unified representation learning for efficient medical image analysis}}.
\newblock {\emph{\JournalTitle{arXiv preprint arXiv:2006.11223}}}  (\bibinfo{year}{2020}).

\bibitem{han2023explainable}
\bibinfo{author}{Han, L.} \emph{et~al.}
\newblock \bibinfo{title}{An explainable deep framework: Towards task-specific fusion for multi-to-one mri synthesis}.
\newblock In \emph{\bibinfo{booktitle}{International Conference on Medical Image Computing and Computer-Assisted Intervention}}, \bibinfo{pages}{45--55} (\bibinfo{organization}{Springer}, \bibinfo{year}{2023}).

\bibitem{zhou2022conditional}
\bibinfo{author}{Zhou, K.}, \bibinfo{author}{Yang, J.}, \bibinfo{author}{Loy, C.~C.} \& \bibinfo{author}{Liu, Z.}
\newblock \bibinfo{title}{Conditional prompt learning for vision-language models}.
\newblock In \emph{\bibinfo{booktitle}{Proceedings of the IEEE/CVF conference on computer vision and pattern recognition}}, \bibinfo{pages}{16816--16825} (\bibinfo{year}{2022}).

\bibitem{jiang2017artificial}
\bibinfo{author}{Jiang, F.} \emph{et~al.}
\newblock \bibinfo{journal}{\bibinfo{title}{Artificial intelligence in healthcare: past, present and future}}.
\newblock {\emph{\JournalTitle{Stroke and vascular neurology}}} \textbf{\bibinfo{volume}{2}} (\bibinfo{year}{2017}).

\bibitem{sinha2020multi}
\bibinfo{author}{Sinha, A.} \& \bibinfo{author}{Dolz, J.}
\newblock \bibinfo{journal}{\bibinfo{title}{Multi-scale self-guided attention for medical image segmentation}}.
\newblock {\emph{\JournalTitle{IEEE journal of biomedical and health informatics}}} \textbf{\bibinfo{volume}{25}}, \bibinfo{pages}{121--130} (\bibinfo{year}{2020}).

\bibitem{le2025moma}
\bibinfo{author}{Le~Vuong, T.~T.} \& \bibinfo{author}{Kwak, J.~T.}
\newblock \bibinfo{journal}{\bibinfo{title}{Moma: momentum contrastive learning with multi-head attention-based knowledge distillation for histopathology image analysis}}.
\newblock {\emph{\JournalTitle{Medical Image Analysis}}} \textbf{\bibinfo{volume}{101}}, \bibinfo{pages}{103421} (\bibinfo{year}{2025}).

\bibitem{huang2025learnable}
\bibinfo{author}{Huang, K.} \emph{et~al.}
\newblock \bibinfo{journal}{\bibinfo{title}{Learnable prompting sam-induced knowledge distillation for semi-supervised medical image segmentation}}.
\newblock {\emph{\JournalTitle{IEEE Transactions on Medical Imaging}}}  (\bibinfo{year}{2025}).

\bibitem{xie2024mh}
\bibinfo{author}{Xie, L.} \emph{et~al.}
\newblock \bibinfo{journal}{\bibinfo{title}{Mh-pflid: Model heterogeneous personalized federated learning via injection and distillation for medical data analysis}}.
\newblock {\emph{\JournalTitle{arXiv preprint arXiv:2405.06822}}}  (\bibinfo{year}{2024}).

\bibitem{shi2023deep}
\bibinfo{author}{Shi, H.}, \bibinfo{author}{Ren, S.}, \bibinfo{author}{Zhang, T.} \& \bibinfo{author}{Pan, S.~J.}
\newblock \bibinfo{title}{Deep multitask learning with progressive parameter sharing}.
\newblock In \emph{\bibinfo{booktitle}{Proceedings of the IEEE/CVF International Conference on Computer Vision}}, \bibinfo{pages}{19924--19935} (\bibinfo{year}{2023}).

\bibitem{liu2023enhancing}
\bibinfo{author}{Liu, P.}, \bibinfo{author}{Gao, Z.-F.}, \bibinfo{author}{Chen, Y.}, \bibinfo{author}{Zhao, W.~X.} \& \bibinfo{author}{Wen, J.-R.}
\newblock \bibinfo{title}{Enhancing scalability of pre-trained language models via efficient parameter sharing}.
\newblock In \emph{\bibinfo{booktitle}{Findings of the Association for Computational Linguistics: EMNLP 2023}}, \bibinfo{pages}{13771--13785} (\bibinfo{year}{2023}).

\bibitem{qiu2024learning}
\bibinfo{author}{Qiu, Z.} \emph{et~al.}
\newblock \bibinfo{journal}{\bibinfo{title}{Learning co-plane attention across mri sequences for diagnosing twelve types of knee abnormalities}}.
\newblock {\emph{\JournalTitle{Nature Communications}}} \textbf{\bibinfo{volume}{15}}, \bibinfo{pages}{7637} (\bibinfo{year}{2024}).

\bibitem{kim2020changes}
\bibinfo{author}{Kim, H.-E.} \emph{et~al.}
\newblock \bibinfo{journal}{\bibinfo{title}{Changes in cancer detection and false-positive recall in mammography using artificial intelligence: a retrospective, multireader study}}.
\newblock {\emph{\JournalTitle{The Lancet Digital Health}}} \textbf{\bibinfo{volume}{2}}, \bibinfo{pages}{e138--e148} (\bibinfo{year}{2020}).

\bibitem{tschandl2020human}
\bibinfo{author}{Tschandl, P.} \emph{et~al.}
\newblock \bibinfo{journal}{\bibinfo{title}{Human--computer collaboration for skin cancer recognition}}.
\newblock {\emph{\JournalTitle{Nature medicine}}} \textbf{\bibinfo{volume}{26}}, \bibinfo{pages}{1229--1234} (\bibinfo{year}{2020}).

\bibitem{zhang2023mani}
\bibinfo{author}{Zhang, Z.}, \bibinfo{author}{Chai, W.} \& \bibinfo{author}{Wang, J.}
\newblock \bibinfo{journal}{\bibinfo{title}{Mani-gpt: A generative model for interactive robotic manipulation}}.
\newblock {\emph{\JournalTitle{Procedia Computer Science}}} \textbf{\bibinfo{volume}{226}}, \bibinfo{pages}{149--156} (\bibinfo{year}{2023}).

\bibitem{cao2023diaggpt}
\bibinfo{author}{Cao, L.}
\newblock \bibinfo{journal}{\bibinfo{title}{Diaggpt: An llm-based chatbot with automatic topic management for task-oriented dialogue}}.
\newblock {\emph{\JournalTitle{arXiv preprint arXiv:2308.08043}}}  (\bibinfo{year}{2023}).

\bibitem{shi2024general}
\bibinfo{author}{Shi, R.} \emph{et~al.}
\newblock \bibinfo{journal}{\bibinfo{title}{From general to specific: Tailoring large language models for personalized healthcare}}.
\newblock {\emph{\JournalTitle{arXiv preprint arXiv:2412.15957}}}  (\bibinfo{year}{2024}).

\bibitem{shenfeld2025language}
\bibinfo{author}{Shenfeld, I.}, \bibinfo{author}{Faltings, F.}, \bibinfo{author}{Agrawal, P.} \& \bibinfo{author}{Pacchiano, A.}
\newblock \bibinfo{journal}{\bibinfo{title}{Language model personalization via reward factorization}}.
\newblock {\emph{\JournalTitle{arXiv preprint arXiv:2503.06358}}}  (\bibinfo{year}{2025}).

\end{thebibliography}

\end{document}